# On the development of OpenFOAM solvers for simulating MHD micropolar fluid flows with or without the effect of micromagnetorotation

Kyriaki-Evangelia Aslani[1] *, Ioannis E. Sarris[2] and Efstratios Tzirtzilakis[3]

[1]*Department of Civil Engineering, University of the Peloponnese, 26334 Patras, Greece*
[2]*Department of Mechanical Engineering, University of West Attica, 12244 Athens, Greece*
[3]*Department of Civil Engineering, University of the Peloponnese, 26334 Patras, Greece*

* Corresponding author: k.aslani@go.uop.gr

**Abstract**: Any micropolar fluid containing magnetic particles, such as blood and ferrofluids, under the influence of an applied magnetic field experiences a magnetic torque resulting from the misalignment between the magnetization of these particles and the magnetic field, called micromagnetorotation (MMR). Although critical in such fluids, MMR remains underexplored in blood flows, where erythrocyte magnetization is often neglected. To address this, two transient OpenFOAM solvers were developed: epotMicropolarFoam, for incompressible, laminar MHD micropolar flows, and epotMMRFoam, which extends the former by incorporating MMR. In epotMicropolarFoam, the PISO algorithm is used for pressure-velocity coupling, while the low-magnetic-Reynolds-number approximation is adopted for the MHD phenomena simulation. Micropolar effects are included by incorporating the microrotation–vorticity difference in the momentum equation, and solving the internal angular momentum equation. EpotMMRFoam also uses the PISO algorithm and the low-magnetic-Reynolds-number approximation with the MMR term included in the internal angular momentum equation. In this solver, a constitutive magnetization equation is also solved. Validation against the analytical MHD micropolar Poiseuille flow showed excellent accuracy (error <2%). Including MMR caused notable reductions in velocity (up to 40%) and microrotation (up to 99.9%), especially under strong magnetic fields and high hematocrit values. Without MMR, magnetic effects were minimal due to the blood's low electrical conductivity. The simulations of 3D MHD artery and 2D MHD aneurysm flows supported these results. Especially in the aneurysm, MMR suppressed any recirculation cores, highlighting its stabilizing and shear-dampening effects. The solvers show strong promise for biomedical applications such as magnetic hyperthermia and targeted drug delivery.

**Program summary**

*Program Title*: epotMicropolarFoam, epotMMRFoam.

*Developer's repository link (epotMicropolarFoam)*:
https://github.com/KEAslani/epotMicropolarFoam.

*Developer's repository link (epotMMRFoam)*: https://github.com/KEAslani/epotMMRFoam.

*Licensing provisions*: GPLv3

*Programming language*: C++

*Nature of problem*: There are no numerical codes for simulating micropolar flows with or without magnetic particles under the influence of applied magnetic fields. When magnetic particles exist in the fluid (e.g., ferrofluids and blood), the effect of micromagnetorotation should be included in the code, i.e., the magnetic torque that arises from the misalignment between the magnetization and the applied magnetic field.

*Solution method*: Two transient OpenFOAM solvers were created, i.e., epotMicropolarFoam and epotMMRFoam, to simulate magnetohydrodynamics micropolar flows without or with magnetic particles, respectively. Both solvers utilize the PISO algorithm for pressure-velocity coupling, while adopting the low-magnetic-Reynolds-number approximation (electric potential formulation) for the MHD simulations. In the case of epotMMRFoam, the micromagnetorotation term is also included in the change of internal angular momentum equation, while a constitutive equation for the magnetization is also included.

## 1. Introduction

A micropolar fluid is a type of fluid that contains small, rigid particles that move independently within the mother-liquid. The theory of micropolar fluids was developed by Eringen in 1966 and is an extension of the Navier-Stokes equations, which describe Newtonian fluids [1]. This theory differs from the Navier-Stokes equations due to the presence of the antisymmetric part of the stress tensor, which is absent in the latter. The antisymmetric part of the stress tensor contributes to the conservation of angular momentum, which is not identically satisfied as in the Navier-Stokes equations. As a result, in micropolar fluid theory, another equation must be solved, i.e., the equation of change of internal angular momentum [2]. Based on this, a micropolar fluid is characterized by both its axial velocity and microrotation $\boldsymbol{\omega}$. Microrotation is defined as the averaged angular velocity of the particles within the fluid. Furthermore, microrotation must differ from the vorticity of the fluid, expressed as $\boldsymbol{\Omega} = \frac{1}{2}\,\boldsymbol{\nabla} \times \boldsymbol{v}$. If they coincide, the fluid can still be classified as Newtonian, even when microrotation is not zero. This coincidence is rare due to the introduction of an additional stress tensor associated with microrotation diffusion, known as the couple stress tensor. As a result of the addition of the couple stress tensor, a new type of viscosity is introduced, namely the spin viscosity $\gamma$ related to the flux of the internal angular momentum [3, 4].

Over the years, micropolar fluid theory has been used to model various fluids, including exotic lubricants [5], colloidal suspensions – e.g. ferrofluids [6], liquid crystals [7], and blood [8]. Micropolar fluid theory can be combined with magnetohydrodynamics (MHD), as proposed by Eringen [9, 10]. This model is based on Maxwell's equations and the micropolar balance laws. In this scenario, the micropolar fluid must also possess finite electrical conductivity and can be influenced by the Lorentz force. This force emerges when electric currents are generated in the direction of the magnetic field applied to the flow. MHD micropolar fluid theory has found various applications, particularly in ferrofluids and biomedical engineering, where magnetic fields are applied to blood flow.

An important class of micropolar fluids is ferrofluids, which consist of ferromagnetic nanoparticles, such as magnetite, suspended in a liquid carrier like water or kerosene [11]. Ferrofluids find applications in engineering and biomedicine, primarily due to their rheological properties, which an externally applied magnetic field can easily control. A key characteristic of ferrofluids is the magnetic polarization of the particles (magnetization) $\boldsymbol{M}$, which tends to 'relax' and align with the external magnetic field $\boldsymbol{H}$. This alignment creates a magnetic moment $\boldsymbol{M} \times \boldsymbol{H}$ that contributes to the change in the internal angular momentum and can significantly affect the microrotation of the ferromagnetic nanoparticles in the fluid. This phenomenon was extensively studied by Shliomis [12] and Rosensweig [13, 14], who introduced the theory of ferrohydrodynamics (FHD).

At this point, the distinction between ferrohydrodynamics and magnetohydrodynamics should be emphasized; the latter refers to the rheology of conducting fluids affected by the Lorentz force. FHD and MHD can be combined in the case of a conducting ferrofluid. This was accomplished by Shizawa & Tanahashi [15, 16], who presented a comprehensive mathematical model for ferrofluids using micropolar fluid theory and Maxwell's equations, while introducing a new equation for the change in magnetization that considers the existing microrotation of the ferromagnetic nanoparticles (micromagnetorotation-MMR). The flow equations were derived using the methodology of irreversible thermodynamics, which ensures that the dissipation function is always positive.

Blood is also classified as a micropolar fluid due to the presence of blood cells in the plasma [17-19]. The experimental study by Ariman, Turk, and Sylvester on steady and pulsatile blood flows demonstrated strong agreement with the predictions made by micropolar fluid theory for these flows [20, 21]. Furthermore, some studies suggest that blood should be treated as a ferrofluid when subjected to an external magnetic field, such as during magnetic resonance imaging (MRI). Blood contains hemoglobin molecules in erythrocytes (red blood cells), which is an iron oxide that acts like a magnetic particle, while blood plasma serves as the liquid carrier [22]. Consequently, the applied magnetic field may affect the microrotation of erythrocytes due to the magnetization of hemoglobin, subsequently influencing blood viscosity and velocity. Experimental studies showed statistically significant evidence of symptoms such as vertigo, nausea, and a metallic taste related to magnetic field intensities of 1.5 and 4T, associated with a reduction in blood velocity [23-26]. Moreover, these observations cannot be attributed solely to the impact of the Lorentz force on blood flow, as blood's relatively low electrical conductivity prevents it from being significantly influenced by the Lorentz force.

Based on the above, it is evident that studying blood as an MHD micropolar fluid without disregarding the MMR term is essential. To this end, several analytical studies conducted by Aslani et al. aimed to investigate the influence of MMR on micropolar MHD flows, such as blood, using the mathematical model of Shizawa and Tanahashi [3, 27-29]. These studies concentrated on establishing the differences that arise from acknowledging versus ignoring the MMR term while consistently including the effect of the Lorentz force. In summary, it was shown that the MMR term significantly impacts both velocity and microrotation. In the planar MHD micropolar Poiseuille flow, deceleration reached 16%, while the reduction in microrotation was as much as 99%. Furthermore, it was demonstrated that the MMR decreases heat transfer by suppressing convection, leading to a temperature drop of up to 8.5% due to the velocity reduction. Additionally, it was shown that an externally applied magnetic field on blood flow influences the microrotation of erythrocytes via the MMR term, which subsequently affects blood velocity through vorticity, a phenomenon documented in numerous experimental studies that the effect of the Lorentz force cannot solely explain.

After all these analytical studies, the next step was the investigation of more complex blood flows, such as 3D arteries, stenoses, and aneurysms. For this reason, the development of computational code using the well-known OpenFOAM library was chosen. OpenFOAM is a powerful, free, open-source CFD platform that has become a reliable tool for investigating fluid dynamics and complex flows [30, 31]. OpenFOAM has a wide range of solvers, including solvers for incompressible and compressible flows, multiphase flows, turbulence, heat transfer, MHD, complex fluid flows, and more. Mesh handling is easy in OpenFOAM as it supports complex geometries with structured, unstructured, and polyhedral mesh types. OpenFOAM also has high customizability, as its highly modular C++ codebase allows users to develop and add custom solvers, boundary conditions, and utilities [32].

Several studies have examined micropolar and biomagnetic flows in various configurations that need to be discussed. Asadi et al. [33] developed a finite-element method for incompressible micropolar flows and applied it to classical benchmark problems to demonstrate how micropolar effects influence the velocity and microrotation fields. Concerning micropolar blood flows, Bourantas [34] combined micropolar fluid theory and classical ferrohydrodynamics (without taking into account the MMR term) to examine stenosed arteries using a meshless collocation method. The results showed that microrotation and magnetization can significantly modify the structure of the flow. A classical numerical reference for numerical micropolar flow studies is the paper of Hogan and Henriksen [35], who used the micropolar model to evaluate blood flow through an idealized stenosis. They demonstrated that even when velocity fields appear similar to the Newtonian case, wall shear stresses can differ by 25–50%, especially near the stenosis throat. Their results highlighted the importance of using micropolar theory in flows with high shear and strong geometric disturbances. Manolis and Koutsoukos [36] were the first to develop an OpenFOAM solver for micropolar flows. They named the solver micropolarFoam. MicropolarFoam is a modification of the icoFoam, a well-established, incompressible, laminar flow solver for transient simulations of Newtonian fluids, to incorporate the force term arising from the microrotation–vorticity difference and to solve the internal angular momentum equation. MicropolarFoam has since been used in later studies—including the work of Karvelas et al. [37] who explored the micropolar effects in realistic vascular geometries without magnetic fields. Although these works show the importance of micropolar rheology and magnetic effects, no open-source numerical tool currently exists for fully coupled MHD micropolar flows that include a constitutive magnetization equation and the micromagnetorotation (MMR) term.

In this manner, this paper presents two OpenFOAM solvers developed for the study of MHD micropolar flows with magnetic particles, such as blood, while either ignoring or including the effect of micromagnetorotation. The first solver, epotMicropolarFoam, combines the micropolar formulation of micropolarFoam with the electric-potential MHD approach of epotFoam. This allows the simulation of classical MHD micropolar flows without the MMR term under the low-magnetic-Reynolds-number approximation. The second solver, epotMMRFoam, extends this framework by solving the Shizawa–Tanahashi magnetization equation [15] and by including the MMR torque in the internal angular-momentum equation. In this way, epotMMRFoam accounts for the combined effects of the Lorentz force, microrotation–vorticity coupling, and magnetic torque due to particle magnetization. To the authors' knowledge, no other open-source solver currently available in OpenFOAM can handle MHD micropolar flows with the full MMR model. This work aims to describe the mathematical model and numerical implementation of the two solvers, to validate them against analytical solutions, and to demonstrate their use in representative biomedical flows, such as arterial and aneurysmal geometries.

This paper is organized as follows: The mathematical model is described in Section 2. The algorithm and the implementation of the epotMicropolarFoam and epotMMRFoam solvers are presented in Section 3. Validation of the solvers and additional results for the simulations of the 3D MHD micropolar artery flow and the MHD micropolar blood flow through a 2D aneurysm are shown in Section 4. Finally, conclusions are drawn in Section 5.

## 2. Governing equations

According to Shizawa and Tanahashi [15], the governing equations for modelling an MHD micropolar flow with ferromagnetic particles, such as blood, are:

$$\rho \frac{D\boldsymbol{v}}{Dt} = -\boldsymbol{\nabla} p + \mu \boldsymbol{\nabla}^2 \boldsymbol{v} + 2\mu_r \boldsymbol{\nabla} \times (\boldsymbol{\omega} - \boldsymbol{\Omega}) + (\boldsymbol{M} \cdot \boldsymbol{\nabla})\boldsymbol{H} + \mu_0 (\boldsymbol{j} \times \boldsymbol{H}), \tag{1}$$

$$\rho j \frac{D\boldsymbol{\omega}}{Dt} = 4\mu_r (\boldsymbol{\Omega} - \boldsymbol{\omega}) + \gamma \boldsymbol{\nabla}^2 \boldsymbol{\omega} + \boldsymbol{M} \times \boldsymbol{H}, \tag{2}$$

$$\boldsymbol{\nabla} \cdot \boldsymbol{B} = 0, \tag{3}$$

$$\boldsymbol{\nabla} \times \boldsymbol{H} = \boldsymbol{j}, \tag{4}$$

$$\boldsymbol{j} = \sigma (\boldsymbol{v} \times \boldsymbol{B}), \tag{5}$$

$$\boldsymbol{M} = \frac{M_0}{H} [\boldsymbol{H} - \tau (\boldsymbol{H} \times \boldsymbol{\omega})], \tag{6}$$

$$\boldsymbol{\nabla} \cdot \boldsymbol{v} = \boldsymbol{0}, \tag{7}$$

where $\boldsymbol{v}$ is the velocity vector, $\rho$ is fluid's density, $t$ is time, $p$ is pressure, $\mu$ is the dynamic viscosity, $\mu_r$ is the rotational viscosity, $\boldsymbol{M}$ is the magnetization vector, $\boldsymbol{H}$ is the applied magnetic field from the magnetic flux density $\boldsymbol{B} = \mu_0 \boldsymbol{H} + \boldsymbol{M}$ with $\mu_0$ being the magnetic permeability of free space, $\boldsymbol{j}$ is the current density vector, $j$ is the microinertia coefficient, $\gamma$ is spin viscosity, $\sigma$ is the electrical conductivity, $M_0$ is the equilibrium magnetization and $\tau$ is the magnetization relaxation time. Equation (1) represents the conservation of linear momentum, while Equation (2) describes the change in internal angular momentum, which appears in micropolar fluids and is derived from the antisymmetric part of the stress tensor that contributes to the conservation of total angular momentum. In the first equation, the term $2\mu_r \boldsymbol{\nabla} \times (\boldsymbol{\omega} - \boldsymbol{\Omega})$ represents force due to microrotation-vorticity difference that characterises micropolar fluids, the term $(\boldsymbol{M} \cdot \boldsymbol{\nabla})\boldsymbol{H}$ is the magnetic body force, and the term $\mu_0(\boldsymbol{j} \times \boldsymbol{H})$ is the Lorentz force. In the second equation, the term $4\mu_r(\boldsymbol{\Omega} - \boldsymbol{\omega})$ represents the internal torque due to the microrotation-vorticity difference, the term $\gamma \boldsymbol{\nabla}^2 \boldsymbol{\omega}$ is the microrotation diffusion and the last term $\boldsymbol{M} \times \boldsymbol{H}$ is the magnetic torque generated due to the magnetization of the particles (micromagnetorotation). Equations (3)-(5) represent Gauss law for magnetic monopoles, Ampere's law and Ohm's law, respectively (Maxwell's equations). Equation (6) is the constitutive magnetization equation from the mathematical model of Shizawa and Tanahashi [15]. The vorticity of the fluid remains $\boldsymbol{\Omega} = \frac{1}{2}(\boldsymbol{\nabla} \times \boldsymbol{v})$ as in Newtonian fluids. It is evident from Equations (1)-(2) that when $\mu_r = 0$ or when $\boldsymbol{\omega} = \boldsymbol{\Omega}$, the classical Newtonian hydrodynamic equations are retrieved. Moreover, from Equation (6) it can be seen that when no microrotation exists, i.e., when $\boldsymbol{\omega} = \boldsymbol{0}$, the magnetization attains its equilibrium value $M_0$ parallel to the applied magnetic field $\boldsymbol{H}$. Equation (7) represents the mass conservation law provided that the solvers simulate viscous and incompressible flows. It should be mentioned that $\rho j$ at the left side of Equation (3) can be replaced by $I = \rho j$, where $I$ is the sum of the particles' moment of inertia per unit volume.

The dynamic viscosity $\mu$, the vortex viscosity coefficient $\mu_r$, the spin viscosity $\gamma$ and the moment of inertia $I$ are correlated as [15, 39, 40]:

$$\mu_r = \frac{I}{4\tau_s}, \tag{8}$$

$$\gamma = \mu j \quad or \quad \gamma = \mu \frac{I}{\rho}, \tag{9}$$

where $\tau_s = \frac{\alpha^2 \rho_\alpha}{15\mu}$ is the spin relaxation time, $\alpha$ being the radius of the suspended particles and $\rho_\alpha$ their density. The microinertia coefficient can be calculated as:

$$j = \frac{8}{15}\pi \alpha^5 N \frac{\rho_\alpha}{\rho}, \tag{10}$$

where $N$ is the number (per unit volume) of the suspended particles, respectively. Using the definition of the volume fraction $\varphi = \frac{4}{3}\pi\alpha^3 N$,

$$\varphi = \frac{4}{3}\pi \alpha^3 N, \tag{11}$$

it is derived that $j = \frac{2}{5}\alpha^2 \frac{\rho_\alpha}{\rho}\varphi$. It should be noted that the volume fraction of blood is the hematocrit. In simulations, the hematocrit $\varphi$ will be used to calculate the micropolar properties of blood.

## 3. Algorithm description and implementation

### 3.1. Algorithm structure

As mentioned above, epotMicropolarFoam and epotMMRFoam were developed to simulate MHD micropolar fluid flows with magnetic particles, such as blood, with or without incorporating micromagnetorotation. These two solvers are based on the micropolarFoam and epotFoam solvers. All equations in these solvers are discretized using the finite volume method (FVM) [41]. Both solvers rely on the PISO algorithm. In summary, this algorithm was developed by Issa in 1986 to calculate the pressure-velocity procedure and serves as an extension of the SIMPLE method algorithm. It usually provides more stable results while consuming less CPU time. The algorithm can be summed up as follows:

1. Establish the boundary conditions.
2. Solve the discretized momentum equation to calculate an intermediate velocity field.
3. Calculate the mass flux at the cell faces.
4. Solve the pressure equation.
5. Adjust the mass flux at the cell faces.
6. Update the velocities based on the new pressure field.
7. Refresh the boundary conditions.
8. Repeat from Step 3 for the specified number of iterations.
9. Increase the time step and start again from Step 1.

Figure 1 shows the flow chart of the PISO algorithm.

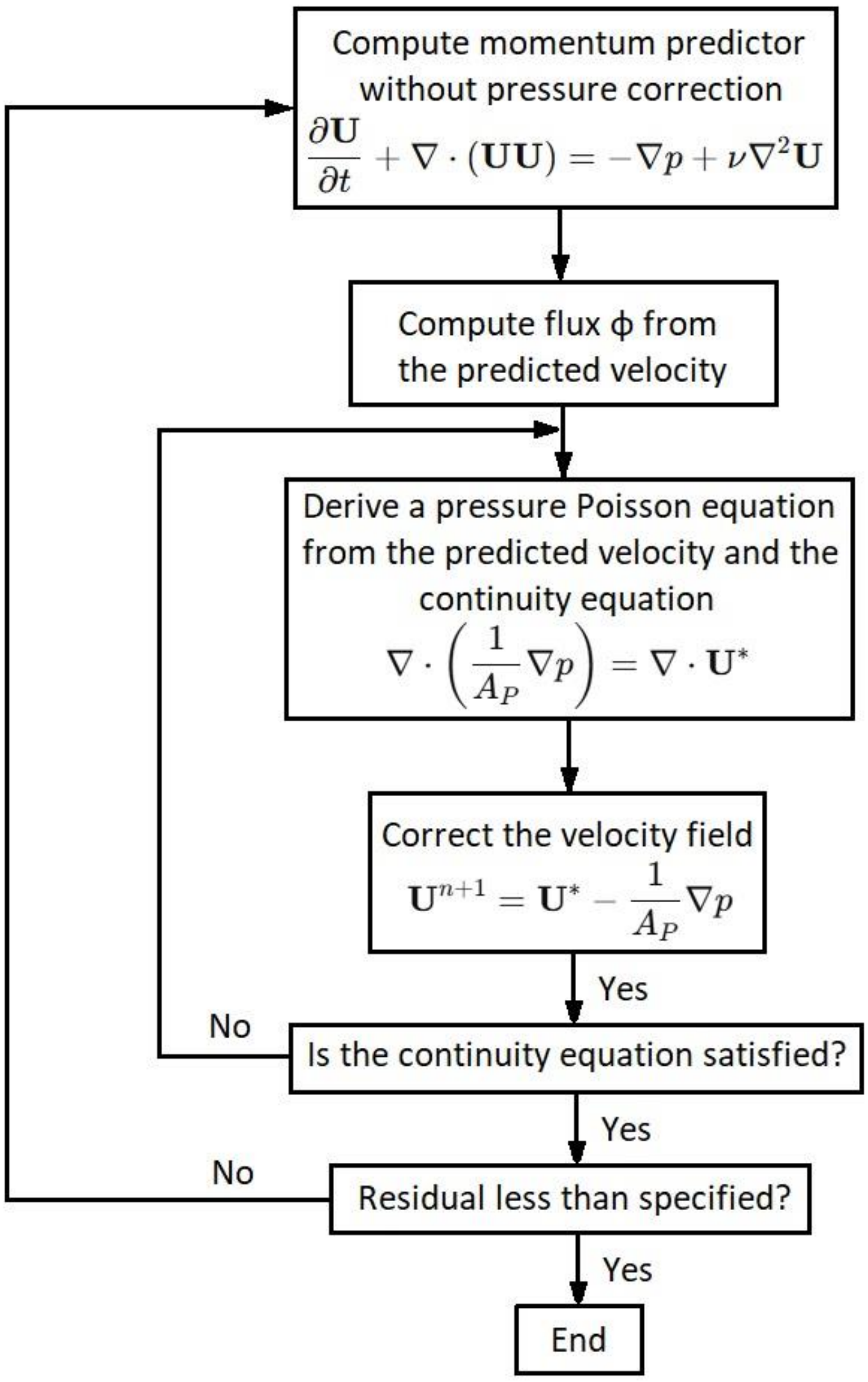

**Figure 1**. Flow chart of the PISO algorithm.

As discussed above, epotFoam is used for incompressible, laminar flows of conducting fluids under the influence of a magnetic field. EpotFoam employs the low-magnetic-Reynolds number approximation ($Re_m \ll 1$, where $Re_m = \sigma\mu_0 H_0 L$, $H_0$ being the applied magnetic field and $L$ the characteristic length of the channel), where the magnetic induction equation is ignored and an electric potential formulation is utilized [38]. This approximation is consistently applicable for many MHD applications, such as fusion reactor blankets, and most MHD numerical studies use the electric potential formula. In epotFoam, the Lorentz force is regarded as the source term, while the current density, $\boldsymbol{j}$, is reconstructed at the cell center using an interpolation technique, ensuring the conservation of momentum and charge. Thus, the algorithm of epotFoam can be summarized in the following steps:

1. Creation of the initial fields of the velocity $\boldsymbol{v}$, pressure $p$, and velocity flux $\varphi_v$.
2. Lorentz force initialization.
3. Discretization of the momentum equation to predict the velocity field for the PISO loop.
4. The PISO loop is executed, and the Poisson equation is solved for a fixed number of iterations to obtain the pressure field, which is then used to correct the previously predicted velocity field.
5. The continuity equation (Equation 7) is solved, and the velocity boundary conditions are updated.
6. The magnetic flux term is computed at the cell faces as:

$$\chi = (\boldsymbol{v} \times \boldsymbol{B})\ \&\boldsymbol{S}_f. \tag{12}$$

7. Next, the Poisson equation for the electric potential $\boldsymbol{E} = -\nabla\varphi$ is solved:

$$\nabla^2\varphi - \nabla \cdot \chi = 0. \tag{13}$$

8. Then, the current density flux from the normal gradient of the electric potential is evaluated at the cell faces along with the magnetic flux:

$$j_f = -\nabla_{sn}\varphi\ \&\boldsymbol{S}_f + \chi. \tag{14}$$

9. The current density is reconstructed at the centroid of the cell as:

$$J_c = \frac{1}{\Omega_c}\sum_{f=1}^{nf} j_f\left(r_f - r_p\right) \cdot \boldsymbol{S}_f \tag{15}$$

10. Finally, the Lorentz force is updated through the final, conservative current density field from the previous step as:

$$\mathcal{L} = \boldsymbol{J}_c \times \boldsymbol{B}. \tag{16}$$

The above algorithm is implemented in the micropolarFoam solver [36] to obtain the epotMicropolarFoam solver. As discussed above, micropolarFoam is a modification of the icoFoam solver to incorporate the force term arising from the microrotation–vorticity difference (see Equation 1) and to solve the internal angular momentum equation (Equation 2). In this manner, the algorithm of the epotMicropolarFoam can be summed up as follows:

1. Creation of the initial fields of the velocity $\boldsymbol{v}$, microrotation $\boldsymbol{\omega}$, pressure $p$ and velocity flux $\varphi_v$.
2. Lorentz force initialization.
3. Discretization of the momentum equation (Equation 1) to predict the velocity field for the PISO loop.
4. Execution of the PISO loop, and solution of the Poisson equation for a fixed number of iterations.

5. Derivation of the pressure field, which is then used to correct the previously predicted velocity field.
6. Solution of the continuity equation (Equation 7) and update velocity boundary conditions.
7. Computation of the magnetic flux at the cell faces.
8. Solution of the Poisson equation for the electric potential $\boldsymbol{E} = -\nabla\varphi$.
9. Evaluation of the current density flux based on the normal gradient of the electric potential at the cell faces, along with the magnetic flux.
10. Reconstruction of the current density at the centroid of the cell.
11. Update the Lorentz force using the final, conservative current density field.
12. Solution of the equation of change of internal angular momentum without the MMR term (Equation 2) to obtain the microrotation field.

In the case of the epotMMRFoam solver, the above algorithm is repeated from Step 1 to Step 11, including the MMR term in Equation (2). The magnetization is computed using the constitutive equation (Equation 6). In this manner, one more step is included after Step 11.

12. Computation of the constitutive magnetization equation (Equation 6).
13. Solution of the change of internal angular momentum equation, including the MMR term (Equation 2), to obtain the microrotation field.

### 3.2. Implementation in OpenFOAM

The algorithm for epotMicropolarFoam is implemented in OpenFOAM by creating a new file named epotMicropolarFoam.C. First, the Lorentz force field is initialized before the time loop starts. Just after the IO printing add the following lines are added:

```
//Lorentz force term initialization
volVectorField L = sigma * (-fvc::grad(PotE) ^ B0) + sigma * ((U ^ B0) ^ B0);
```

The Lorentz force is described as a volume vector field object, which is computed as the cross product $\boldsymbol{j} \times \boldsymbol{B}$, where the current density distribution is obtained by Ohm's Law (Equation 5). Now that the Lorentz force is defined, it should be included in the momentum equation. Moreover, the micropolar source term term $2\mu_r \boldsymbol{\nabla} \times (\boldsymbol{\omega} - \boldsymbol{\Omega})$ should be also added, using the definition of the vorticity $\boldsymbol{\Omega} = \frac{1}{2}(\boldsymbol{\nabla} \times \boldsymbol{v})$ and the identity $\boldsymbol{\nabla} \times (\boldsymbol{\nabla} \times \boldsymbol{v}) = \boldsymbol{\nabla}(\boldsymbol{\nabla} \cdot \boldsymbol{v}) - \boldsymbol{\nabla}^2 \boldsymbol{v}$, where $\boldsymbol{\nabla} \cdot \boldsymbol{v} = 0$ due to mass conservation (Equation 7). Inside the time cycle, the UEqn is altered accordingly to include the source terms.

```
//Momentum predictor
fvVectorMatrix UEqn
(
      fvm::ddt(U)
   + fvm::div(phi, U)
   - fvm::laplacian(nu+ku, U)
   - 2.0 * ku * fvc::curl(N)
   - (1.0/rho) * L
);
```

It should be noted that microrotation is denoted as N in the solver. Now, the velocity field is estimated during the momentum predictor step of the PISO loop, taking into account the influence of the magnetic field and microrotation. No further modification is necessary for the velocity-pressure coupling, and the code remains the same as in icoFoam.

To calculate the electric potential distribution necessary for determining the Lorentz force, the Poisson equation (Equation 12) must be solved, which includes the cross product $\boldsymbol{v} \times \boldsymbol{B}$. By using the updated velocity values after completing the PISO loop, the quantity psiub will be estimated by interpolating the cross-product results at the cell faces. Consequently, the following lines are added:

```
//Interpolating cross product u x B over mesh faces
surfaceScalarField psiub = fvc::interpolate(U ^ B0) & mesh.Sf();
```

The matrix PotEEqn is then introduced, which, once solved, will provide the electric potential distribution. Since it is defined similarly to the pressure field, it requires a reference cell to serve as the null potential location. This will be specified later in the createFields.H file. Just after the psiub line, the following lines are added.

```
//Poisson equation for electric potential
fvScalarMatrix PotEEqn
(
   fvm::laplacian(PotE)
   ==
   fvc::div(psiub)
);

//Reference potential
PotEEqn.setReference(PotERefCell, PotERefValue);

//Solving Poisson equation
PotEEqn.solve();
```

The current density at the cell center is calculated from the electric potential distribution by first computing the current density fluxes on the cell faces and then interpolating to obtain the desired value. Consequently, the current density distribution is utilized to update the boundary condition. In this way, the following lines are added after the PotEEqn.solve() line.

```
//Computation of current density at cell faces
surfaceScalarField jn = -(fvc::snGrad(PotE) * mesh.magSf()) + psiub;

//Current density at face center
surfaceVectorField jnv = jn * mesh.Cf();

//Interpolation of current density at cell center
volVectorField jfinal = fvc::surfaceIntegrate(jnv) - (fvc::surfaceIntegrate(jn) * mesh.C());

//Update current density distribution and boundary condition
jfinal.correctBoundaryConditions();

//Lorentz force computation
L = sigma * (jfinal ^ B0);
```

Now, the equation of change of internal angular momentum without the MMR term is discretized. This equation is described as a matrix NEqn similar to the momentum predictor UEqn. After the discretization of the equation of change of internal angular momentum, microrotation is obtained. In this way, the following lines are added:

```
// Microrotation equation
fvVectorMatrix NEqn
(
     j*fvm::ddt(N)
  +  j*fvm::div(phi, N)
  -  2*ku*fvc::curl(U)
  +  4*ku*N
```

```
 -  fvm::laplacian(gu, N)
);

NEqn.solve();
```

No further additions or modifications are needed. The epotMicropolarFoam solver is ready.

As mentioned above, the code of the epotMMRFoam solver is similar to the code of the epotMicropolarFoam, except for the inclusion of the constitutive magnetization equation (Equation 6) and the magnetic flux density $\boldsymbol{B} = \mu_0 \boldsymbol{H} + \boldsymbol{M}$ as follows:

```
//Constitutive Magnetization equation M

M = ((( 1.0 / (mag(H0) )) * Mo) * H0) - ((( 1.0 / (mag(H0) )) * (Mo * t)) * (H0 ^ N));

//Magnetic flux density term B
dimensionedVector B = mu * H0;
```

Finally, the MMR term $\boldsymbol{M} \times \boldsymbol{H}$ is added to the equation of change of internal angular momentum.

```
// Microrotation equation
fvVectorMatrix NEqn
(
     j*fvm::ddt(N)
  +  j*fvm::div(phi, N)
  -  2*ku*fvc::curl(U)
  +  4*ku*N
  -  fvm::laplacian(gu, N)
  -  (1.0/rho) * (M ^ H0)
);
```

It should be noted that the effect of the magnetic force term $(\boldsymbol{M} \cdot \boldsymbol{\nabla})\boldsymbol{H}$ in the momentum equation is expected to be small and is therefore ignored, which is a standard practice in many studies regarding micropolar fluids with magnetic particles [42-45].

### 3.3. EpotMicropolarFoam and epotMMRFoam directories

A typical OpenFOAM solver folder breakdown contains a main source code file, e.g., solvername.C, a createFields.H file where all variables related to the mathematical model and their units are defined, and the Make folder, which includes files and options (see Figure 2). The Options file includes which libraries to link against and includes all paths, while files tells the compiler what files to compile and what to call the executable. For the epotMicropolarFoam and the epotMMRFoam, the solvername.C file was analyzed in the previous section. For the complete description of the solver, the createFields.H file should also be described.

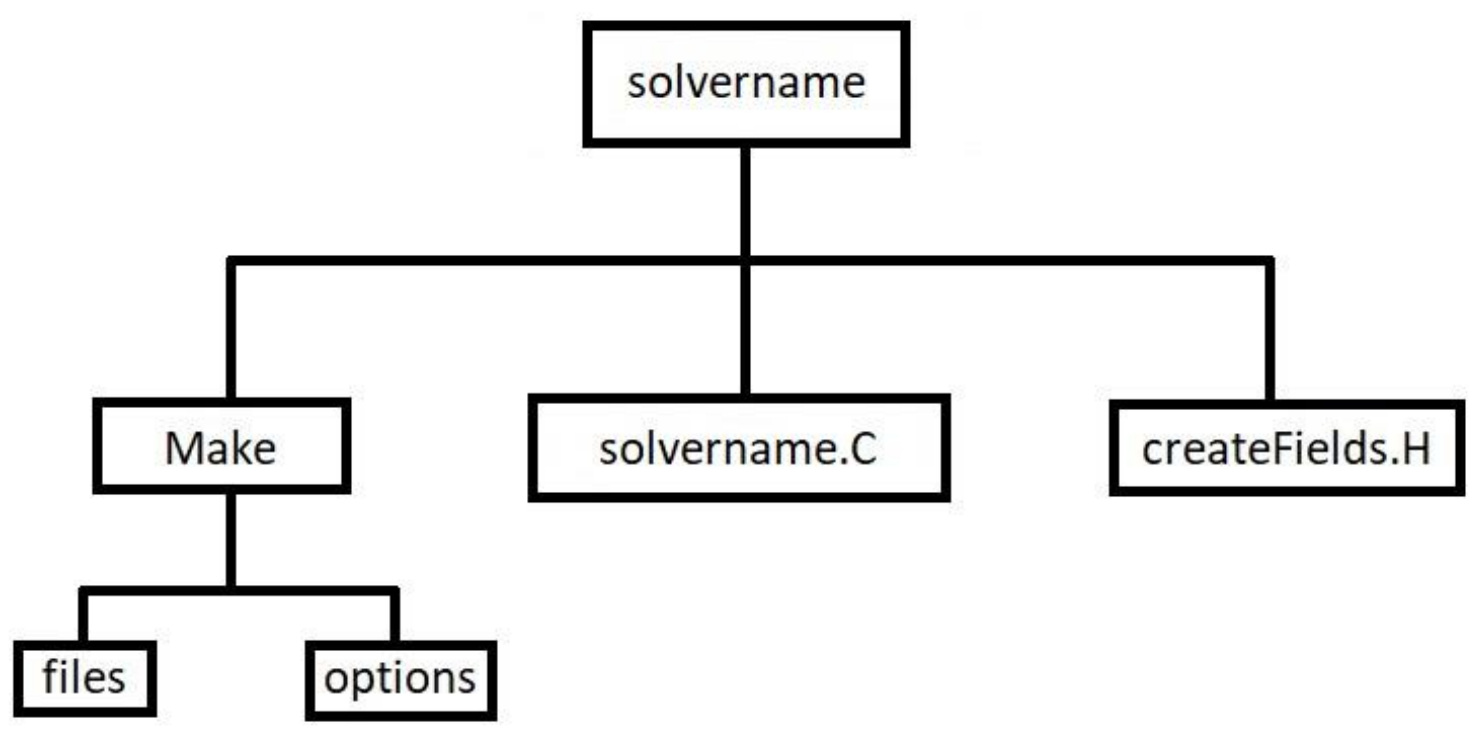

**Figure 2**. OpenFOAM solver directory structure.

For epotMicropolarFoam, five additional properties, ku, gu, j, rho, and sigma (corresponding to the rotational viscosity $\mu_r$, spin viscosity $\gamma$, microinertia coefficient $j$-devided by the fluid's density $\rho$, and electrical conductivity $\sigma$) are added to the transportProperties dictionary, whereas a new one, electromagneticProperties, is created to define the magnetic field B0. Regarding fields, two new fields are added, i.e., PotE, which is a volScalarField object and microrotation N, which is a volVectorField object. The entry for kinematic viscosity serves as a template, with modifications made to the dimensionSet variables.

The createFields.H file for the epotMMRFoam is pretty identical to the epotMicropolarFoam, except for the addition of the variable mu, Mo, and t, i.e., the magnetic permeability of free space $\mu_0$, the equilibrium magnetization $M_0$, and the magnetization relaxation time $\tau$. Moreover, the magnetic flux density B0 is replaced by the intensity of the magnetic field H0.

## 4. Validation and results

After developing the two solvers, the first step is to validate them against the analytical results from Aslani et al. [3] concerning an MHD micropolar Poiseuille blood flow using various hematocrit values and different intensities of the applied magnetic field. The values chosen for the blood's physical properties are derived from various numerical and experimental studies related to biomedical applications. Then, the two solvers were used to study the effect of MMR on a 3D Then, the two solvers were utilized to study the impact of MMR on a 3D MHD blood artery flow.

### 4.1. Validation of the solvers

As mentioned above, the numerical results for an MHD micropolar Poiseuille blood flow, where the effect of MMR is acknowledged and ignored, are analyzed and compared to the corresponding analytical results. For this flow's numerical simulation, epotMicropolarFoam and epotMMRFoam are utilized. To facilitate comparison, the numerical results of Newtonian Poiseuille blood flow and simple micropolar Poiseuille blood flow are also analyzed alongside their corresponding analytical results. For these numerical simulations, the transient solvers icoFoam and micropolarFoam were employed, respectively.

#### 4.1.1. Case setup

In general, OpenFOAM operates in a three-dimensional Cartesian coordinate system $(x, y, z)$. For modeling the Poiseuille blood flow, a rectangular prism domain measuring $0.001 \times 0.001 \times 0.02\ m$ was created. Figure 3 illustrates the physical domain of the Poiseuille blood flow.

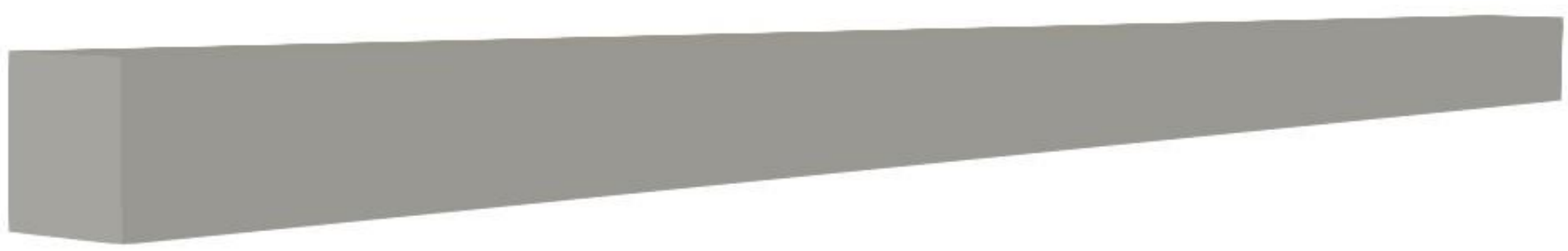

**Figure 3**. Physical domain of the Poiseuille blood flow.

The discretization of the computational domain was achieved using a $20 \times 50 \times 12$ hexagonal mesh (see Figure 4). This mesh is considered fine for Poiseuille flow simulations [36]. The time step for each simulation was selected to ensure that the Courant number is less than one $Co < 1$. The duration of the simulation was chosen to allow the flow to reach a steady state. A good rule for planar laminar flow simulations is that the fluid should pass through the domain 10 times to reach steady state [41]. Here, this rule was applied to all Poiseuille blood flow simulations.

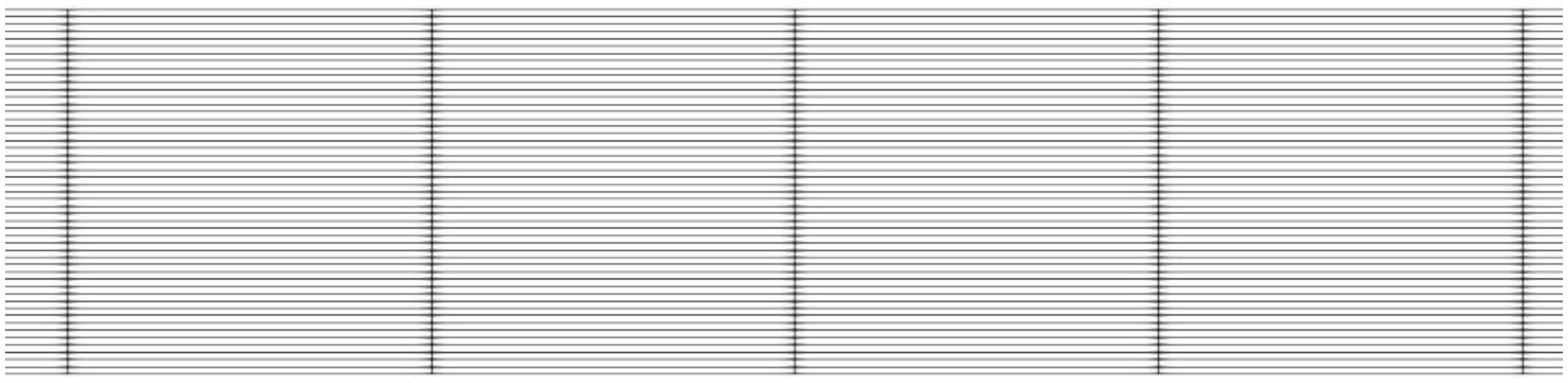

**Figure 4**. Computational hexagonal mesh for the $0.001 \times 0.001 \times 0.02\, m$ domain of the Poiseuille blood flow.

No-slip boundary conditions were applied to the walls of the channel, while zeroGradient boundary conditions were set at the inlet and the outlet. For the sake of comparison, results for hematocrit values at $\varphi = 25\%$ and $\varphi = 45\%$ are presented. Both hematocrit values are realistic; the value of $\varphi = 25\%$ is below the normal range of 40-54% for males and 36-48% for females, but it is a common value for patients with diseases such as cancer or sickle cell anemia [46]. Table 1 shows the physical properties of blood as derived from the studies of Tzirtzilakis [17, 19], Ariman et al. [21] and Pai et al. [26].

**Table 1**. Blood's physical properties.

| Physical properties | Value | |
|---|---|---|
| volume fraction (hematocrit) $\varphi$ (%) | 25 | 45 |
| dynamical viscosity $\mu$ ($Pa \cdot s$) | $4 \cdot 10^{-3}$ | $4 \cdot 10^{-3}$ |
| rotational viscosity $\mu_r$ ($Pa \cdot s$) | $1.5 \cdot 10^{-3}$ | $2.7 \cdot 10^{-3}$ |
| microinertia coefficient $j$ ($m^2$) | $1.75 \cdot 10^{-10}$ | $7.5 \cdot 10^{-10}$ |
| spin viscosity $\gamma$ ($\frac{kg \cdot m}{sec}$) | $7 \cdot 10^{-13}$ | $3 \cdot 10^{-12}$ |
| fluid density $\rho$ ($kg \cdot m^{-3}$) | 1050 | 1050 |
| equilibrium magnetization $M_0$ ($A \cdot m^{-1}$) | 100 | 100 |
| electrical conductivity $\sigma$ ($S \cdot m^{-1}$) | 0.7 | 0.7 |
| magnetization relaxation time $\tau$ ($sec$) | 0.001 | 0.001 |

As in the analytical study by Aslani et al. [3], three values of the applied magnetic field intensity are used, which are commonly found in experimental and numerical studies related to biomedical applications: $H_0 = 795{,}774.72\ A/m$, $H_0 = 2{,}387{,}324.15\ A/m$ and $H_0 = 6{,}366{,}197.72\, A/m$ [17, 23, 24]. Given that $H_0 = B_0/\mu_0$, it follows that $H_0 = 795{,}774.72\ A/m$ corresponding to $1\,T$, $H_0 = 2{,}387{,}324.15\ A/m$ corresponding to $3\,T$ and $H_0 = 6{,}366{,}197.72\, A/m$ corresponding to $8\,T$.

For the simulation of a blood flow, a suitable pressure gradient $G$ or an inlet velocity must be specified. In order for the numerical simulations to resemble a realistic human blood flow, no value for the inlet velocity was applied. Instead, a suitable maximum velocity value was selected from the bibliography based on the channel's height $2L = 0.001\, m$ [47, 48]. In this manner, the corresponding pressure gradient was calculated using the relation $v_{x_{max}} = {GR^2}/{2\mu}$. For an arteriole with a diameter of $2L = 0.001\, m$, $v_{x_{max}} = 0.001\ m/sec$ was selected. Consequently, it was derived that $G = 32\ Pa/m$. The Reynolds number was calculated using the relation $Re = {\rho v_{x_{max}} L}/{\mu}$, and $Re = 0.13125$. Considering that the length of the channel is $l = 0.02\ m$, and using the relation ${P_2 - P_1}/{l} = G$, where $P_1$ is the inlet pressure and $P_2$ is the outlet pressure, it was derived that $P_1 = 0.64\, Pa$ (setting $P_2 = 0\, Pa$). Finally, the inlet pressure $P_1$ values was divided by blood's density $\rho = 1050\ kg/m^3$, because OpenFOAM uses kinematic pressure $P/\rho$ as input.

#### 4.1.2. Results

Figure 5 presents the analytical and numerical results for the velocity upsilon $\upsilon_x$ of the Newtonian Poiseuille blood flow. Due to the Newtonian nature of the flow, the transient solver icoFoam was employed, which is a standard OpenFOAM solver. The hematocrit does not affect the flow characteristics since the Poiseuille blood flow is modeled as Newtonian. As expected, numerical and analytical results align very well, with a calculated error of $0.05\%$. It is evident that the OpenFOAM simulation of Newtonian blood flow produces accurate results, and the hexagonal mesh used is of high quality.

In Figure 6, the analytical and numerical results of velocity and microrotation for the micropolar Poiseuille blood arteriole flow, without an applied magnetic field, with hematocrit values of $\varphi = 25\%$, and $\varphi = 45\%$ are presented. Here, the simulations were conducted using the solver micropolarFoam. It is evident that the numerical results for both velocity and microrotation are in good agreement, with an error of $ER_\upsilon = 0.55\ \%$ for the velocity and $ER_\omega = 1.2\ \%$ for the microrotation. Notably, the difference between the velocity of the micropolar Poiseuille blood flow and the Newtonian flow is not very large, reaching a maximum value of $2\%$ at the center of the channel (i.e., $x = 0$) for $\varphi = 25\%$, and $3.5\%$ for $\varphi = 45\%$. These small differences are expected because the channel is relatively wide; therefore, the microstructure has limited ability to influence the flow. As hematocrit increases, more particles are present in the fluid, which slightly enhances microrotation diffusion and reduces the velocity near the centerline.

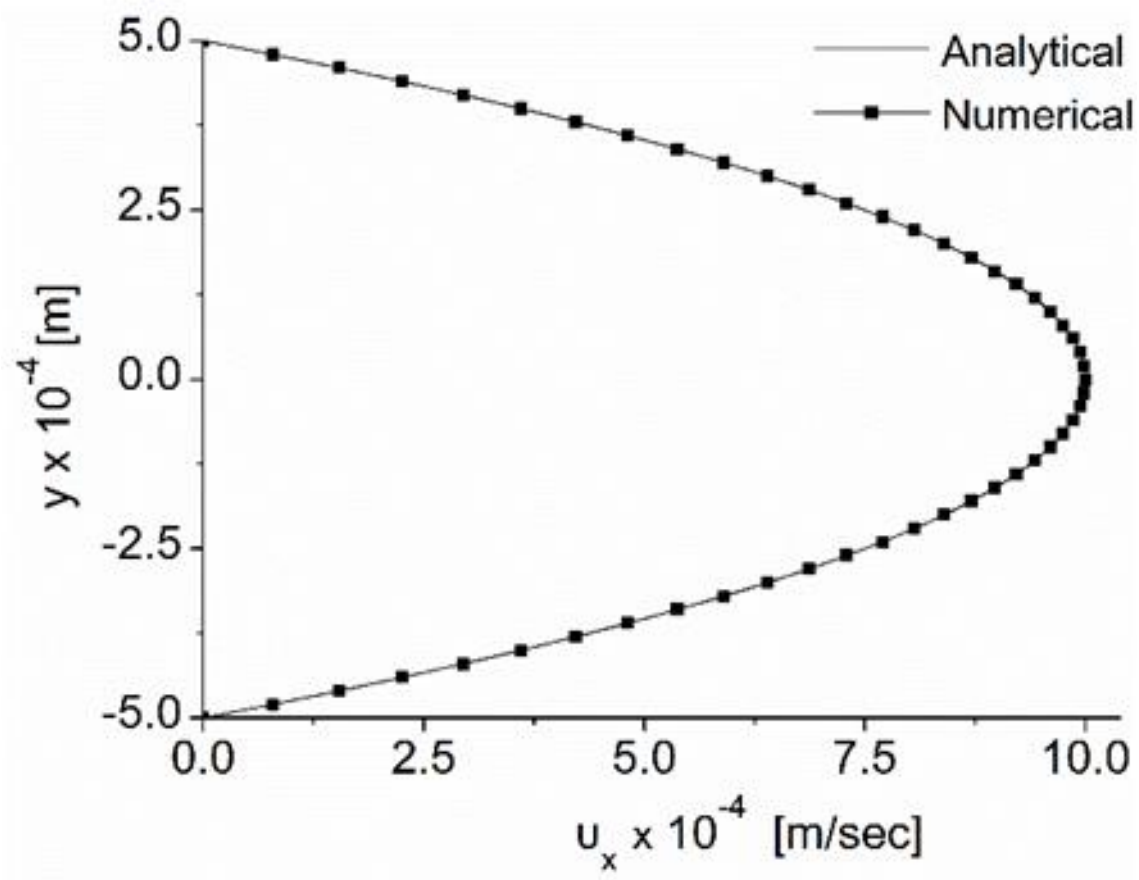

**Figure 5.** Analytical and numerical results for the Newtonian Poiseuille blood flow with channel height of $2L = 0.001\ m$ (arteriole).

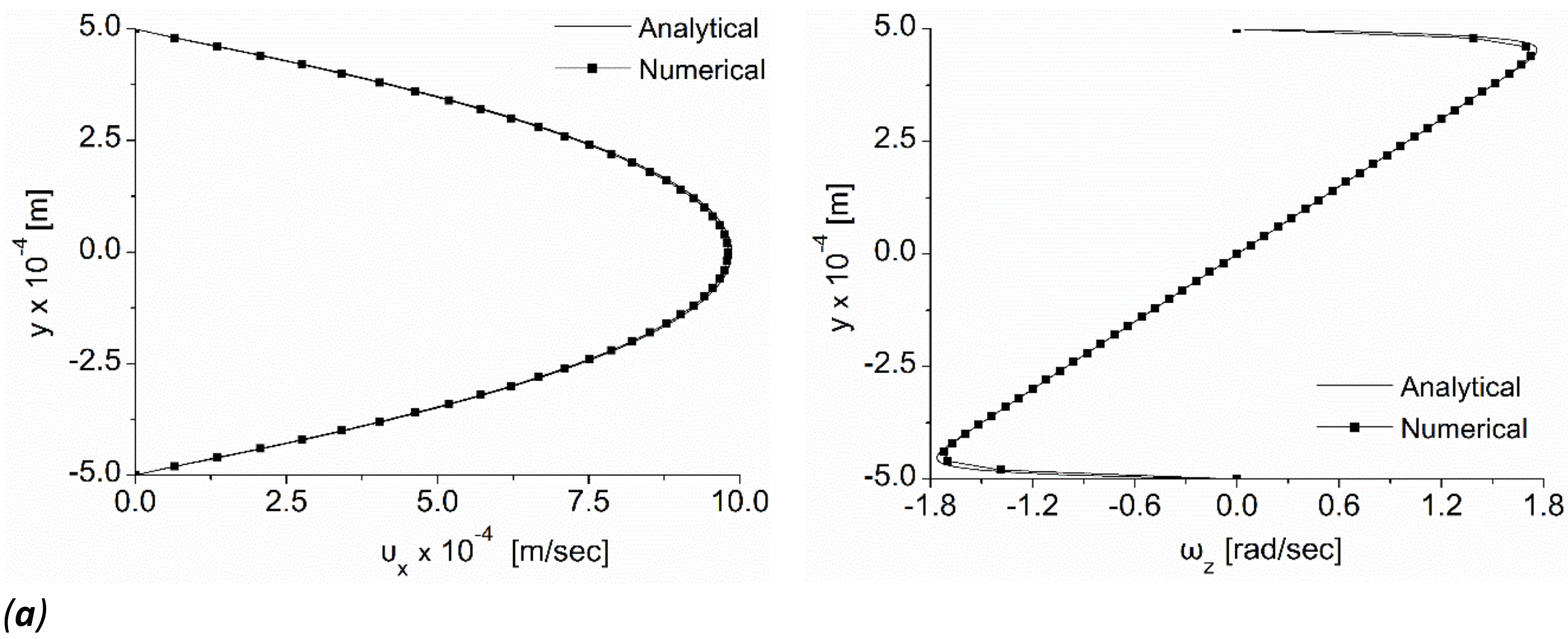

***(a)***

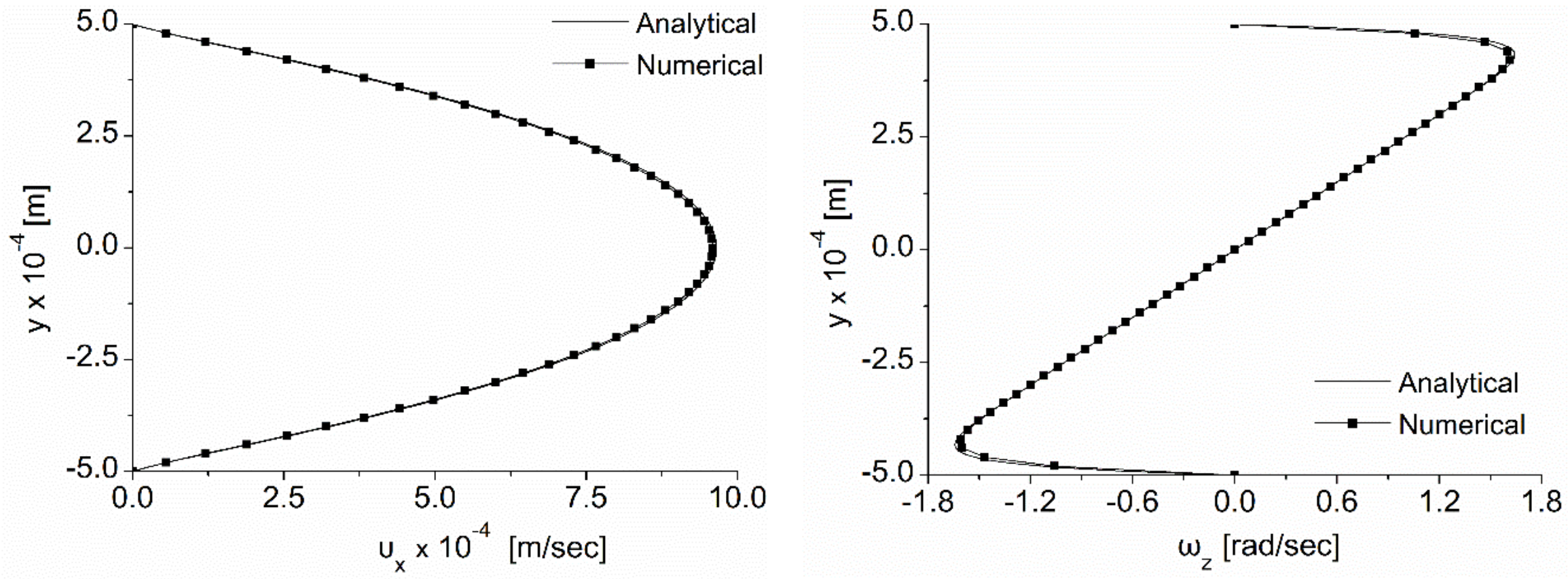

***(b)***

**Figure 6.** Analytical and numerical results for the velocity $\upsilon_x$ (left) and microrotation $\omega_z$ (right) of the micropolar Poiseuille blood flow with hematocrit (a) $\varphi = 25\%$, and (b) $\varphi = 45\%$, and channel height of $2L = 0.001\ m$ (arteriole).

Figure 7 shows the analytical and numerical results for the MHD micropolar Poiseuille blood arteriole flow with a hematocrit of $\varphi = 25\%$, where the effect of micromagnetorotation is ignored (acronym MHD) and where that effect is included (acronym MMR). As discussed previously, the results were generated for three different values of the applied magnetic field: $1\ T$ , $3\ T$, and $8\ T$. It is evident that the numerical and analytical results align closely for both velocity and microrotation, whether MMR is ignored or included, across all intensities of the applied magnetic field. The errors are:

- $ER_\upsilon = 0.52$ % and $ER_\omega = 1.17$ % for the MHD case and magnetic field of $1\ T$.
- $ER_\upsilon = 0.054$ % and $ER_\omega = 2$ % for the MMR case and magnetic field of $1\ T$.
- $ER_\upsilon = 0.52$ % and $ER_\omega = 1.17$ % for the MHD case and magnetic field of $3\ T$.
- $ER_\upsilon = 0.023$ % and $ER_\omega = 0.73$ % for the MMR case and magnetic field of $3\ T$.
- $ER_\upsilon = 0.51$ % and $ER_\omega = 1.17$ % for the MHD case and magnetic field of $8\ T$.
- $ER_\upsilon = 0.012$ % and $ER_\omega = 0.29$ % for the MMR case and magnetic field of $8\ T$.

It is evident that the epotMicropolarFoam and epotMMRFoam solvers produce accurate results, while the $20 \times 50 \times 12$ hexagonal mesh for the arteriole is sufficiently fine to yield numerical results with minimal error.

It is also noteworthy from Figure 7 that the magnetic field, without considering the effect of MMR, has a very small influence on the micropolar Poiseuille blood flow, due to its relatively low electrical conductivity and small channel size. On the other hand, the effect of MMR on the MHD micropolar Poiseuille blood flow is significant, with a velocity reduction of 26% for a magnetic field intensity of $8\ T$, and an almost 99.9% reduction in microrotation. This physically means that the erythrocytes are polarized in the direction of the externally applied magnetic field, and no internal rotation is permitted.

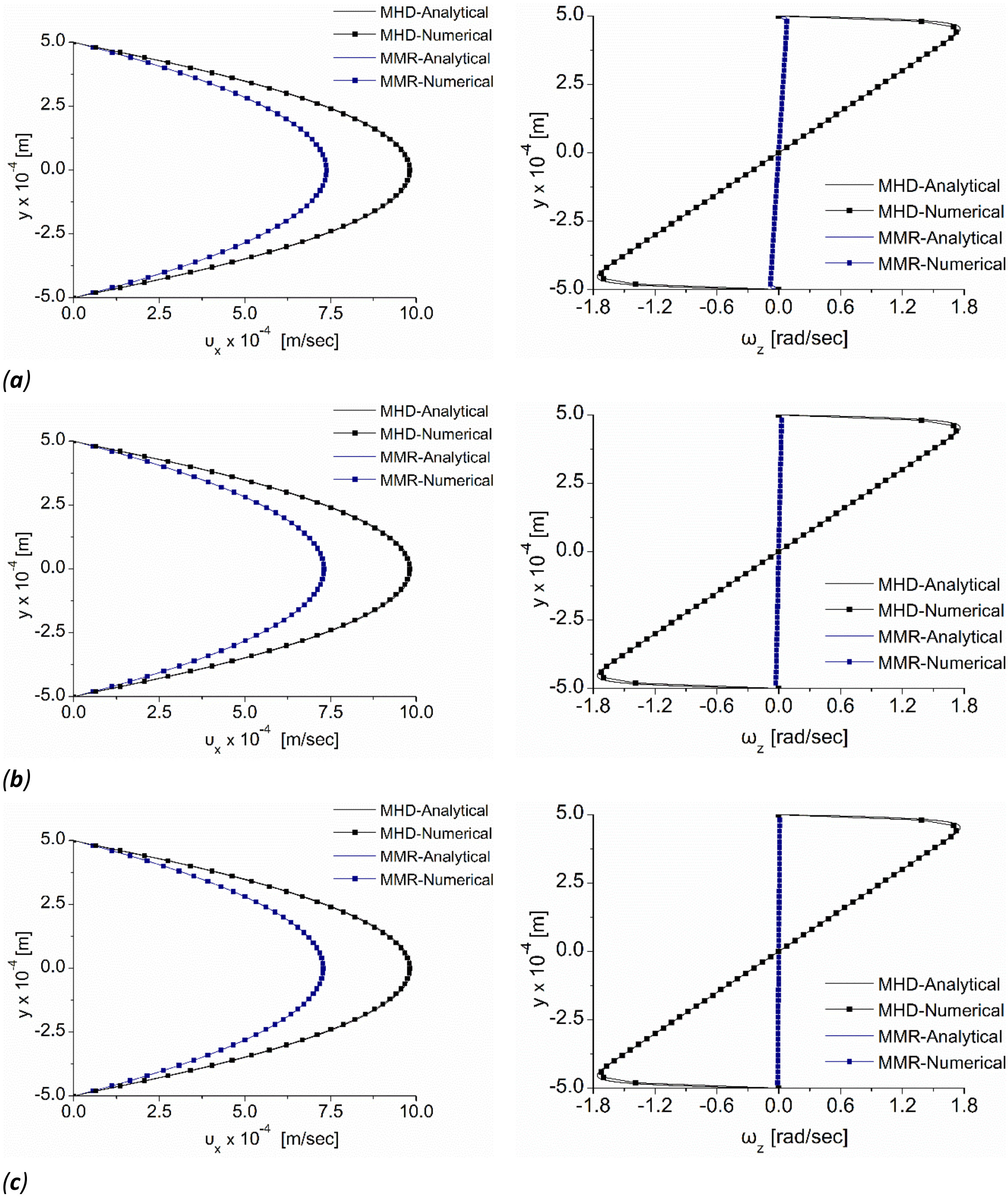

**Figure 7.** Analytical and numerical results for the velocity $\upsilon_x$ (left) and microrotation $\omega_z$ (right) of the MHD and MMR micropolar Poiseuille blood flows with hematocrit $\varphi = 25\%$, channel height of $2L = 0.001\ m$ (arteriole) and applied magnetic field of (a) $1\ T$, (b) $3\ T$ and (c) $8\ T$.

Figure 8 illustrates the analytical and numerical results for the MHD micropolar Poiseuille blood arteriole flow with a hematocrit of $\varphi = 45\%$, while ignoring the effect of micromagnetorotation (acronym MHD) and considering its effect (acronym MMR). The results are presented for three different values of the applied magnetic field: $1\ T$ , $3\ T$, and $8\ T$. It is clear that the numerical and analytical results show very good agreement for both velocity and microrotation, irrespective of whether MMR is considered or ignored, across all intensities of the applied magnetic field. The errors are:

- $ER_\upsilon = 0.6\ \%$ and $ER_\omega = 1.24\ \%$ for the MHD case and magnetic field of $1\ T$.
- $ER_\upsilon = 0.1\ \%$ and $ER_\omega = 0.48\ \%$ for the MMR case and magnetic field of $1\ T$.

- $ER_{\upsilon} = 0.6$ % and $ER_{\omega} = 1.24$ % for the MHD case and magnetic field of 3 $T$.
- $ER_{\upsilon} = 0.045$ % and $ER_{\omega} = 0.46$ % for the MMR case and magnetic field of 3 $T$.
- $ER_{\upsilon} = 0.6$ % and $ER_{\omega} = 1.24$ % for the MHD case and magnetic field of 8 $T$.
- $ER_{\upsilon} = 0.022$ % and $ER_{\omega} = 0.41$ % for the MMR case and magnetic field of 8 $T$.

It is evident that the epotMicropolarFoam and epotMMRFoam solvers provide accurate results with minimal error, even for higher hematocrit levels values.

It is also noteworthy that the magnetic field, when excluding the effect of MMR, has a minimal influence on micropolar Poiseuille blood flow, even at higher hematocrit values. Conversely, the impact of MMR on the MHD micropolar Poiseuille blood flow is significant, resulting in a 38% velocity reduction at a magnetic field intensity of $8\,T$, and an almost 99.99% reduction in microrotation. The very small influence of the magnetic field in the MHD (no-MMR) case is consistent with the low electrical conductivity of blood, which makes the Lorentz force weak. In contrast, when MMR is included, the magnetic torque aligns the particles with the field direction. This suppresses internal rotation and increases the effective flow resistance, which explains the noticeable drop in velocity and microrotation.

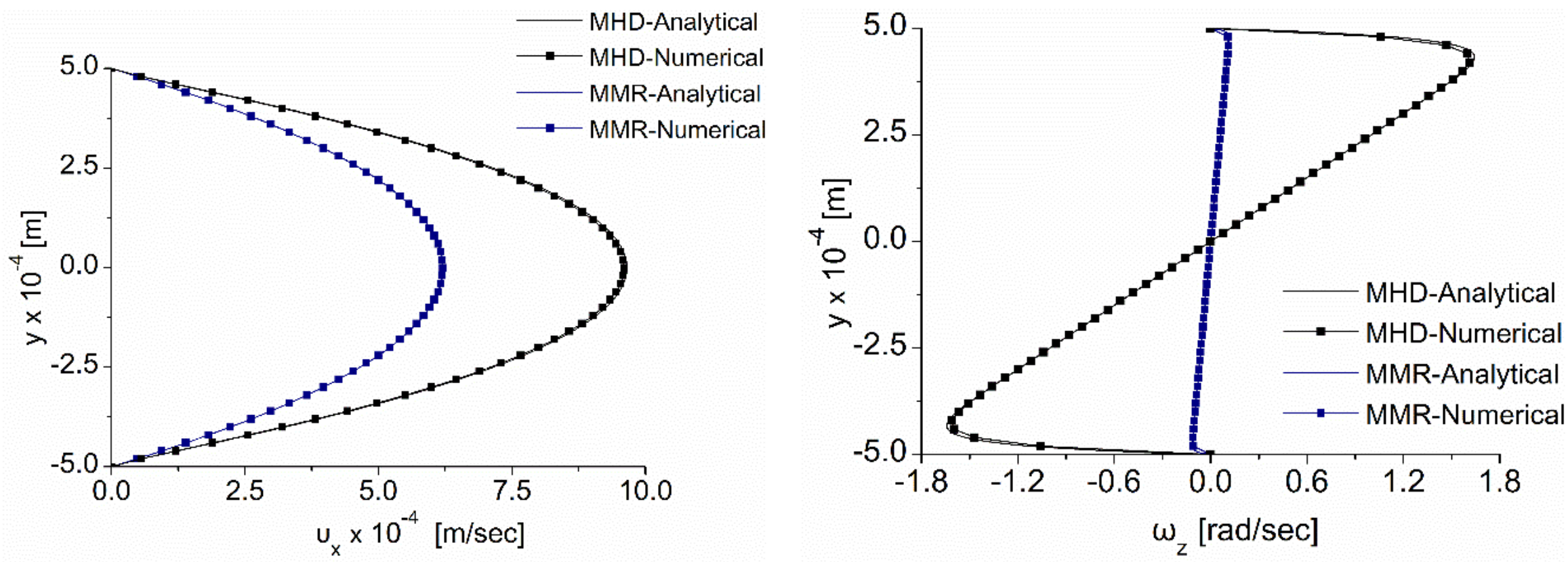

*(a)*

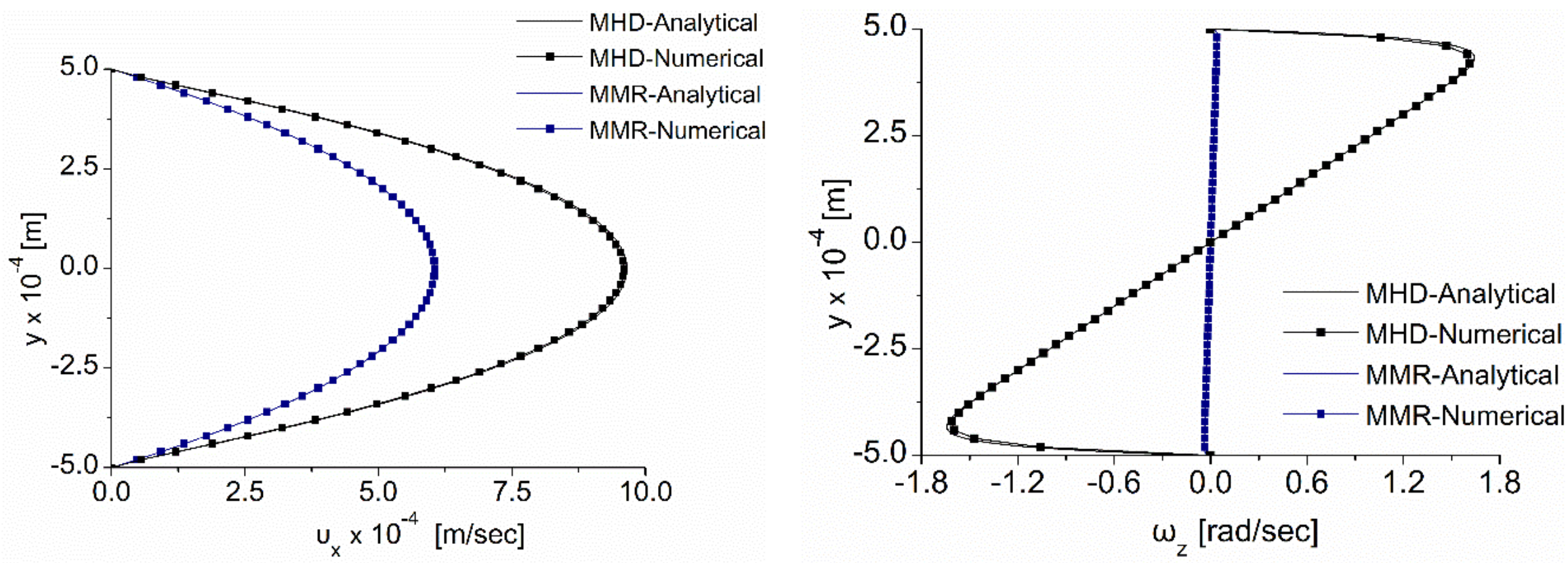

*(b)*

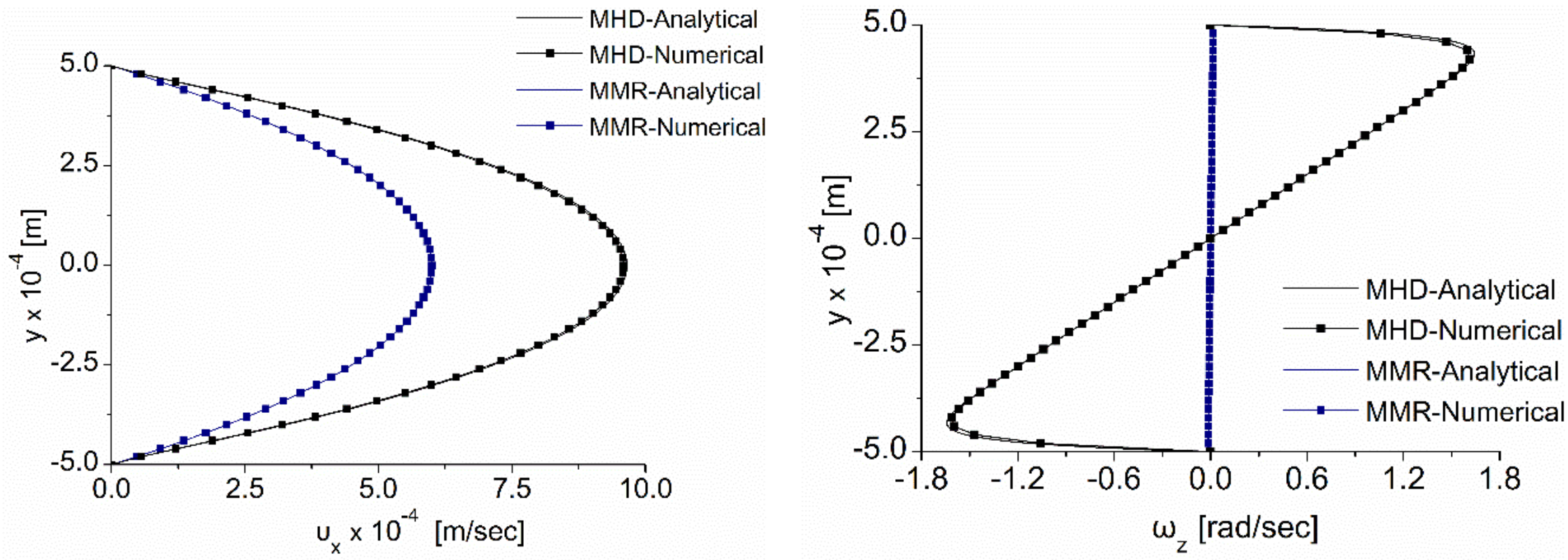

*(c)*

**Figure 8.** Analytical and numerical results for the velocity $\upsilon_x$ (left) and microrotation $\omega_z$ (right) of the MHD and MMR micropolar Poiseuille blood flows with hematocrit $\varphi = 45\%$, channel height of $2L = 0.001\ m$ (arteriole) and applied magnetic field of (a) $1\ T$, (b) $3\ T$ and (c) $8\ T$.

## 4.2. Effect of MMR on an MHD blood flow through a 3D artery

Following the validation of the OpenFOAM solvers, numerical results for MHD micropolar blood flow in an artery are presented using two hematocrit values ($\varphi = 25\%$ and $\varphi = 45\%$) and varying intensities of the applied magnetic field ($1\ T$ and $5\ T$). Emphasis is given to the effect of micromagnetorotation (MMR) on characteristic flow variables, such as velocity and microrotation. To facilitate comparison, these variables are derived for Newtonian blood flow, the corresponding micropolar blood flow without an applied magnetic field, and MHD blood flow, both with and without considering the MMR term. For these simulations, the solvers icoFoam, micropolarFoam, epotMicropolar, and epotMMRFoam are used. The geometric configuration of the artery was chosen as the simplest model for blood flow, using an appropriate pressure gradient.

### 4.2.1. Case setup

As mentioned above, the numerical results for MHD micropolar blood flow in a simple artery, accounting for the effect of micromagnetorotation, are presented here. A simple cylinder with a radius of $R = 0.0015\ m$ (where the diameter $D = 2R = 0.003\ m$) and a length of $l = 0.06\ m$ is employed. Figure 9 shows the domain of the cylinder as depicted in Paraview. The magnetic field $H_0$ is applied transverse to the direction of the flow. The 3D cylindrical coordinate system $(r, \theta, z)$ is utilized, where $z$ is the axial coordinate, $r$ is the radial coordinate, and $\theta$ is the azimuthal angle. It is known that for a Haagen-Poiseuille flow in a cylinder, which is the configuration used here for modelling the artery flow, the following assumptions are made:

- The radial and azimuthal velocity components are zero, i.e., $\upsilon_r = \upsilon_\theta = 0$.
- The flow is axisymmetric, i.e., $\frac{\partial}{\partial \theta} = 0$.

In this manner, the components of the velocity and microrotation are $\boldsymbol{\upsilon} = 0, 0, \upsilon_z(r)$ and $\boldsymbol{\omega} = 0, \omega_\theta(r), 0$, respectively. No-slip boundary conditions are imposed for the velocity and Condiff-Dahler conditions are applied for microrotation, i.e., $\upsilon_z(R) = 0$ and $\omega_\theta(R) = 0$.

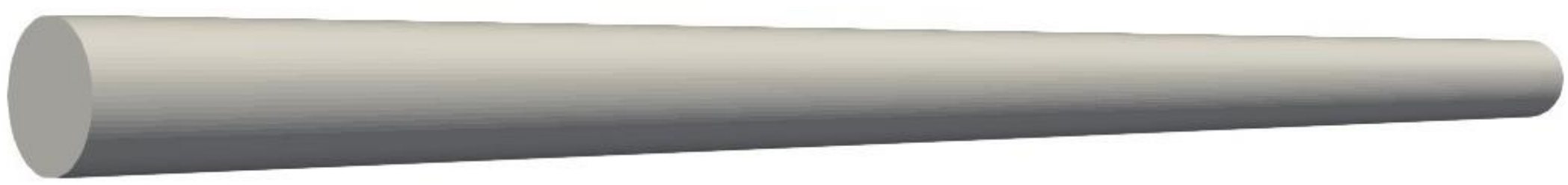

**Figure 9.** Geometrical configuration of the artery in Paraview.

The discretization was achieved using a hexagonal O-grid mesh. The computational mesh was created with the open-source platform Salome. Figure 10 illustrates the computational mesh as seen from Paraview.

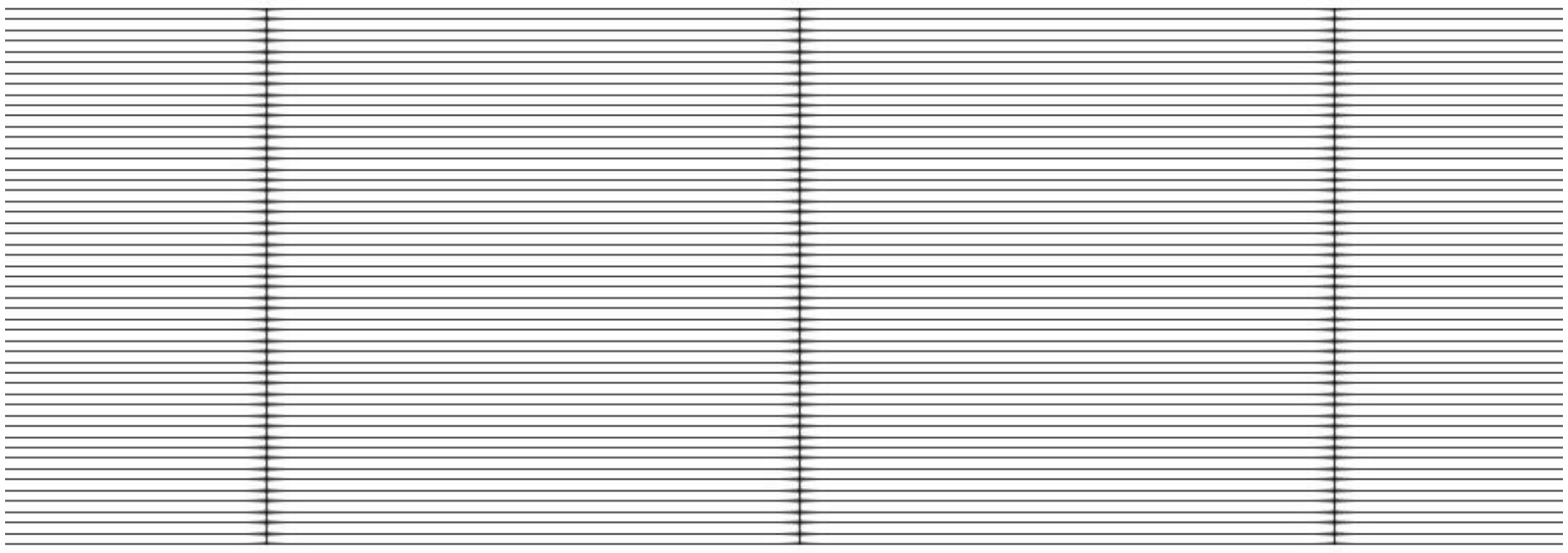

***(a)***

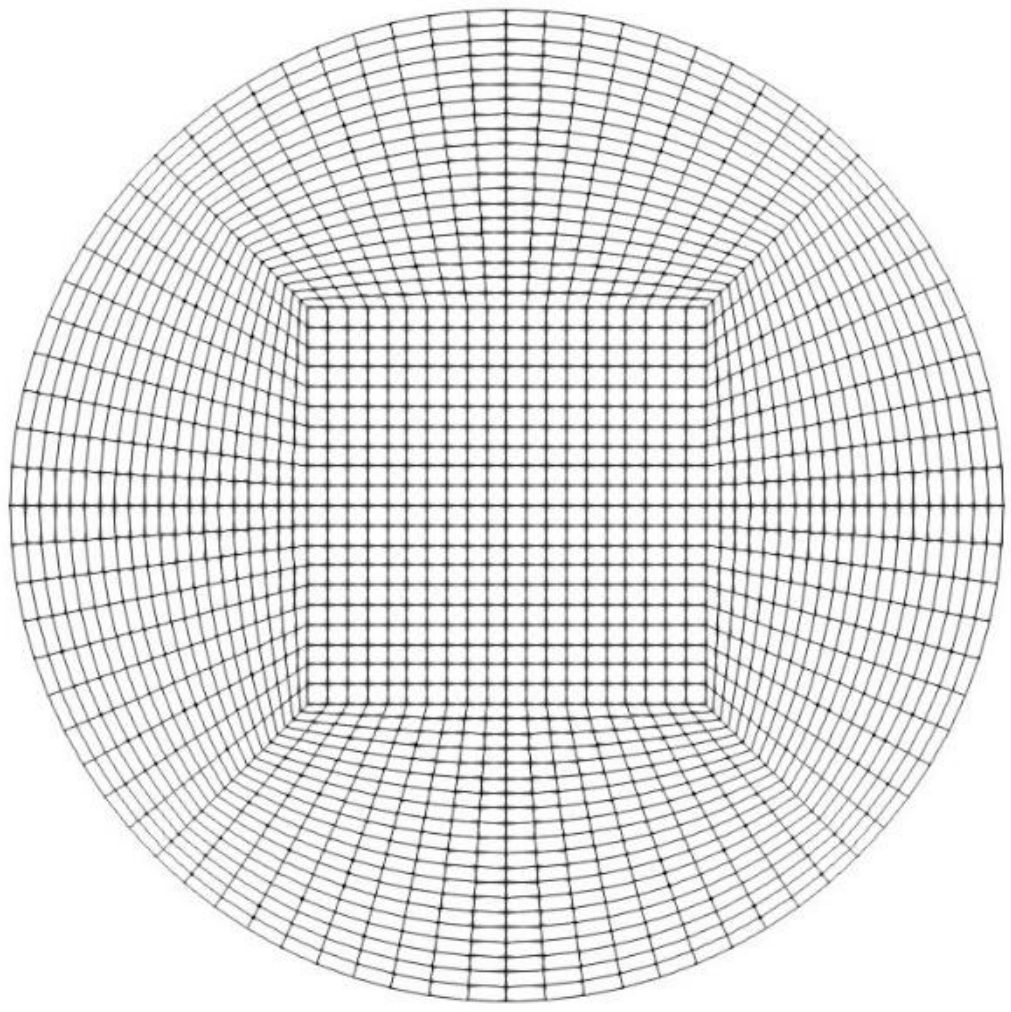

***(b)***

**Figure 10.** Computational mesh for the artery as it is depicted in Paraview from (a) the lengthwise position and (b) the diameter.

The independence of the mesh was tested by comparing the maximum velocity, vorticity, and microrotation values for all flow cases - Newtonian, micropolar, and MHD - by ignoring and considering the effect of the MMR term using different cell densities. Starting from a coarse mesh of 36 cells along the diameter and 16 cells along the artery, the grid was refined until the numerical results reached an error smaller than 2% (which was the error between the analytical and numerical results for the case of Poiseuille flow). In the cases of the 60 × 20 and 75 × 25 grid densities, the corresponding numerical errors were calculated at approximately 0.4%, 1.8%, and 2.4%. Due to the increase in computational time with the increase in grid density, the 60 × 20 mesh was found to be the optimal choice. The computational time for the 75 × 25 mesh was too large; thus, this mesh was not selected.

As in the simulation of planar Poiseuille blood flow, here, the time step for each simulation was chosen to ensure that the Courant number remained less than one $Co < 1$. The duration of the simulation was selected to allow the flow to reach a steady state and become fully developed, i.e., $\partial v_z / \partial z = 0$. The same criterion of the fluid passing through the domain at least 10 times was applied again, consistent with the planar Poiseuille blood flow simulations.

Similar to the planar Poiseuille MHD blood flow, two values for the hematocrit are used, i.e., $\varphi = 25\%$ and $\varphi = 45\%$. Considering the physical properties of blood, these values were again derived from the work of Tzirtzilakis [17, 19], Ariman et al. [21] and Pai et al. [26] and they are tabulated in Table 1. In this study, two values of the applied magnetic field 's intensity are used, which were also used in the experiment of Pai et al. [26], i.e., $H_0 = 795{,}774.72\ A/m$ and $H_0 = 3{,}978{,}873.58\ A/m$. Reminding that $H_0 = B_0/\mu_0$ which leads to $H_0 = 795{,}774.72\ A/m$ corresponding to $1\ T$ and $H_0 = 3{,}978{,}873.58\ A/m$ corresponding to $5\ T$.

Similar to the planar Poiseuille MHD micropolar flow of the previous section, here, a suitable pressure gradient $G$ was applied to resemble a realistic human blood flow. The maximum velocity value of $v_{z_{max}} = 0.1\ m/sec$ was selected and the corresponding pressure gradient was calculated using the relation $v_{z_{max}} = {GR^2}/{4\mu}$. In this manner, it was derived that $G = 355.556\ Pa/m$. Considering that $l = 0.06$ and ${P_2 - P_1}/{l} = G$, and that the outlet pressure $P_2 = 0\ Pa$, the inlet pressure is $P_1 = 21.336\ Pa$. It should be noted again that the pressure values should be divided by density ($\rho = 1050\ kg/m^3$), because OpenFOAM uses kinematic pressure $P/\rho$ as input.

#### 4.2.2. Results

First, numerical results for the Newtonian artery flow and the micropolar artery flow using two different hematocrit values, i.e., $\varphi = 25\%$ and $\varphi = 45\%$ are analyzed and compared. For the Newtonian simulation, the icoFoam solver was used, while for the micropolar simulation, the micropolarFoam solver was applied. Figure 11 illustrates the velocity and microrotation for the Newtonian and micropolar blood flows in the artery. As expected, the microrotation field is not defined for the Newtonian blood flow and is not depicted in Figure 11b. As can be seen in Figure 11a, the velocity of the micropolar blood flow in the artery is slightly less than that of the Newtonian flow; the velocity for $\varphi = 45\%$ is slightly smaller than the one of $\varphi = 25\%$. A maximum difference of $\Delta v_{z_{max}} = 1.1\%$ is calculated between the Newtonian velocity and the velocity for $\varphi = 25\%$ at the center of the diameter of the artery, while $\Delta v_z = 2\%$ is calculated between the Newtonian velocity and the velocity for $\varphi = 45\%$. Because the artery diameter is much larger than the microstructural length scale, micropolar effects become weak. This is why the velocity difference between Newtonian and micropolar flow is small, and microrotation remains low for both hematocrit values. (see paper by Aslani et al. [3]). In this manner, there is a small difference between the microrotation profiles for $\varphi = 25\%$ and $\varphi = 45\%$ with $\Delta\omega_{\theta_{max}} = 2.8\%$.

Figure 12 illustrates the numerical results for the MHD micropolar blood flow in an artery with hematocrit $\varphi = 25\%$, both when the micromagnetorotation effect is ignored (referred to as MHD) and when it is included (referred to as MMR). It is evident that the influence of the applied magnetic field on the velocity and microrotation of the blood flow is minimal when the MMR term is disregarded. The maximum velocity and microrotation difference between the simple micropolar and the MHD micropolar blood artery flow is $\Delta v_{z_{max}} = 1.11\%$ and $\Delta\omega_{\theta_{max}} = 3.2\%$ for the magnetic field of $1\ T$, whereas these differences are $\Delta v_{z_{max}} = 1.19\%$ and $\Delta\omega_{\theta_{max}} = 3.9\%$ for the magnetic field of $5\ T$. The impact of the Lorentz force is small due to the relatively low electrical conductivity of blood and the small diameter of the artery.

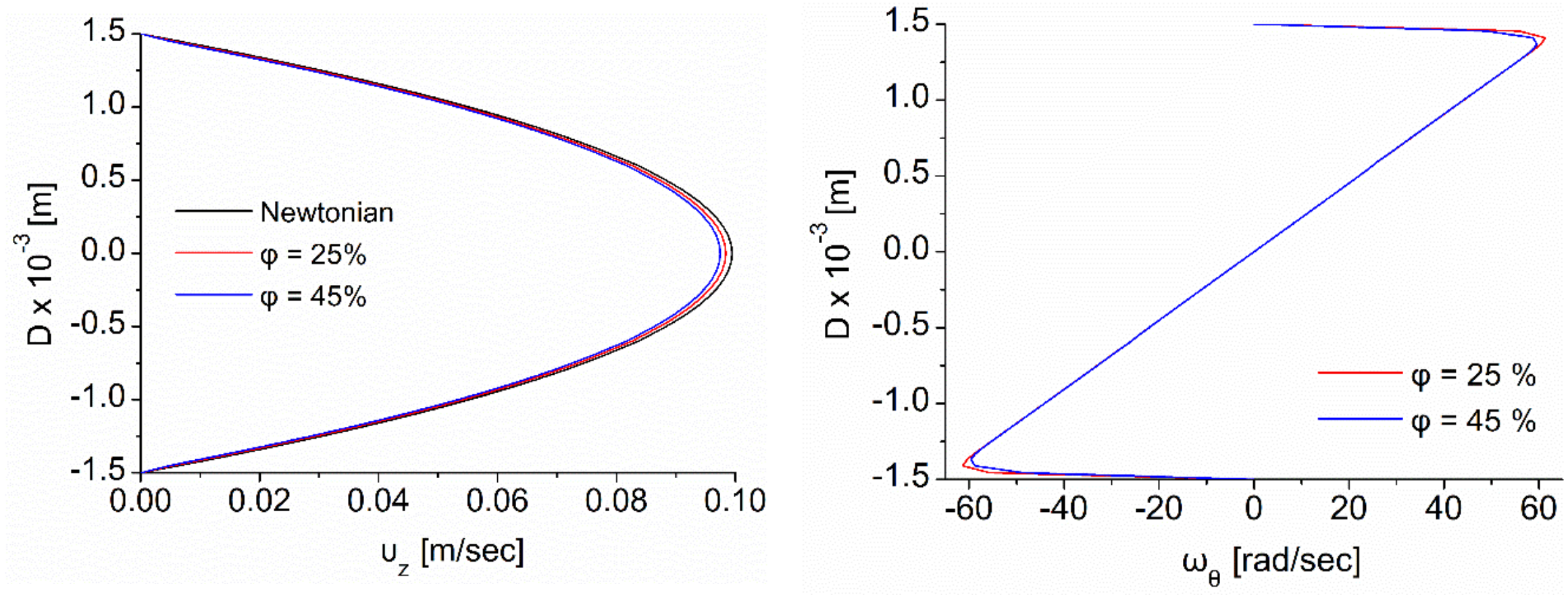

*(a)* *(b)*

**Figure 11.** Numerical results for the (a) velocity $v_z$ and (b) microrotation $\omega_\theta$ of the Newtonian and micropolar Poiseuille blood flows in an artery with hematocrit $\varphi = 25\%$ and $45\%$.

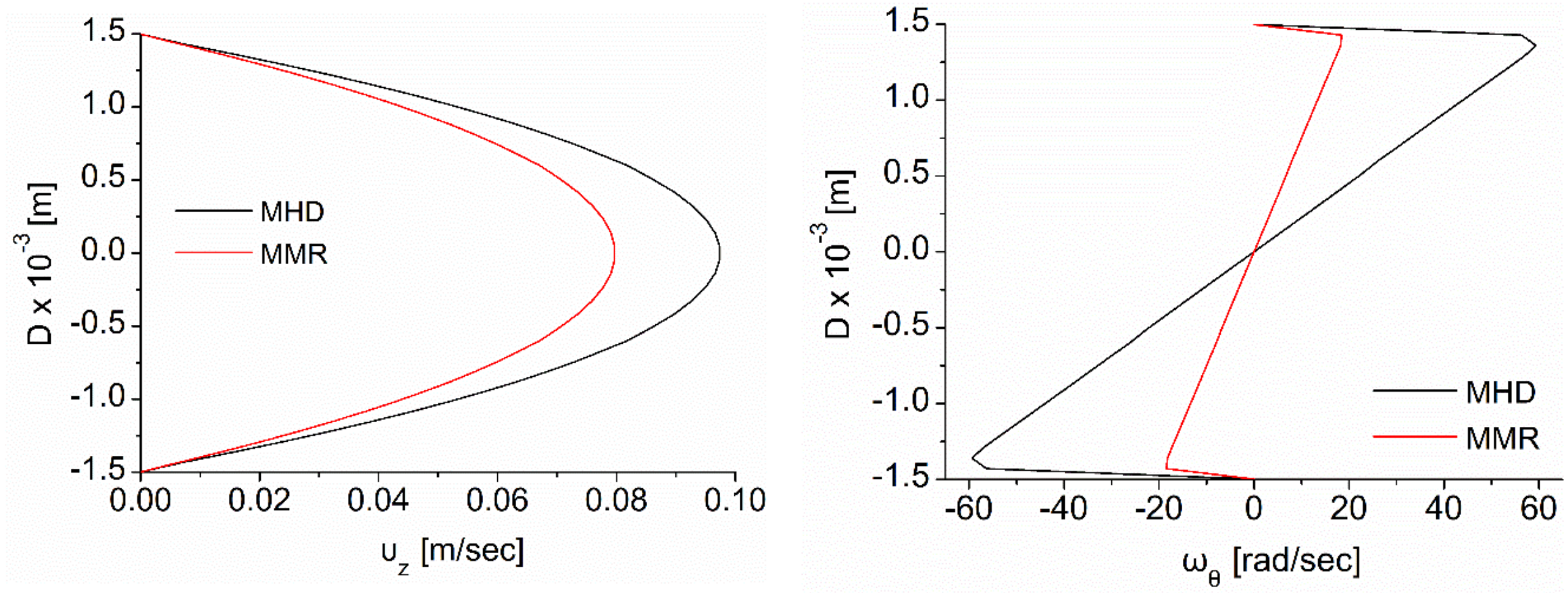

*(a)*

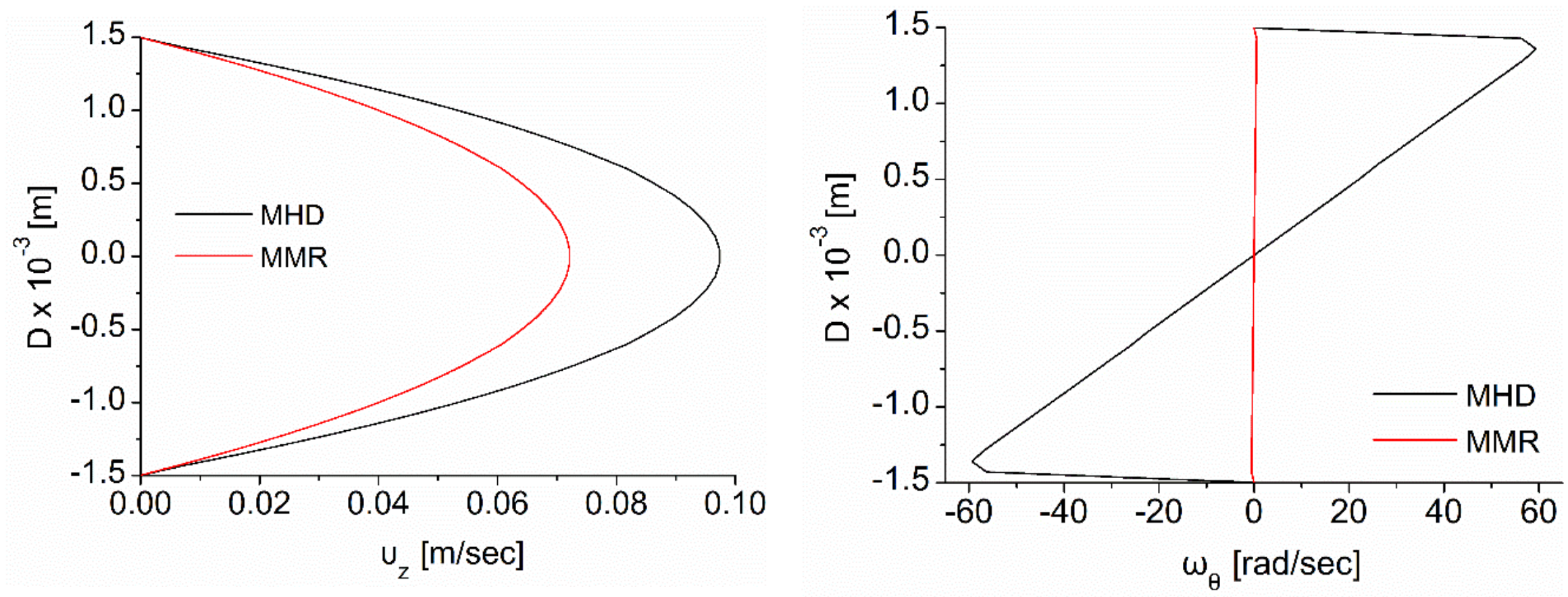

*(b)*

**Figure 12.** Numerical results for the velocity $v_z$ (left) and microrotation $\omega_\theta$ (right) of the MHD and MMR micropolar artery blood flows with hematocrit $\varphi = 25\%$ and applied magnetic field of (a) $1\ T$ and (b) $5\ T$.

As expected from the MHD Poiseuille blood flow results, the influence of the applied magnetic field on micropolar blood artery flow is significant when the effect of the MMR term is included. In this case, the maximum velocity and microrotation difference between the simple micropolar and the MMR micropolar blood artery flow is $\Delta v_{z_{max}} = 19\%$ and $\Delta \omega_{\theta_{max}} = 70\%$ for the magnetic field of 1 $T$, while these differences are $\Delta v_{z_{max}} = 26.7\%$ and $\Delta \omega_{\theta_{max}} = 99.1\%$ for the magnetic field of 5 $T$. Practically, as the intensity of the applied magnetic field increases, the internal rotation of erythrocytes nearly “freezes” because they are oriented parallel to the applied magnetic field. This also reduces the velocity of blood significantly.

Similar to the previous figure, Figure 13 illustrates the numerical results for the MHD micropolar artery flow with hematocrit value of $\varphi = 45\%$ and two different values for the applied magnetic field, i.e., 1 $T$ and 5 $T$, when the effect of the micromagnetorotation is ignored (acronym MHD) and the effect of the latter is included (acronym MMR). As expected from the case of $\varphi = 25\%$, the effect of the applied magnetic field on the velocity and the microrotation of the blood flow is barely noticeable, when the MMR term is ignored. The maximum velocity and microrotation difference between the simple micropolar and the MHD micropolar blood artery flow is $\Delta v_{z_{max}} = 1.29\%$ and $\Delta \omega_{\theta_{max}} = 2.85\%$ for the magnetic field of 1 $T$, while these differences are $\Delta v_{z_{max}} = 1.35\%$ and $\Delta \omega_{\theta_{max}} = 3.26\%$ for the magnetic field of 5 $T$.

Similar to the previous results of this study, the influence of the applied magnetic field on the micropolar blood artery flow is significant when the effect of the MMR term is included. Here, the maximum velocity and microrotation difference between the simple micropolar and the MMR micropolar blood artery flow is $\Delta v_{z_{max}} = 21.2\%$ and $\Delta \omega_{\theta_{max}} = 71.8\%$ for the magnetic field of 1 $T$, while these differences are $\Delta v_{z_{max}} = 38.9\%$ and $\Delta \omega_{\theta_{max}} = 99.7\%$ for the magnetic field of 5 $T$. These differences are more significant compared to the case of $\varphi = 25\%$; increasing hematocrit strengthens the MMR effect because more magnetic particles (erythrocytes) are present in the fluid. As a result, the velocity and microrotation reductions are more pronounced..

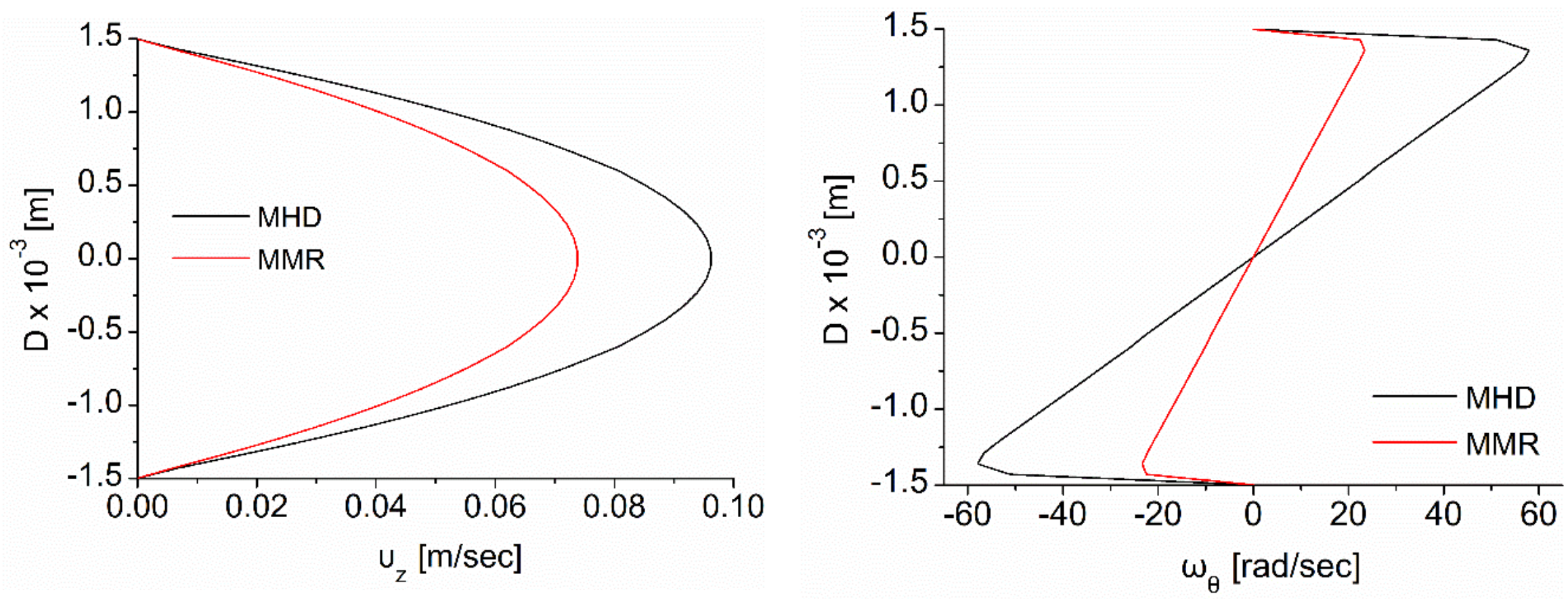

*(a)*

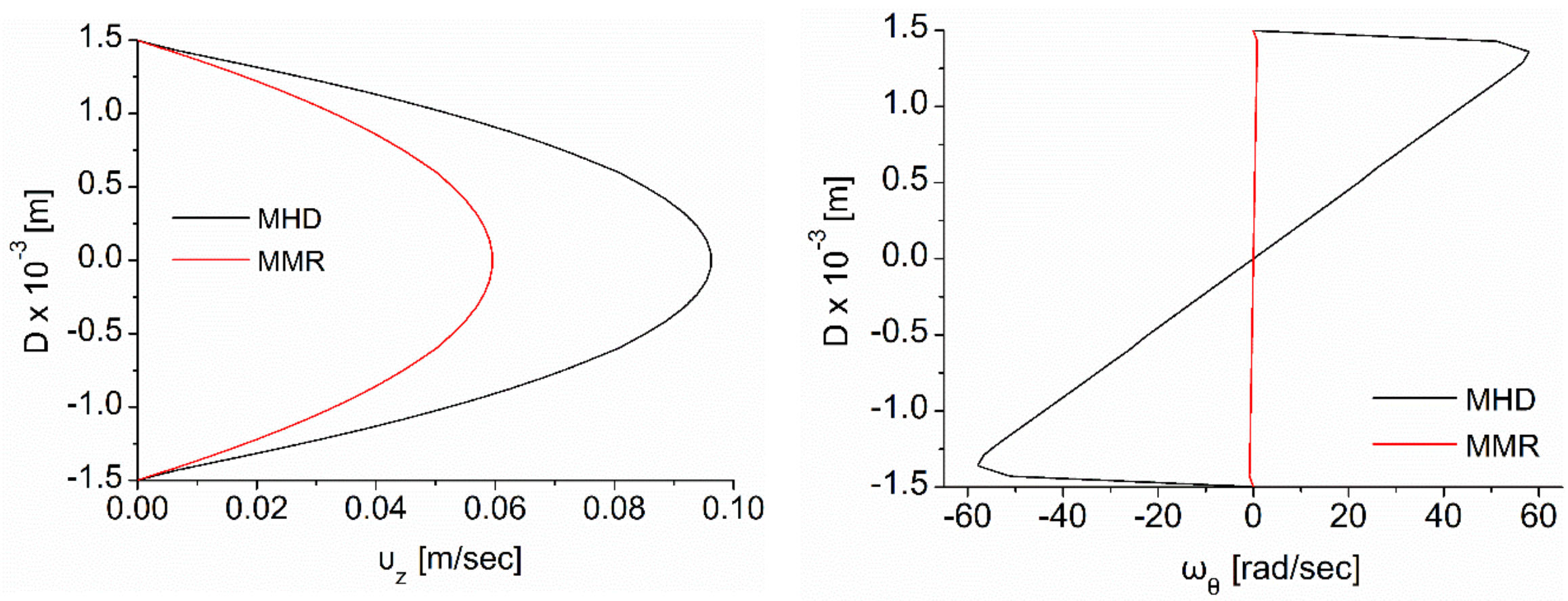

*(b)*

**Figure 13.** Numerical results for the velocity $v_z$ (left) and microrotation $\omega_\theta$ (right) of the MHD and MMR micropolar artery blood flows with hematocrit $\varphi = 45\%$ and applied magnetic field of (a) $1\,T$ and (b) $5\,T$.

## 4.3. Effect of MMR on an MHD blood flow through a 2D aneurysm

Building on the analysis of the MHD blood flow through a simple 3D artery, numerical results for MHD micropolar blood flow in a 2D aneurysm are presented under varying magnetic field intensities ($1\,T$, $3\,T$, and $8\,T$), with the hematocrit level fixed at $\varphi = 45\%$. Particular attention is given to the role of micromagnetorotation (MMR) on key flow characteristics, including streamlines, vorticity and microrotation contours, as well as velocity and microrotation profiles both within and outside the aneurysmal sac. For comparative purposes, these variables are computed for the non-magnetic Newtonian and micropolar blood flows, and the MHD micropolar blood flow with and without the inclusion of the MMR effect. The simulations were carried out using the solvers icoFoam, micropolarFoam, epotMicropolarFoam, and epotMMRFoam. The selected aneurysmal geometry is representative of a clinically significant vascular pathology, thus underscoring the biomedical relevance of the study.

### 4.3.1. Case setup

As mentioned above, in this paper, a 2D micropolar MHD blood through a fusiform aneurysm is numerically examined by acknowledging and ignoring MMR. The Cartesian coordinate system $(x, y, z)$ is used. As can be seen from Figure 14 an external magnetic field $H_0$ is applied transverse to the flow. The velocity components are given as $\boldsymbol{v} = v_x, 0, 0$, while the microrotation components are given as $\boldsymbol{\omega} = 0, 0, \omega_z,$. No-slip boundary conditions are imposed for both velocity and microrotation. The blood flow is driven by a uniform pressure gradient in the $z$ direction. The geometry of the flow is a cylinder of radius $R = 0.0015\,m$ (diameter $D = 2R = 0.003\,m$) and length $l = 0.06\,m$.

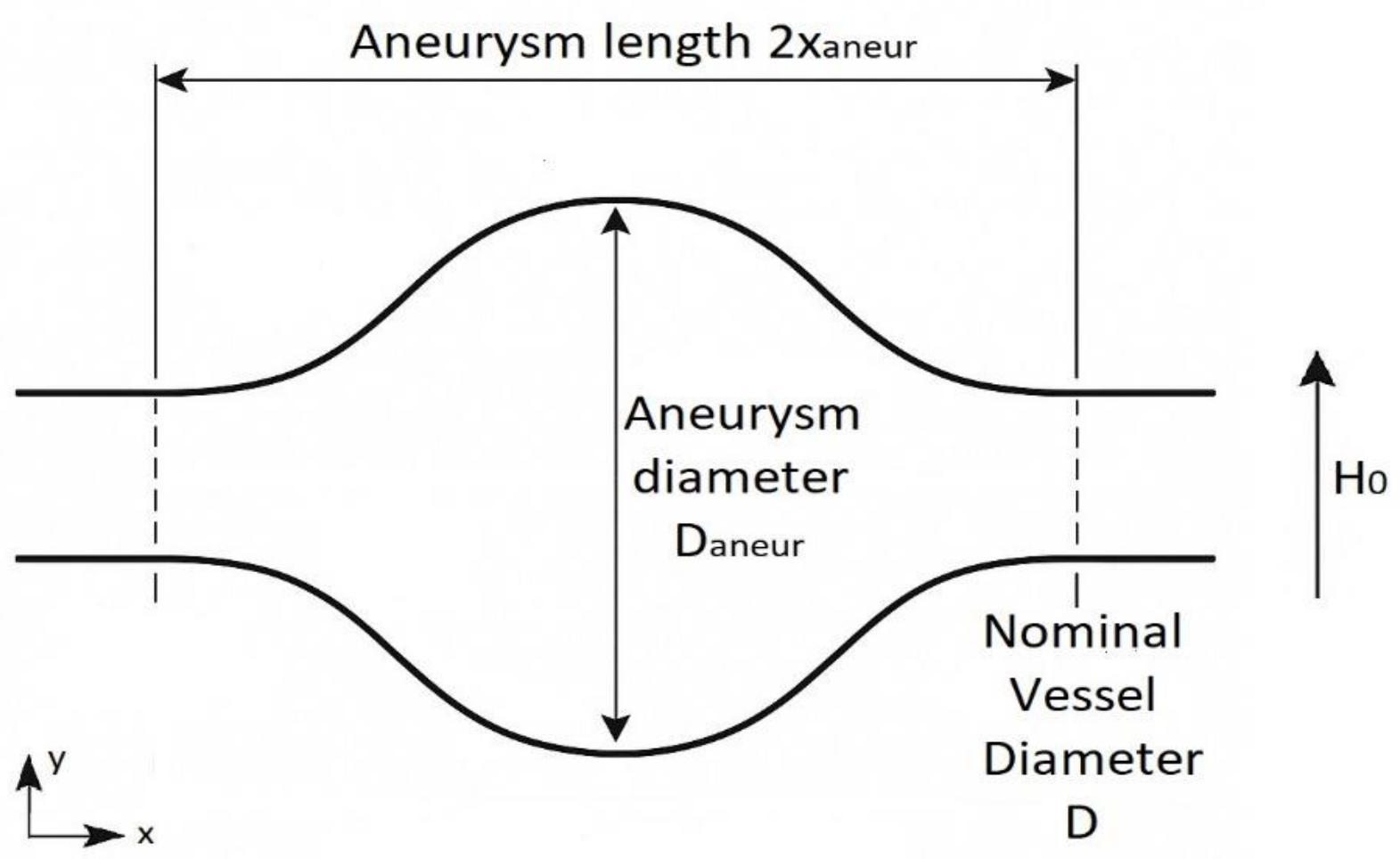

**Figure 14.** Schematic diagram of the 2D aneurysm.

The boundaries of the aneurysm can be drawn with the use of Equation (17):

$$l_{aneur}(x) = R - \frac{R - R_{aneur}}{2}\left(1 + cos\frac{\pi x}{x_{aneur}}\right), \tag{17}$$

where $R_{aneur}$ is the radius of the aneurysm (aneurysm's diameter $D_{aneur} = 2R_{aneur}$) and $2x_{aneur}$ is the total length of the aneurysm. Two aneurysm sizes have been considered, one with a diameter equal to $166\%$ of the nominal vessel diameter ($D_{aneur} = 1.666D = 0.005\,m$), and another with a diameter equal to $200\%$ of the nominal diameter ($D_{aneur} = 2D = 0.006\,m$). The first aneurysm is considered moderate according to the bibliography, while the second aneurysm is considered severe [49-51]. The corresponding total length for the two aneurysms is $2x_{aneur} = 0.005\ m$ for the $166\%$ aneurysm and $2x_{aneur} = 0.0075\ m$ for the $200\%$ aneurysm. The center of the $166\%$ aneurysm is located at $x = 0.0275\ m$ (considering that $x = 0$ at the inlet), while the center of the $200\%$ aneurysm is located at $x = 0.02625\ m$.

The discretization was performed using a hexagonal mesh comprising 40 cells across the vessel diameter and 100 cells along its length. As with the simple artery case, the computational mesh was created using the open-source platform Salome. Figure 15 presents the mesh visualization as rendered in ParaView.

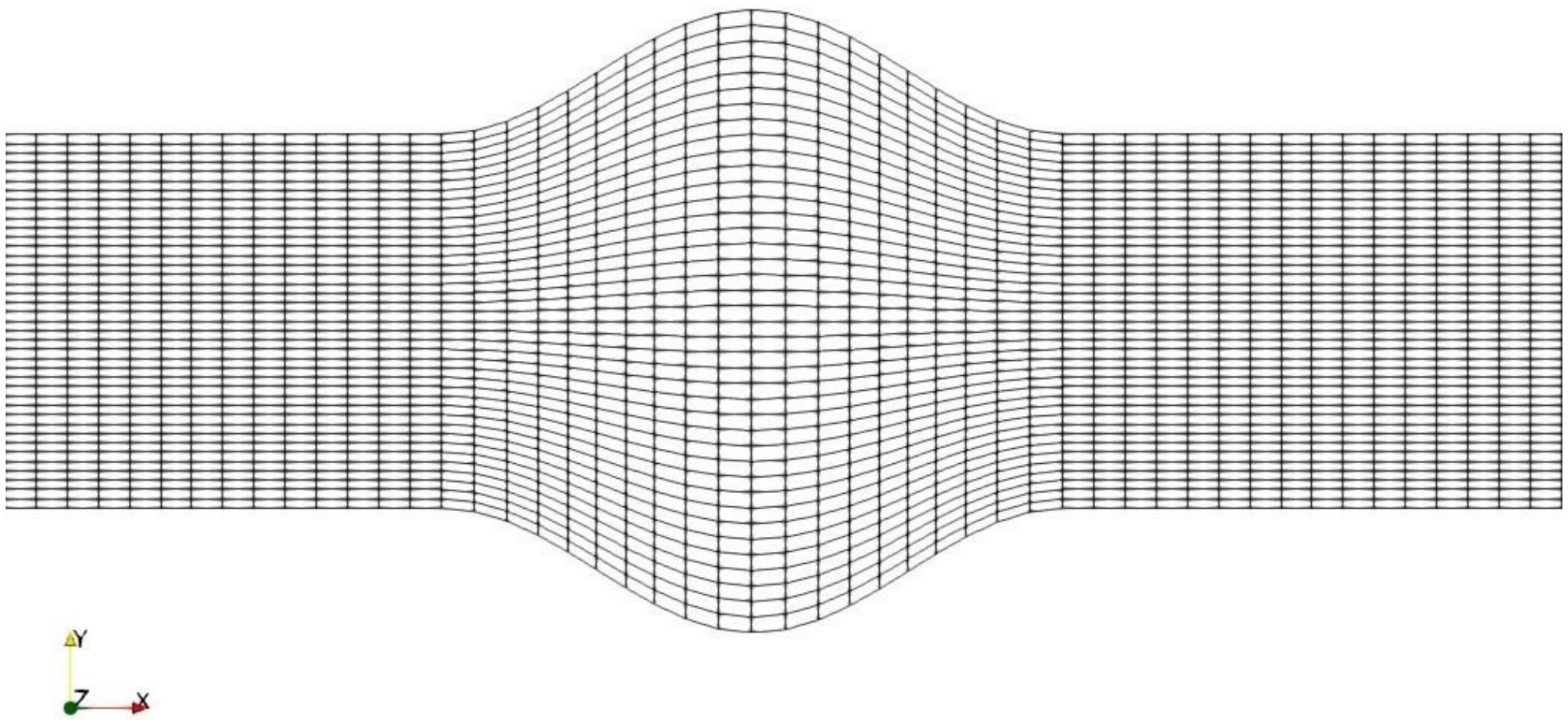

**Figure 15.** Computational mesh for the aneurysm as it is depicted in Paraview.

As in the case of the simple artery, the mesh independence was conducted by comparing the maximum values of velocity, vorticity, and microrotation across all flow cases—Newtonian, micropolar, and MHD—with and without the inclusion of the MMR term, using various cell densities. The process began with a coarse mesh consisting of 30 cells across the vessel diameter and 80 cells along its length, and the grid was progressively refined until the numerical results exhibited an error of less than $2\%$, which corresponds to the discrepancy observed between the analytical and numerical solutions for Poiseuille flow. For the $40 \times 100$ and $50 \times 120$ mesh densities, the respective numerical errors were approximately $0.2\%$, $1\%$, and $2.1\%$. Considering the significant increase in computational time associated with higher mesh densities, the $40 \times 100$ grid was deemed optimal. The $50 \times 120$ mesh was not selected due to its prohibitively high computational cost.

As in the previous two simulations, here, the time step for each simulation was chosen to ensure that the Courant number remained less than one $Co < 1$. The duration of the simulation was selected to allow the flow to reach a steady state and become fully developed, i.e., ${\partial v_x}/{\partial x} = 0$. The same criterion of the fluid passing through the domain at least 10 times was applied again, consistent with the planar Poiseuille and artery blood flow simulations.

As mentioned above, one value for the hematocrit was used, i.e., $\varphi = 45\%$. The physical properties of blood were again derived from the papers of Tzirtzilakis [17, 19], Ariman et al. [21], and Pai et al. [26], as summarized in Table 1. For the present simulation, the same three magnetic field intensities as in the Poiseuille blood flow case were applied, i.e., $1\ T$, $3\ T$, and $8\ T$ due to their consistency with the experiments. Additionally, an appropriate pressure gradient $G$ was imposed to approximate realistic human blood flow. Here, the maximum velocity value of $v_{x_{max}} = 0.3\ m/sec$ was selected and using the same mathematical relations as in the case of the Poiseuille blood flow, the corresponding pressure gradient was calculated as $G = 933.33\ \ Pa/m$. Given that $l = 0.06$ and that the outlet pressure $P_2 = 0\ Pa$, the inlet pressure is $P_1 = 56\ Pa$.

#### 4.3.2. Results

#### 4.3.2.1. Results for the aneurysm with a 166% dilation

Figure 16 presents the streamlines for the Newtonian and the micropolar blood flow through a 166% aneurysm with no external magnetic field applied to the flow. One can immediately notice that as the fluid approaches the aneurysmal bulge, the streamlines remain closely packed and nearly parallel, indicating a laminar, plug-like behavior. This region is characterized by minimal disturbances, consistent with a parabolic velocity profile. Upon entering the aneurysm, the sudden radial expansion causes a local deceleration near the walls. This leads to flow separation and the emergence of two large symmetric recirculation zones, both in the upper and lower regions of the aneurysmal sac. In the case of the micropolar blood flow, these recirculation zones are slightly shorter and more confined. This is due to microrotation diffusion and the intrinsic angular momentum transport in micropolar fluids, which dampens the formation of vortices. Moreover, there is a small reduction in the maximum velocity value. It should be noted that it was expected that the micropolar effects to be small, due to the relatively large size of the vessel. Moreover, the slightly smaller recirculation zones in the micropolar case occur because microrotation diffuses angular momentum and tends to smooth sharp gradients, which naturally weakens vortex formation.

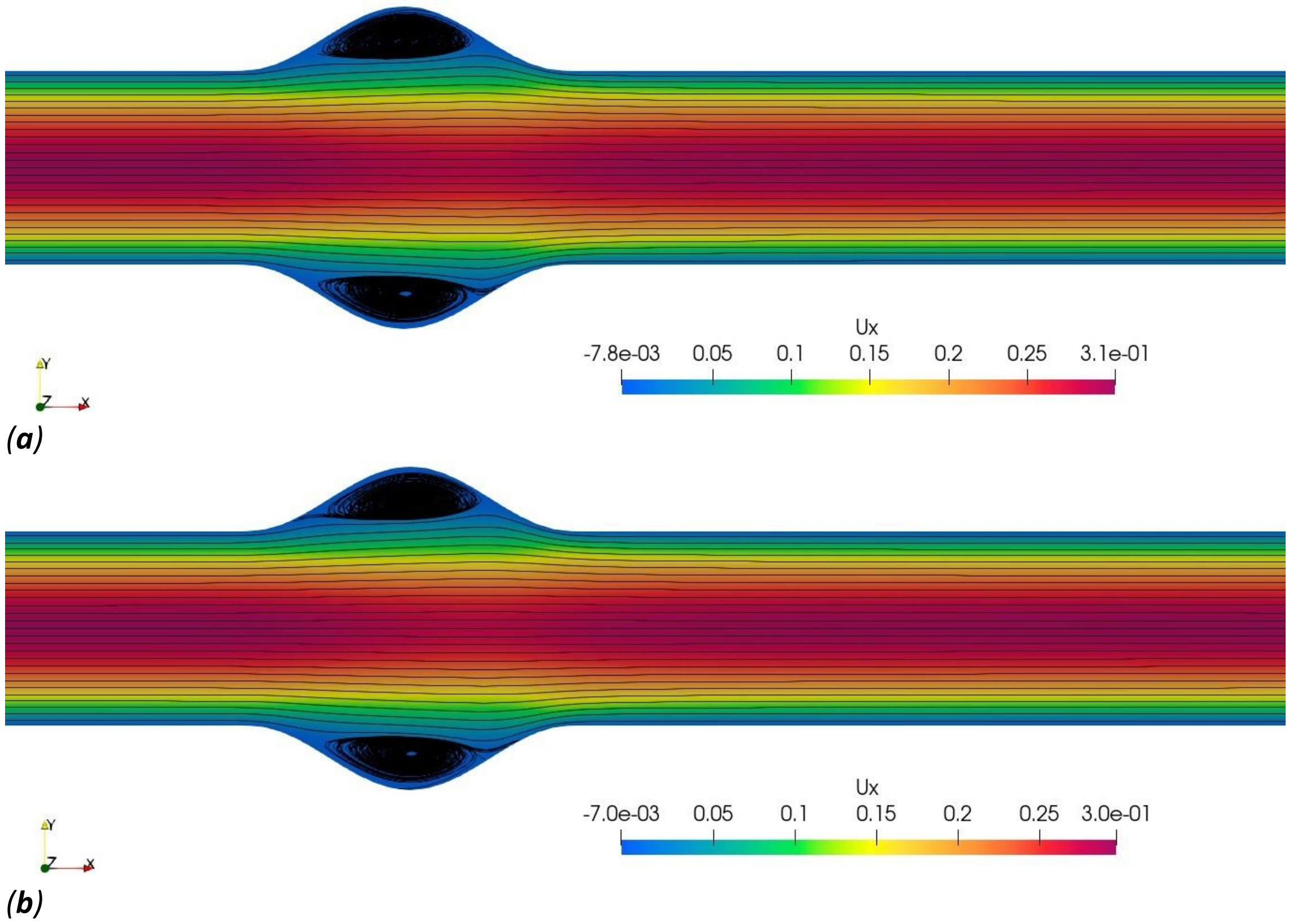

**Figure 16**. Streamlines for blood flow through a 166% aneurysm using (a) Newtonian modelling, and (b) micropolar modelling without an applied magnetic field.

Figure 17 illustrates the vorticity contours for the Newtonian and the micropolar blood flow through a 166% aneurysm with no external magnetic field applied to the flow. It is evident that two intense regions of opposing-sign vorticity are observed within the aneurysmal sac. These correspond directly to the recirculation zones identified in the streamlines, confirming the presence of a symmetric vortex pair. The highest vorticity magnitudes are concentrated near the boundary layers where the flow detaches and reattaches. On the other hand, the central region of the vessel, particularly along the core, shows minimal vorticity (yellow values near zero). In the case of the micropolar blood flow, vorticity gradients are smoothed, indicating a thicker boundary layer near the vessel walls. Moreover, peak vorticity values are noticeably lower in the micropolar flow case. These phenomena were expected, as vorticity should be dampened due to the existence of microrotation, ensuring that the total angular momentum remains constant. The intense vorticity near the aneurysm walls reflects local flow separation caused by the sudden expansion. This behavior is classical in aneurysmal flows and demonstrates that the solver correctly captures shear-driven instabilities.

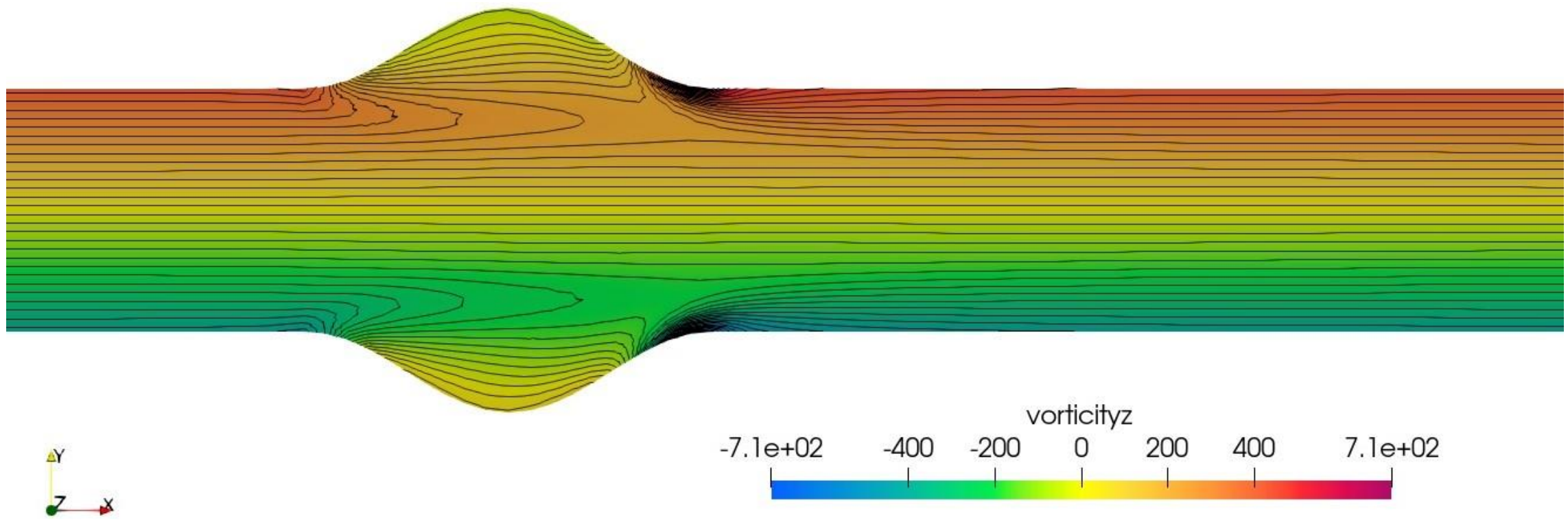

***(a)***

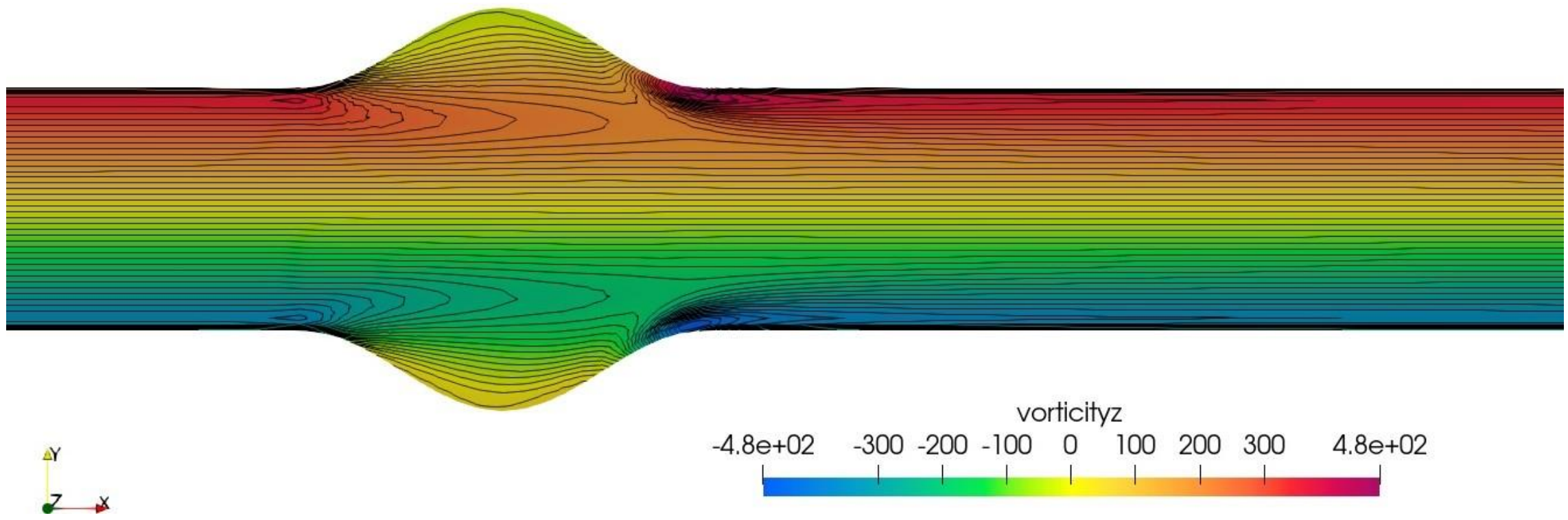

***(b)***

**Figure 17**. Vorticity contour plots for blood flow through a 166% aneurysm using (a) Newtonian modelling, and (b) micropolar modelling without an applied magnetic field.

Figure 18 shows the microrotation contours for the micropolar blood flow through a 166% aneurysm with no external magnetic field applied to the flow. As expected, the microrotation closely resembles the vorticity, exhibiting an antisymmetric pattern about the vessel's centerline, with positive values concentrated in the upper half and negative values in the lower half of the domain. This behavior reflects the rotational response of the fluid microstructure, especially near the vessel walls where velocity gradients are strongest. In the core of the vessel, microrotation values approach zero, consistent with the plug-like velocity profile typical in such flows.

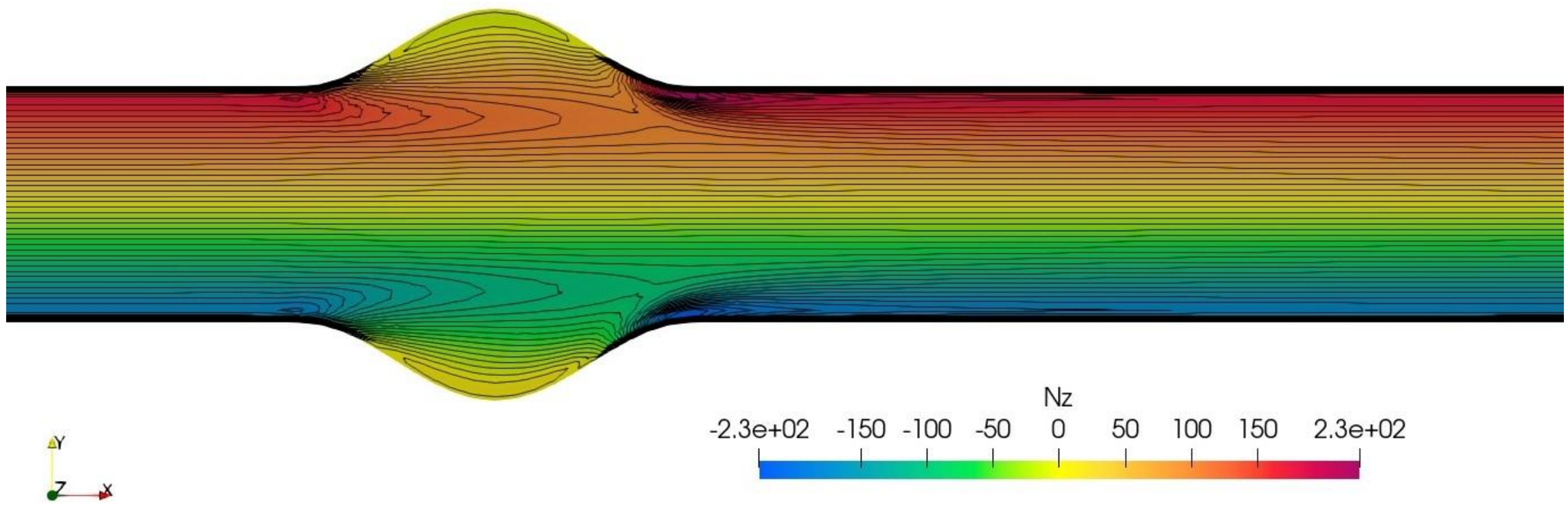

**Figure 18**. Microrotation contour plot for blood flow through a 166% aneurysm using micropolar modelling without an applied magnetic field.

Figure 19 presents the streamlines for the 166% aneurysm based on the MHD micropolar fluid model, both with and without accounting for the micromagnetorotation (MMR) effect. The simulations were conducted under varied magnetic field intensities of $1\ T$, $3\ T$, and $8\ T$. The results indicate that the streamlines for micropolar and MHD micropolar blood flow through the aneurysm exhibit negligible differences when the MMR term is excluded, regardless of the magnetic field strength. This observation was expected, given that the Lorentz force exerts only a minimal influence on the flow due to the relatively low electrical conductivity of blood. However, when the MMR is taken into account, the velocity is significantly diminished, as indicated by the maximum velocity value (approximately $0.19\ m/s$ in $1\ T$ and $0.18\ m/sec$ in $3\ T$ and $8\ T$ compared to $0.3\ m/s$ in the non-magnetic micropolar case). The recirculation zones within the aneurysmal sac persist but appear less compact and intensified. The streamline loops are tighter and more localized compared to the non-magnetic micropolar case, indicating an enhanced vortex confinement. This suggests that micromagnetorotation contributes to vortex stabilization, a result that has been predicted by Aslani et al. [28]. When MMR is included, the magnetic torque aligns the particles and suppresses their internal rotation. This reduces local shear production and prevents the development of strong recirculation cores, explaining why the flow becomes more stable inside the aneurysm.

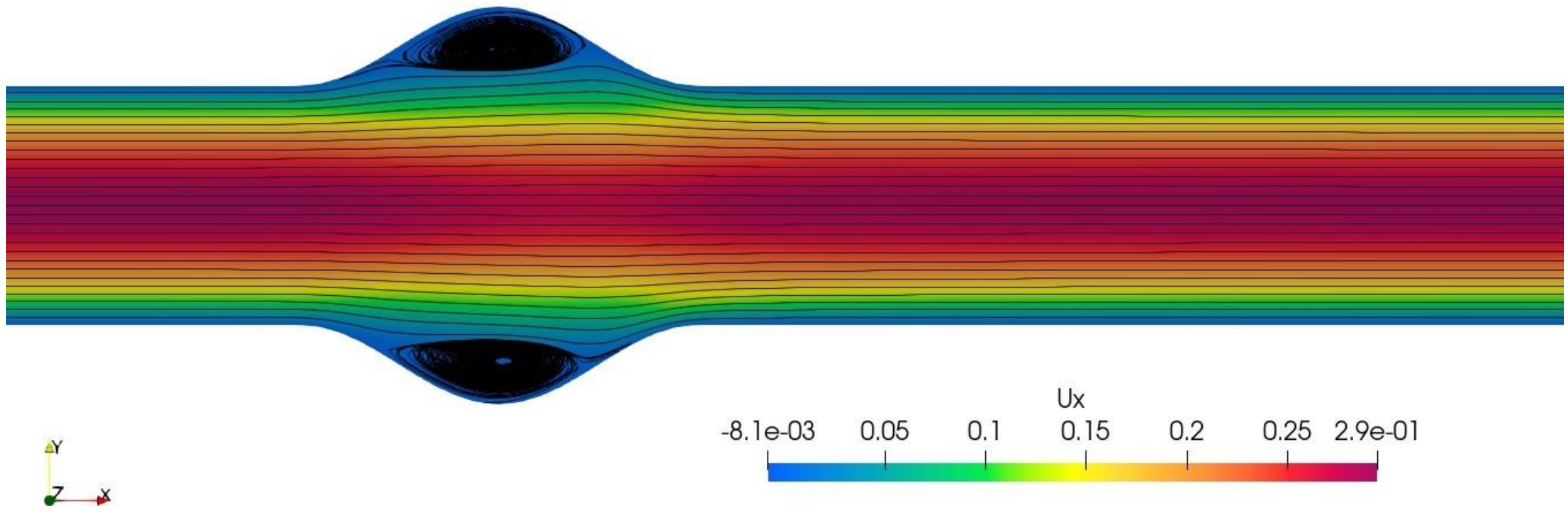

***(a.i)***

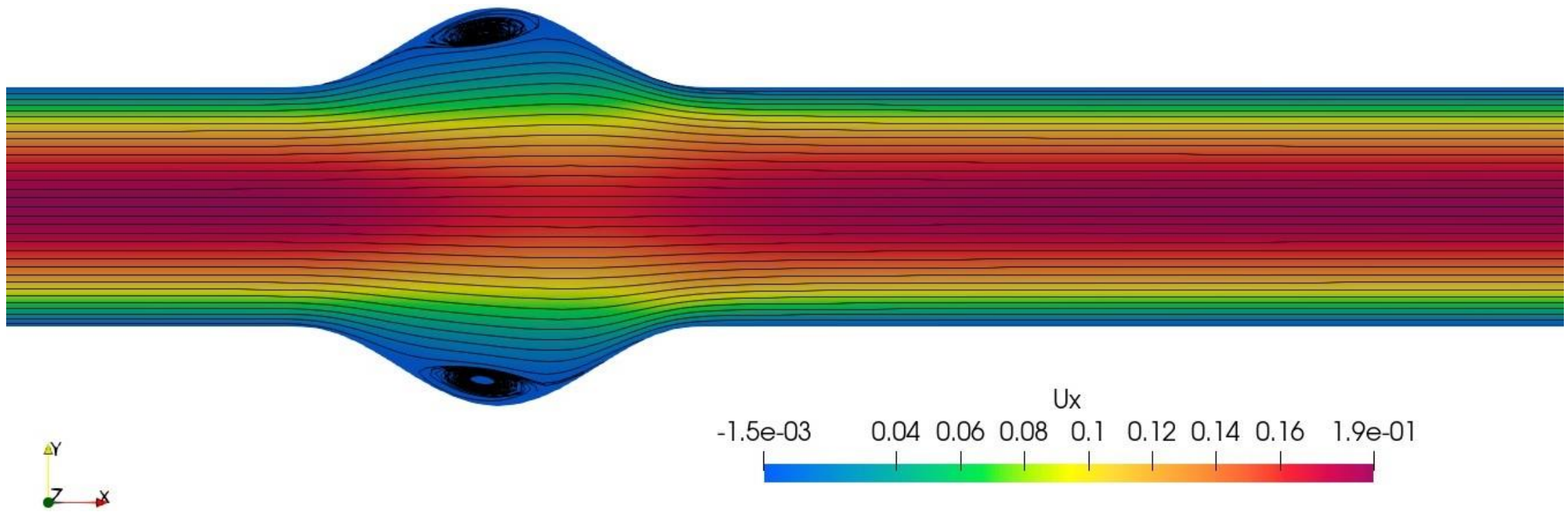

***(a.ii)***

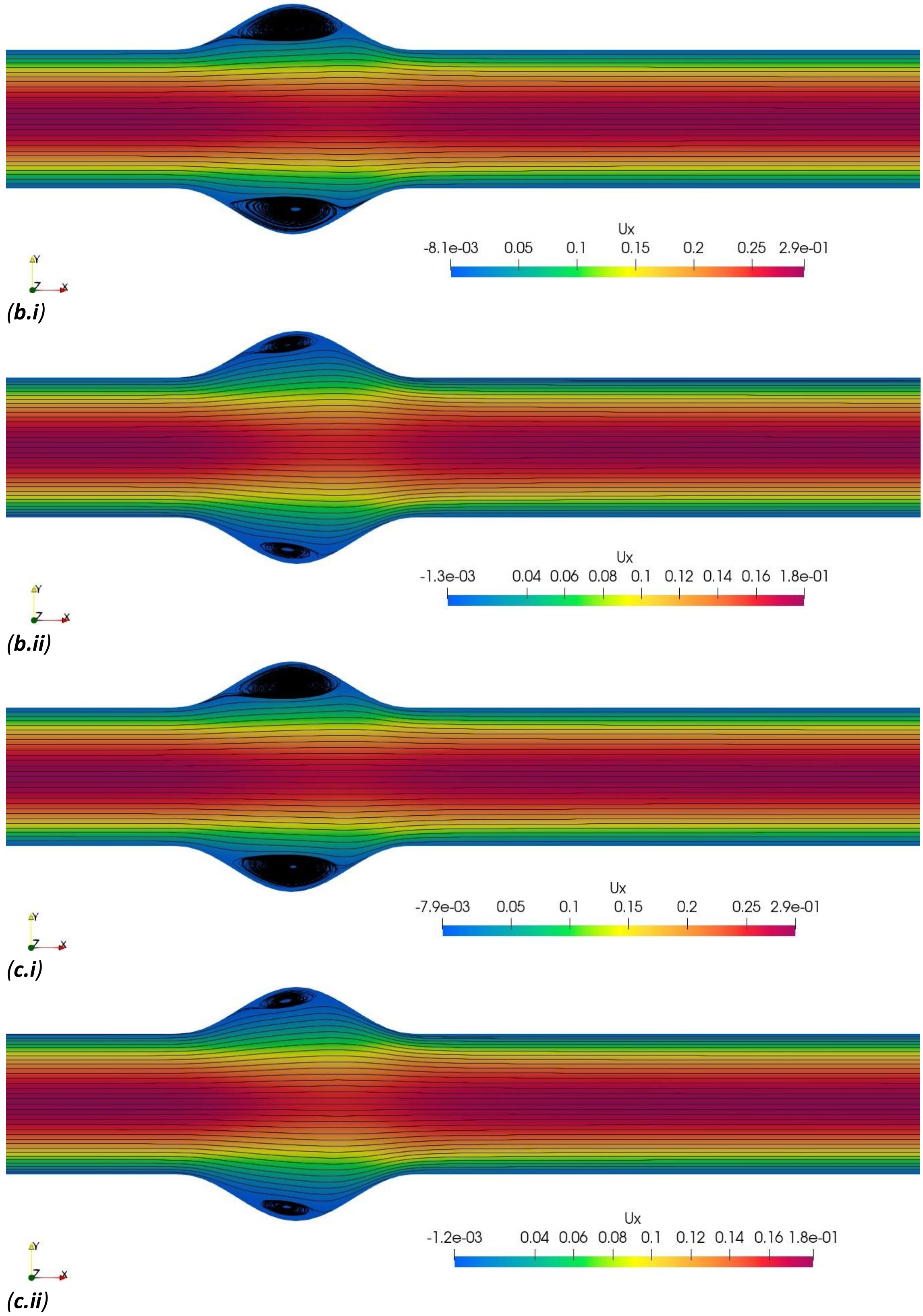

**Figure 19**. Streamlines for blood flow through a $166\%$ aneurysm using MHD micropolar modeling (i) without acknowledging MMR and (ii) considering MMR for an applied magnetic field of (a) 1 $T$, (b) 3 $T$, and (c) 8 $T$.

Figure 20 illustrates the vorticity contours for the 166% aneurysm using the MHD micropolar fluid theory, considering both the absence and presence of the MMR term. Again, the applied magnetic field is varied at 1 $T$, 3 $T$, and 8 $T$. Similar to the streamlines, the vorticity contours for the non-magnetic micropolar blood flow and the MHD micropolar blood flow through the aneurysm, when the MMR term is excluded, exhibit no significant differences, regardless of the magnetic field intensity. However, when the MMR term is included, the maximum vorticity magnitude decreases slightly, with a reduction of 16.67% at 8 $T$. Furthermore, the classic vortex pair structure persists but becomes tighter and less diffusive, consistent with the observations from the streamline plot. MMR seems to dampen large-scale rotational diffusion, leading to vortex sharpening.

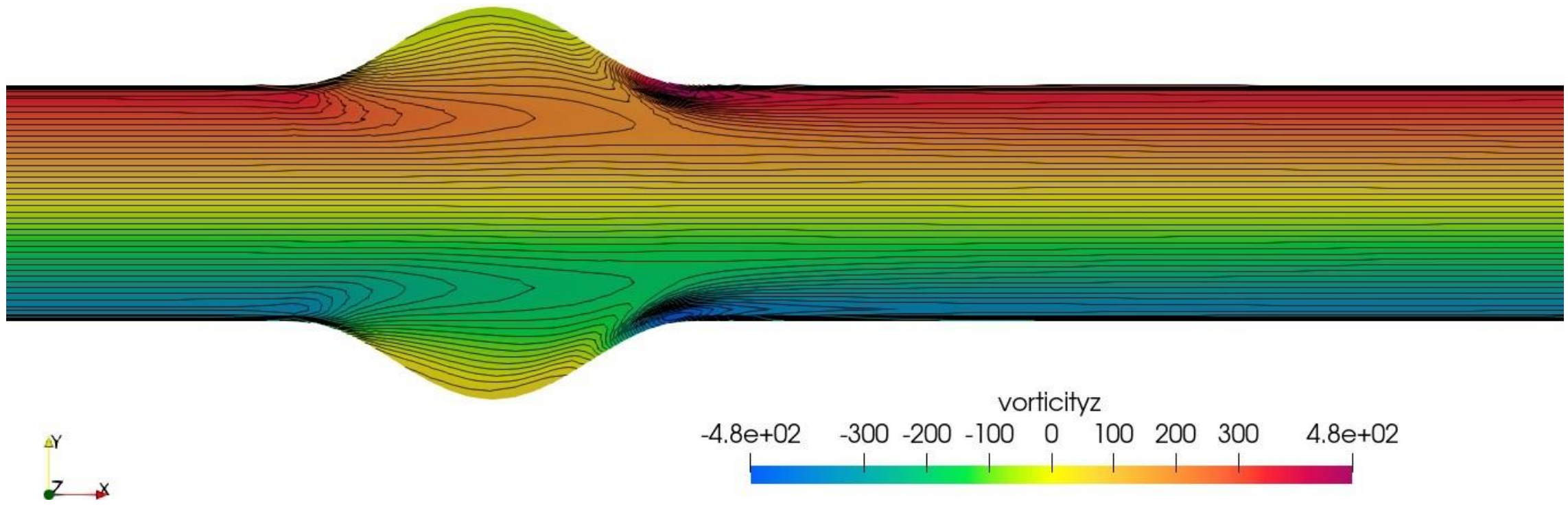

***(a.i)***

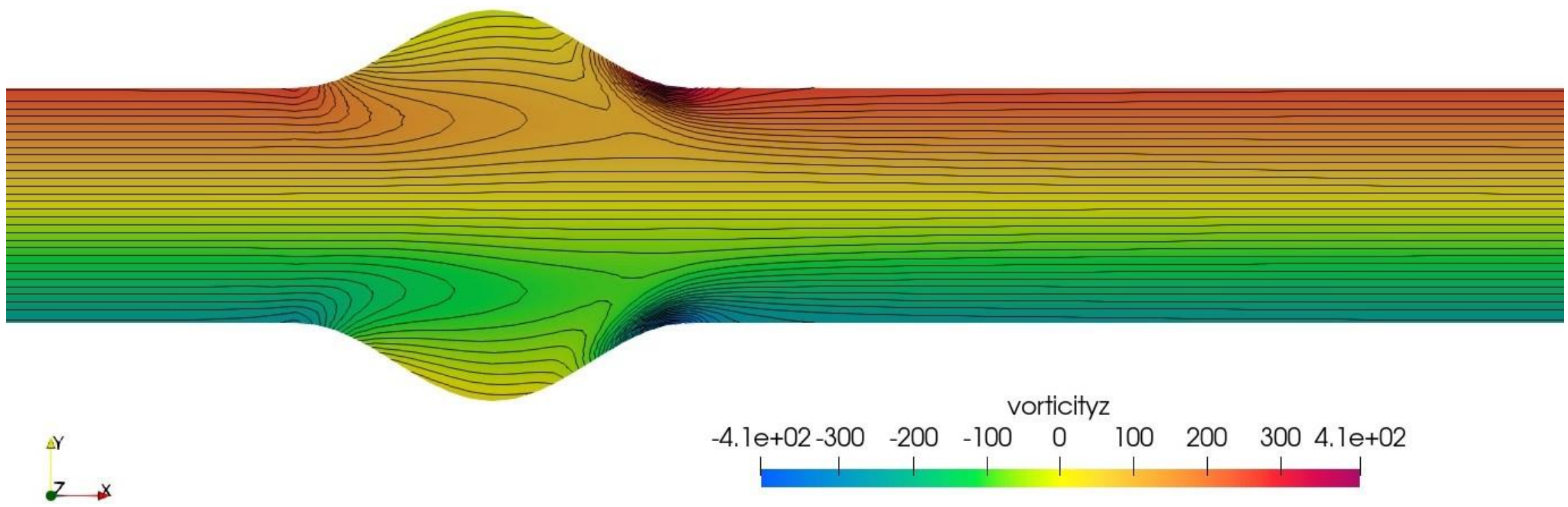

***(a.ii)***

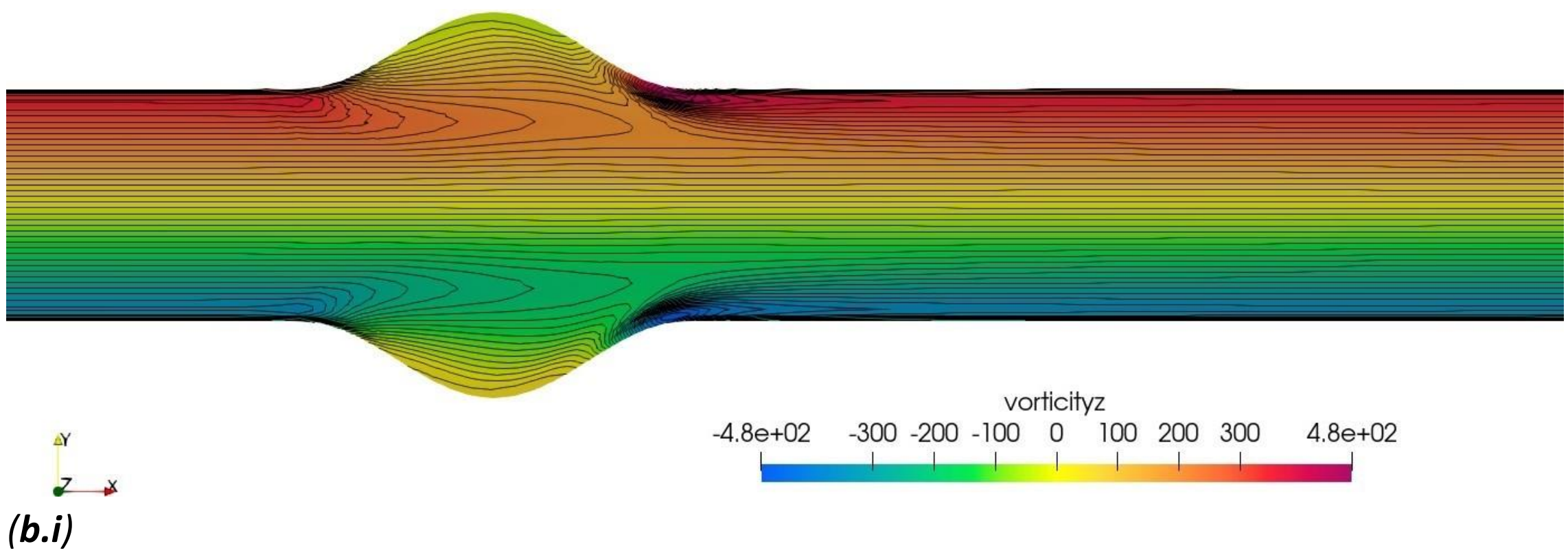

***(b.i)***

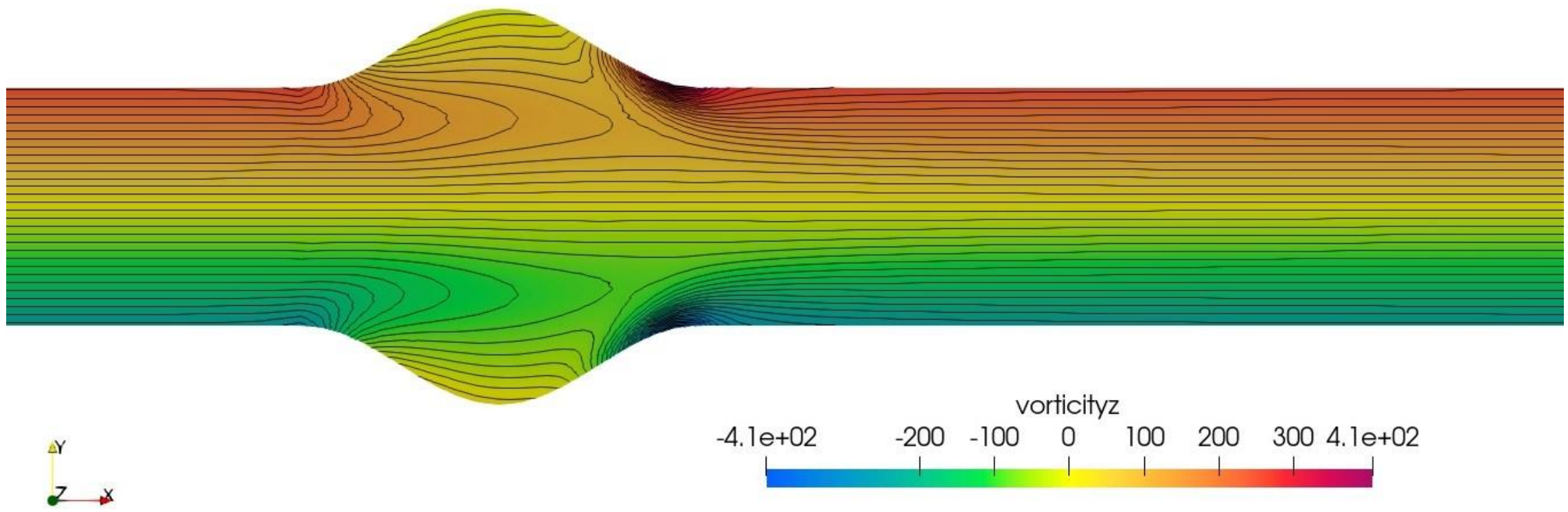

*(b.ii)*

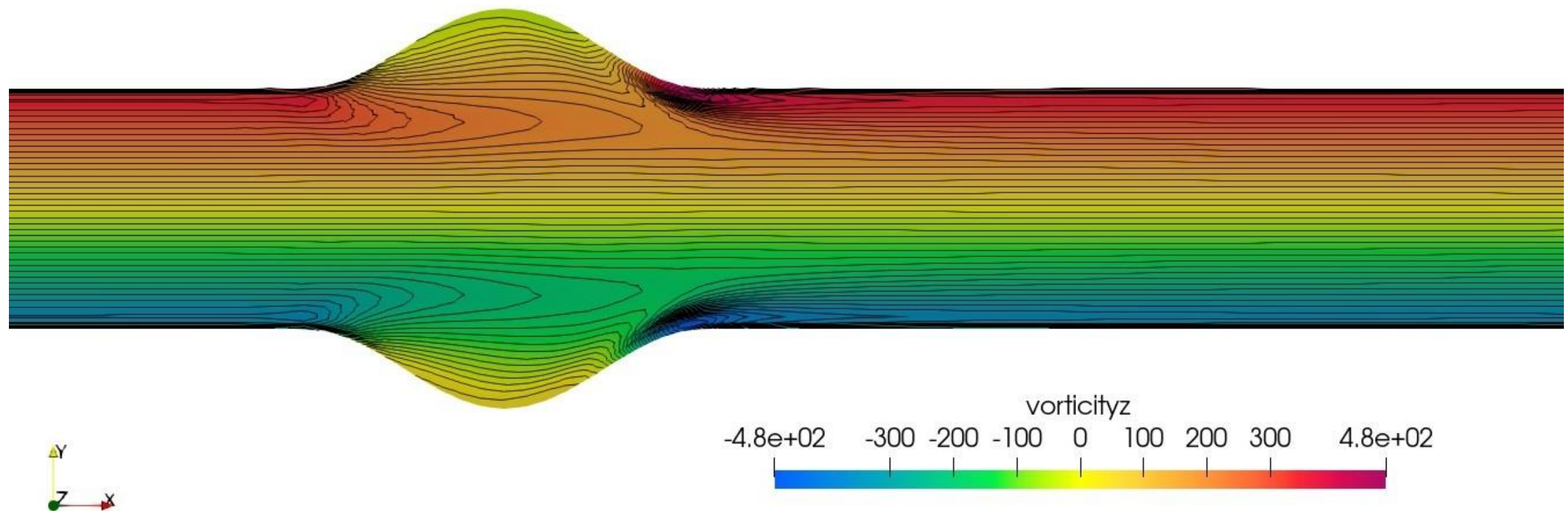

*(c.i)*

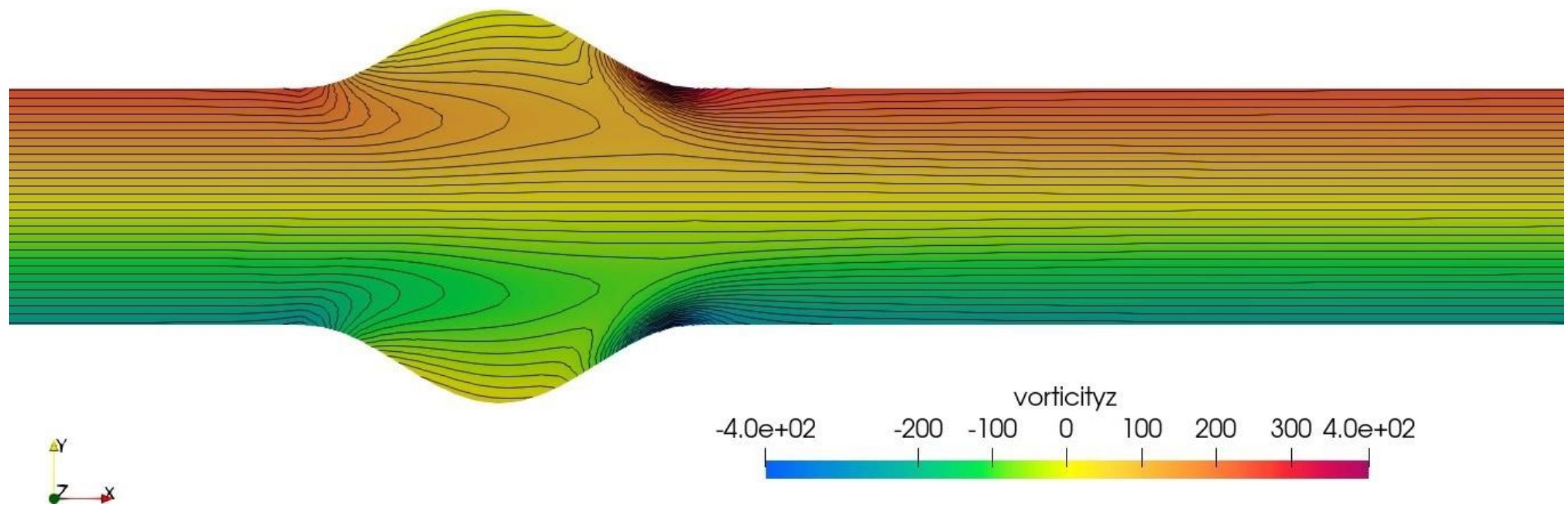

*(c.ii)*

**Figure 20**. Vorticity contour plots for blood flow through a $166\%$ aneurysm using MHD micropolar modeling (i) without acknowledging MMR and (ii) considering MMR for an applied magnetic field of (a) 1 $T$, (b) 3 $T$, and (c) 8 $T$.

Figure 21 shows the microrotation contours for the 166% aneurysm using the MHD micropolar fluid model, both ignoring and acknowledging MMR. Once again, the applied magnetic field is varied at 1 $T$, 3 $T$, and 8 $T$. Similar to the vorticity contours, the microrotation contours for the non-magnetic micropolar blood flow and the MHD micropolar blood flow through the aneurysm without MMR show no significant differences, even as the applied magnetic field intensity increases. However, when the MMR term is included, the maximum and minimum microrotation magnitude values decrease significantly, reducing by 93.4% at 1 $T$, 97.7% at 3 $T$, and 99.1% at 8 $T$, reflecting the strong suppression of the erythrocytes' angular motion. Additionally, the contours display sharper microrotation gradients, suggesting that the magnetic field not only suppresses magnitude but also increases rotational confinement, leading to thinner boundary layers of spin. Physically, these results are attributed to the alignment of erythrocytes with the externally applied magnetic field, which inhibits any internal rotation.

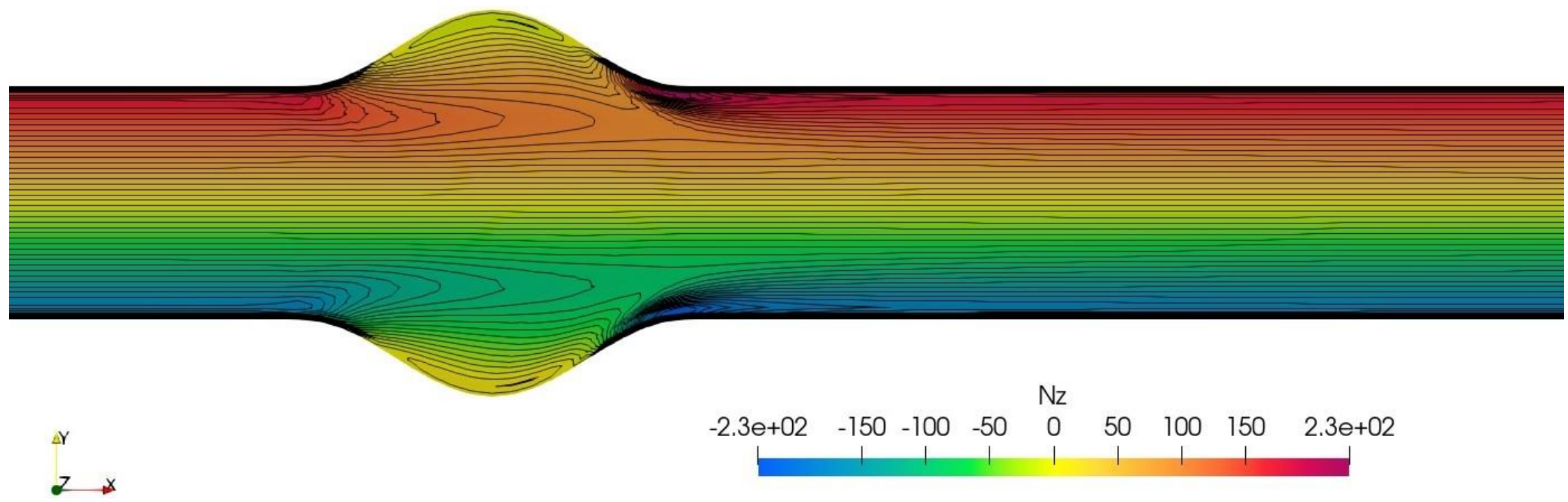

*(a.i)*

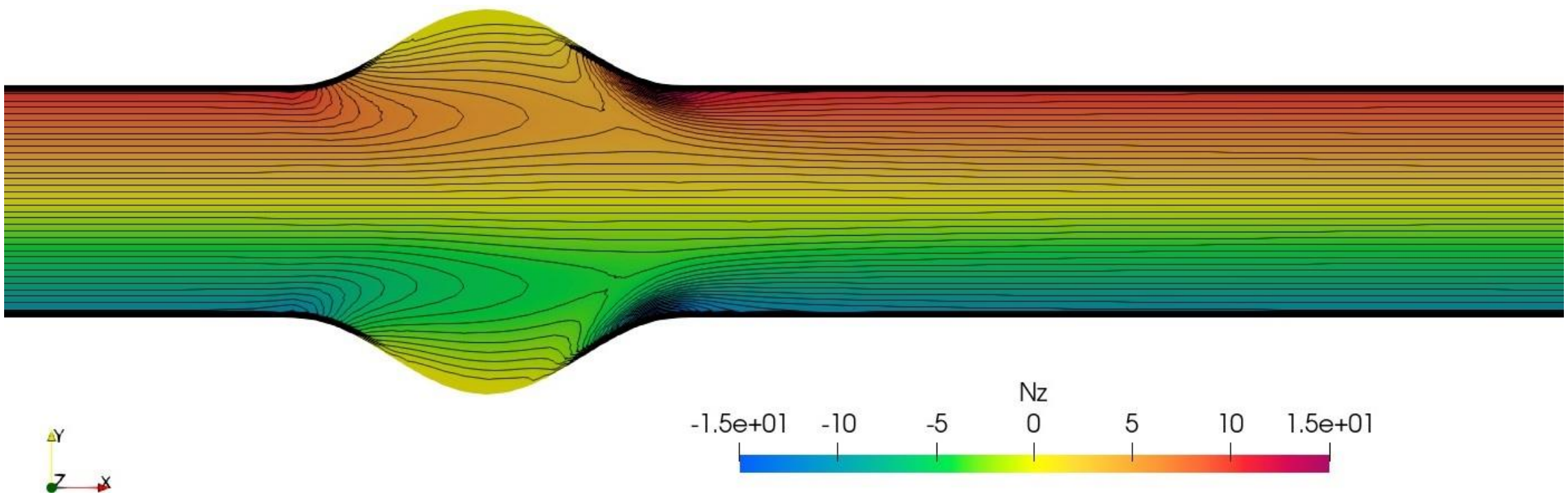

*(a.ii)*

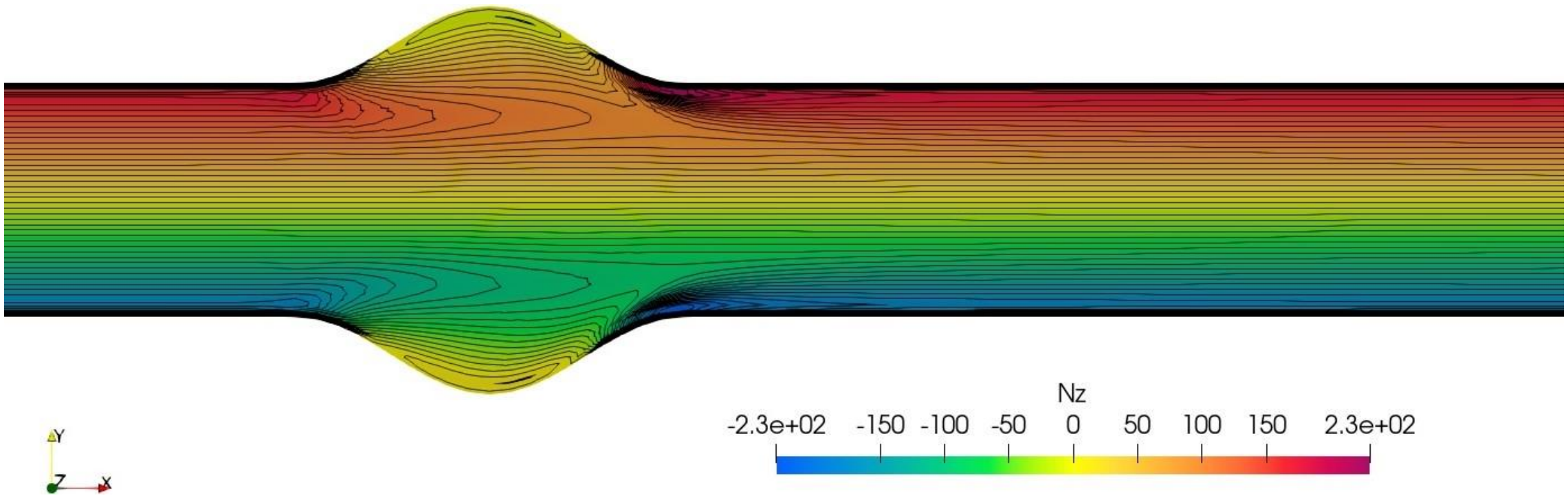

*(b.i)*

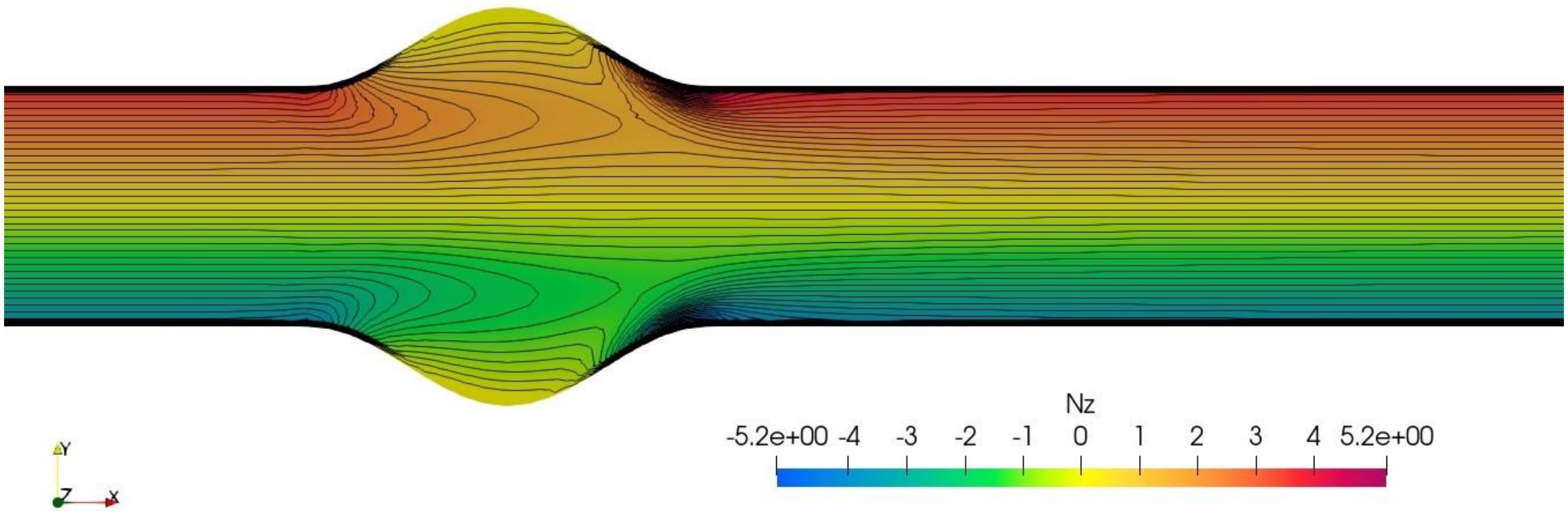

*(b.ii)*

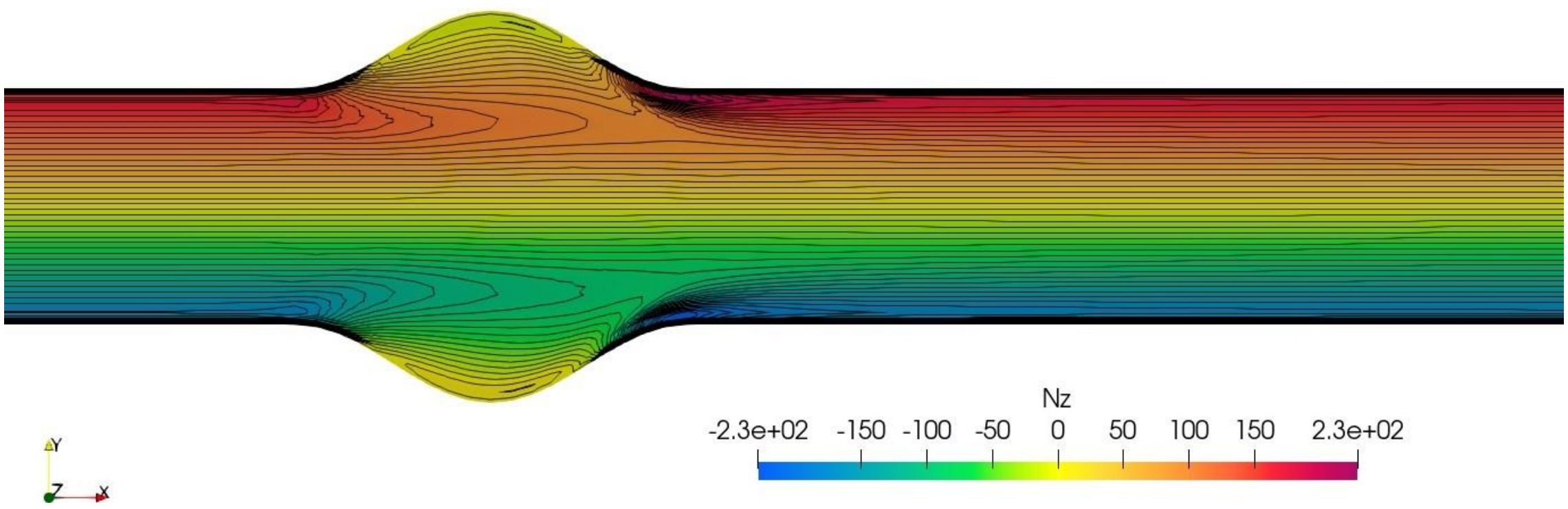

*(c.i)*

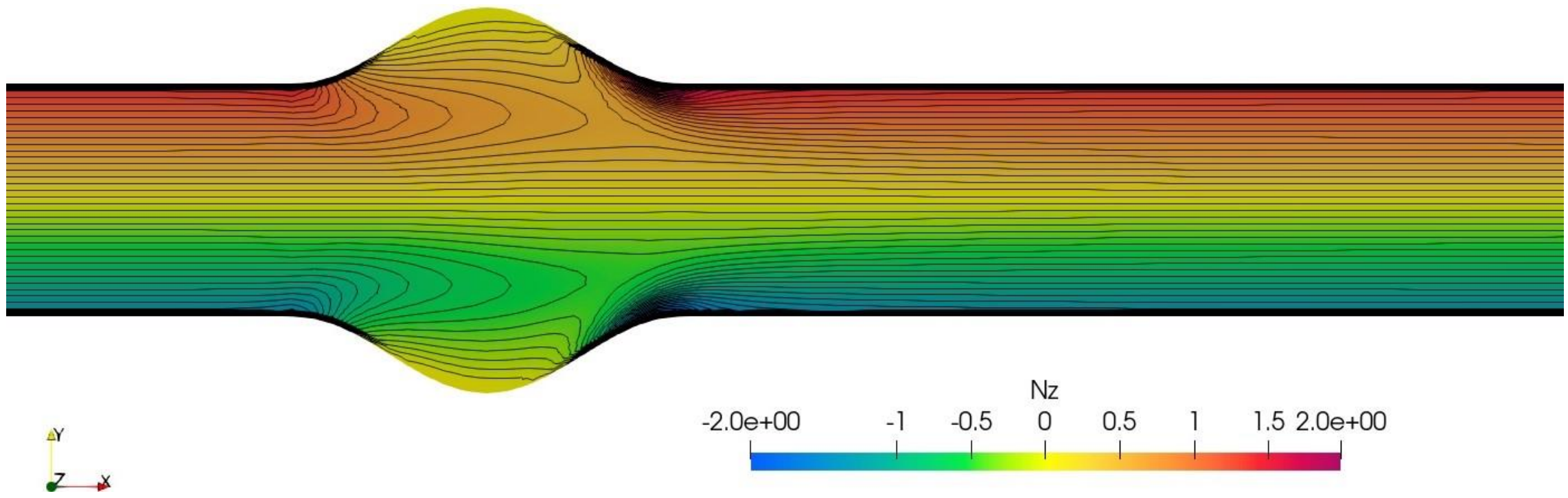

*(c.ii)*

**Figure 21**. Microrotation contour plots for blood flow through a 166% aneurysm using MHD micropolar modeling (i) without acknowledging MMR and (ii) considering MMR for an applied magnetic field of (a) 1 $T$, (b) 3 $T$, and (c) 8 $T$.

In Figure 22, the velocity profiles are presented for a 166% aneurysm inside the aneurysm sac and downstream of it (at $l = 0.05\ m$). The profiles are plotted for the non-magnetic Newtonian and micropolar blood flows, the MHD micropolar blood flow without the MMR effect, and the MHD micropolar blood flow with the MMR effect included. As previously mentioned, the applied magnetic field is varied at 1 $T$, 3 $T$, and 8 $T$. At the center of the aneurysm, the velocity exhibits a characteristic oscillatory profile (see Refs. [49, 51]) due to flow separation and the formation of the recirculating vortices, as shown in the streamline plots. Downstream of the aneurysm sac, the velocity profile has the characteristic parabolic shape. Similar to the streamline plots, the differences observed in velocity between the non-magnetic Newtonian and micropolar profiles are small both inside and downstream of the aneurysm sac, with a velocity reduction up to 6.9%. Once again, no significant differences are observed between the velocity profiles of the non-magnetic micropolar blood flow and the MHD micropolar blood flow without MMR, for all considered values of the applied magnetic field, both within and downstream of the aneurysm sac, mainly due to blood's relatively low electrical conductivity. On the other hand, when the MMR term is included, the velocity profiles become suppressed and structurally compressed, with a maximum reduction of 42.8% at 8 $T$.

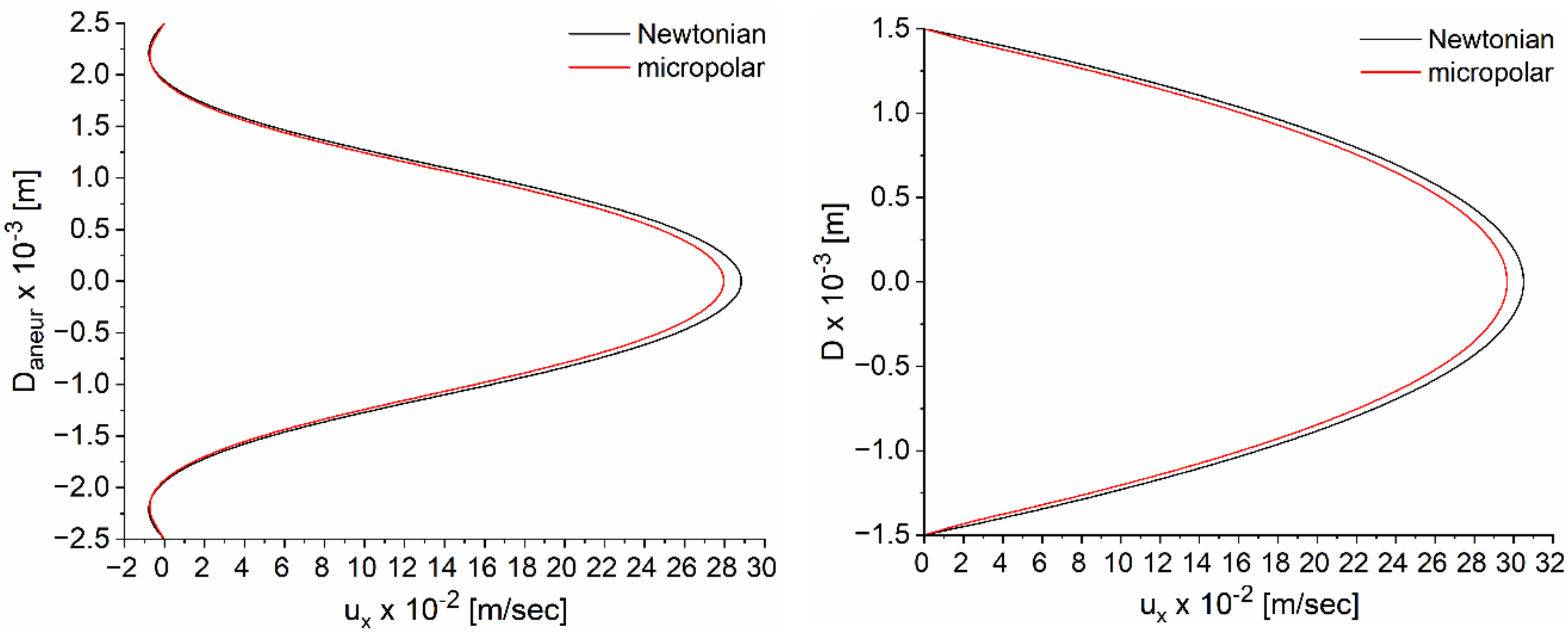

*(a)*

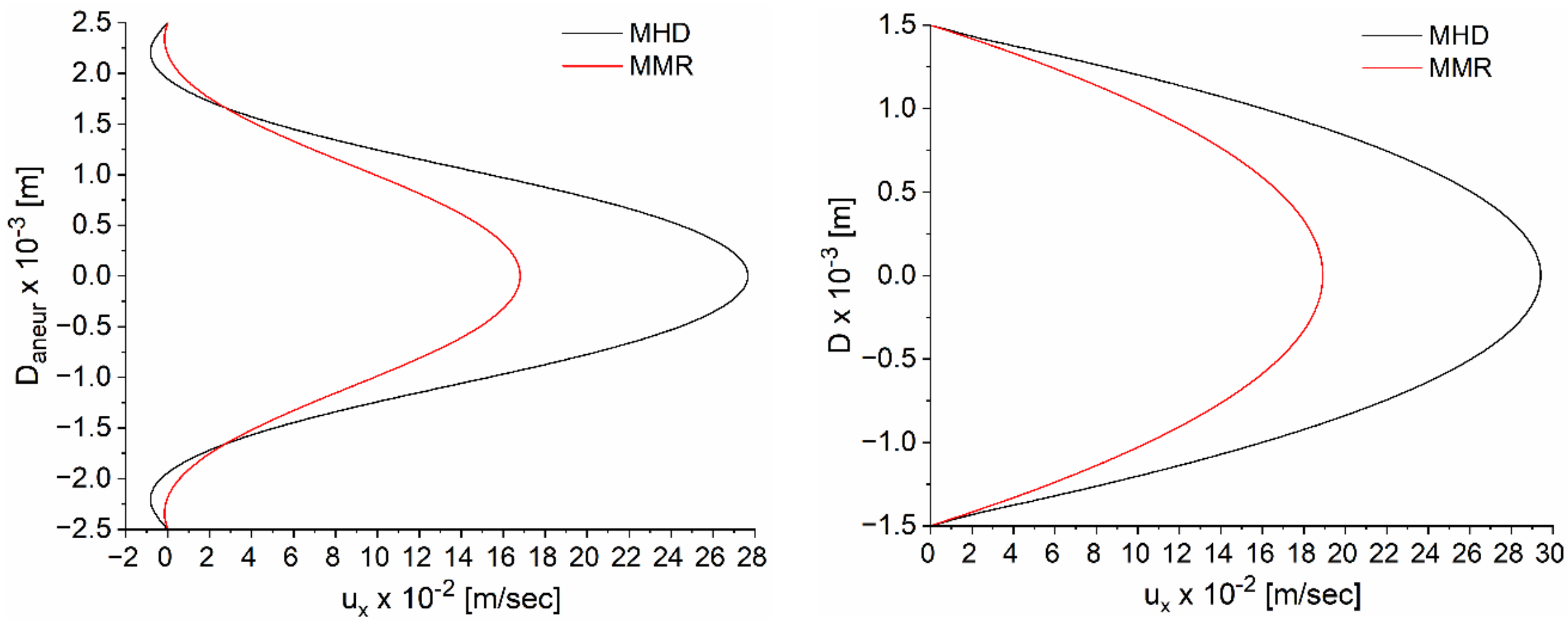

*(b)*

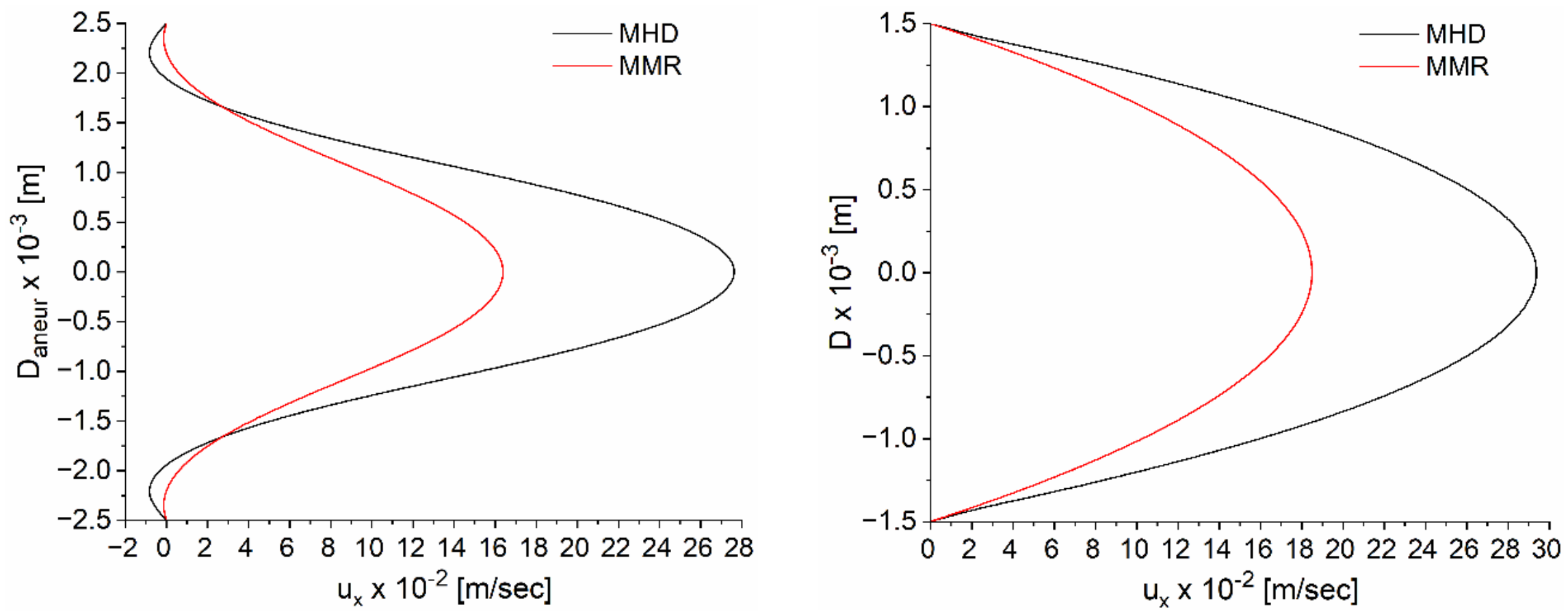

*(c)*

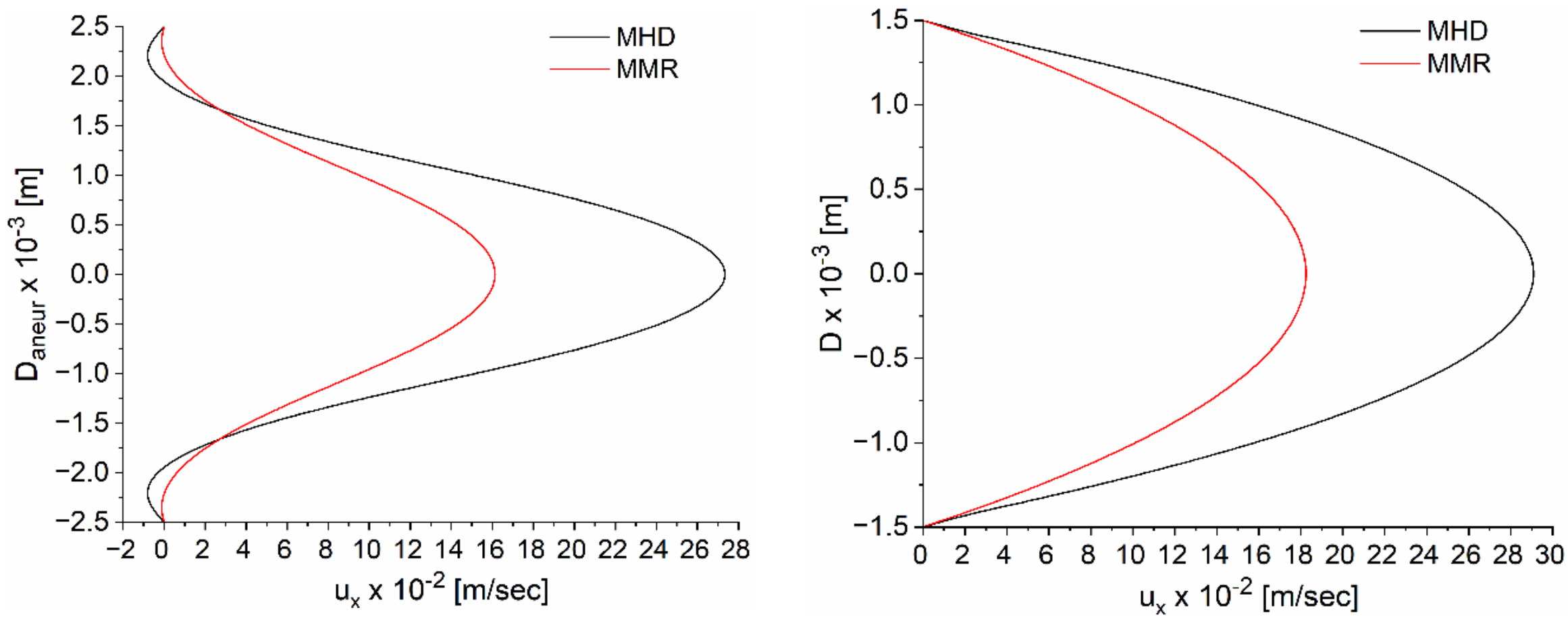

*(d)*

**Figure 22**. Velocity profiles at the center of the aneurysm sac (left) and downstream the latter (right) with a 166% aneurysm for (a) the non-magnetic Newtonian and micropolar blood flows, and (b) the MHD micropolar blood flow by ignoring and considering the MMR term with a magnetic field of 1 $T$ , (c) 3 $T$ and (d) 8 $T$.

In Figure 23, the microrotation profile is illustrated for a 166% aneurysm inside the aneurysm sac and downstream of it (at $l = 0.05\ m$). Similar to the velocity, microrotation is plotted for the ton-magnetic micropolar blood flow, the MHD micropolar blood flow without the MMR effect, and the MHD micropolar blood flow with the MMR effect included. The applied magnetic field is varied at 1 $T$, 3 $T$, and 8 $T$. As expected, the Newtonian model exhibits zero microrotation throughout the domain due to the absence of internal spin degrees of freedom. It is immediately evident that the microrotation profile has a more oscillatory profile inside the aneurysm sac, similar to the velocity profile, due to the presence of the recirculation structures. Downstream of the aneurysm sac, the microrotation profile has its characteristic shape, which is usually seen in parabolic flow profiles. Similar to the velocity, the application of the external magnetic field without considering the MMR term does not result in any noticeable differences in microrotation, due to the small effect of the Lorentz force on the flow. On the other hand, when the impact of MMR term is considered, the microrotation profile is heavily suppressed in both the aneurysm sac and outside of it, with a reduction of nearly 99.99% for the magnetic field of 8 $T$. It is evident that the internal rotation of the erythrocytes nearly “freezes”, as they become aligned parallel to the applied magnetic field.

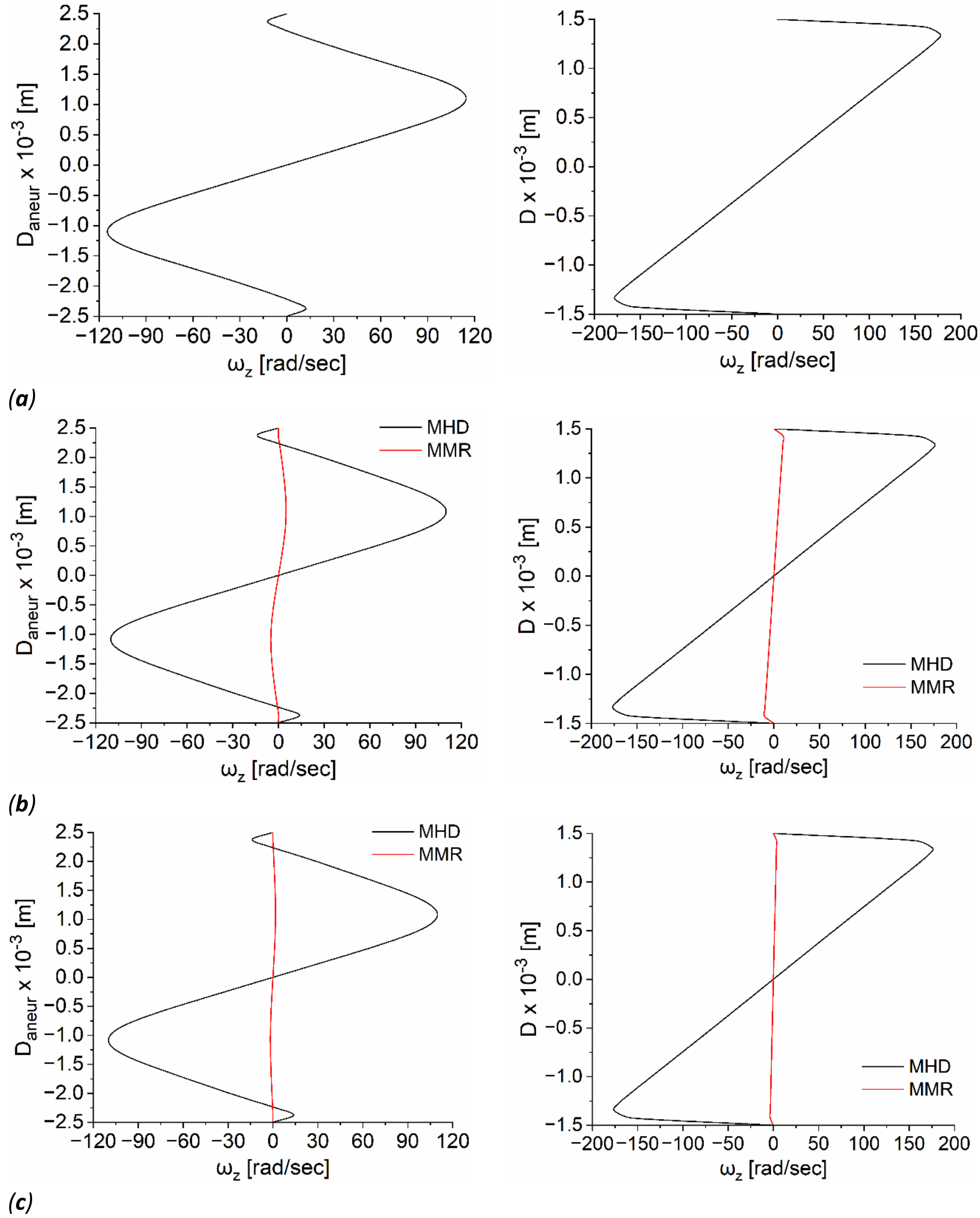

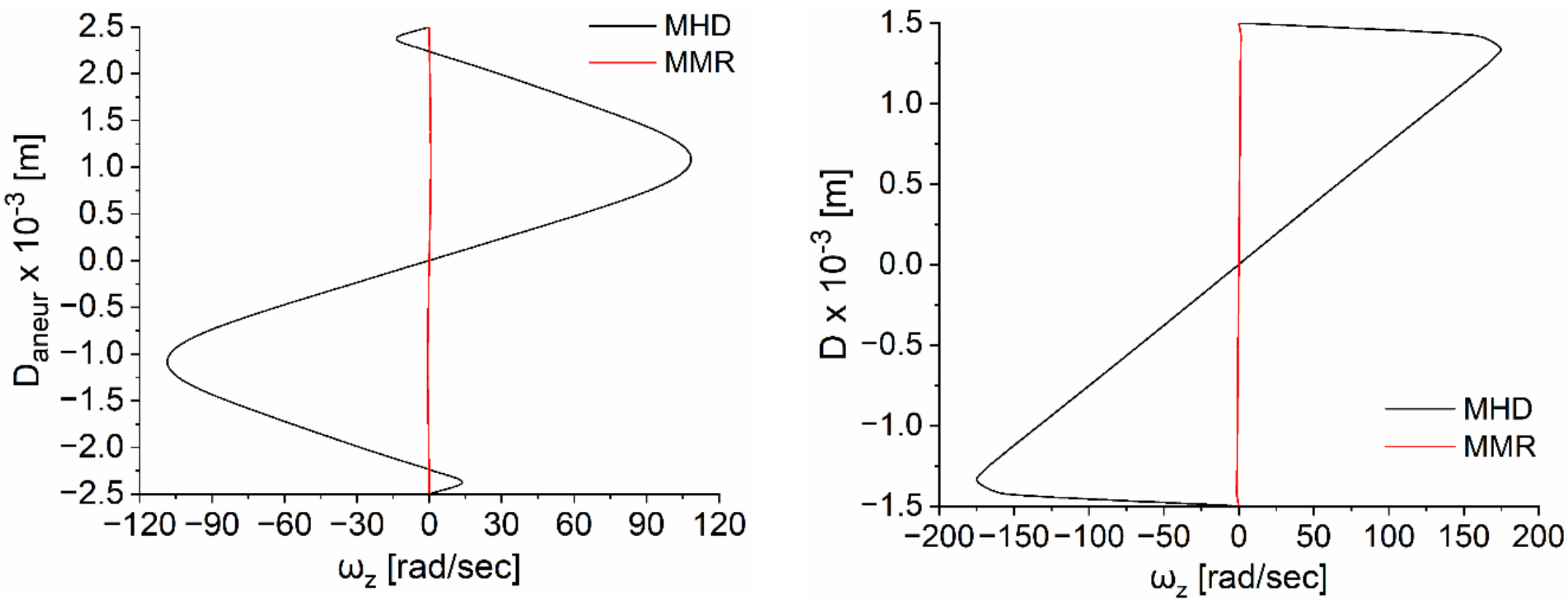

*(d)*

**Figure 23**. Microrotation profiles at the center of the aneurysm sac (left) and downstream the latter (right) with a 166% aneurysm for (a) the non-magnetic micropolar blood flow, and (b) the MHD micropolar blood flow by ignoring and considering the MMR term with a magnetic field of 1 $T$ , (c) 3 $T$ and (d) 8 $T$.

#### 4.3.2.2. Results for the aneurysm with a 200% dilation

Figure 24 presents the streamlines for the non-magnetic Newtonian and micropolar blood flows through a 200% aneurysm. Compared to the previously discussed 166% aneurysm, the increased geometric expansion in the 200% case significantly alters the internal flow and the recirculation structures. The aneurysmal dilation leads to stronger flow separation and the formation of larger recirculation zones within the sac. These vortices are more pronounced than in the 166% case, occupying a greater portion of the aneurysmal cavity and encroaching further toward the centerline, while the core flow is narrower and less energetic. The micropolar effects on this aneurysm are similar to the 166% aneurysm. The recirculation zones persist but are slightly reduced in strength and vertical extent compared to the Newtonian case. Moreover, the maximum velocity value in the core flow is slightly decreased (from $0.31\ m/sec$ to $0.3\ m/sec$).

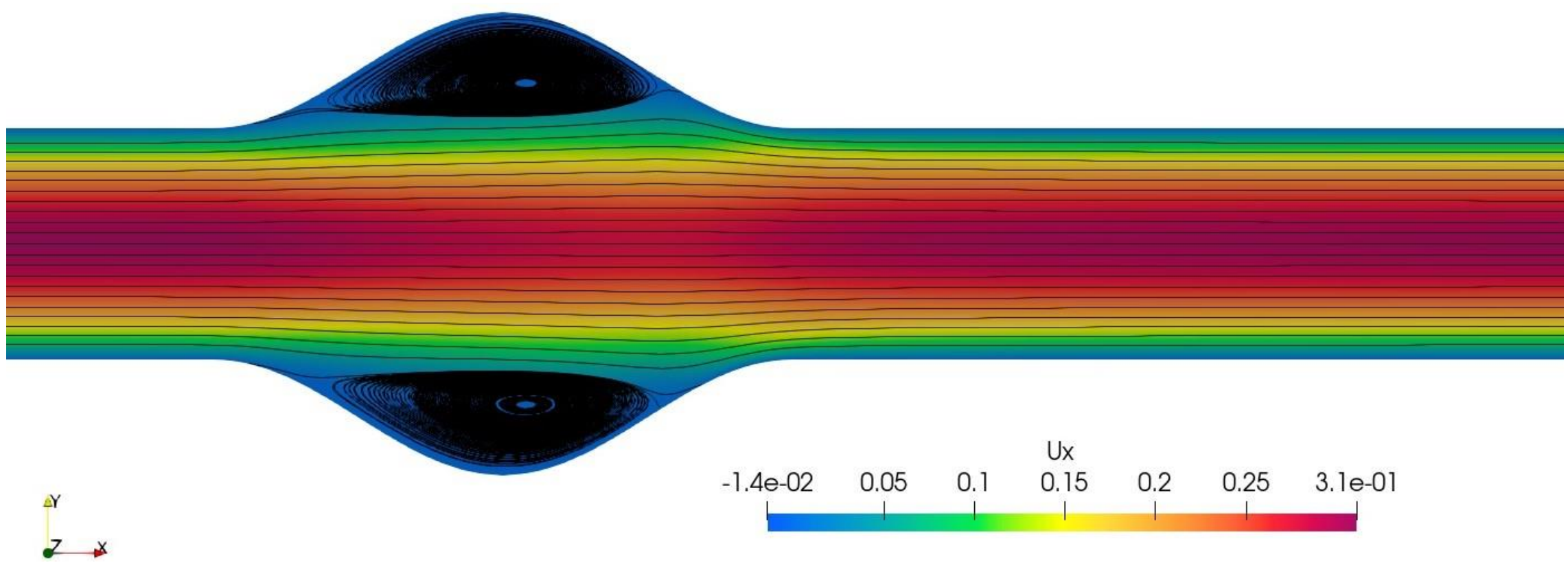

*(a)*

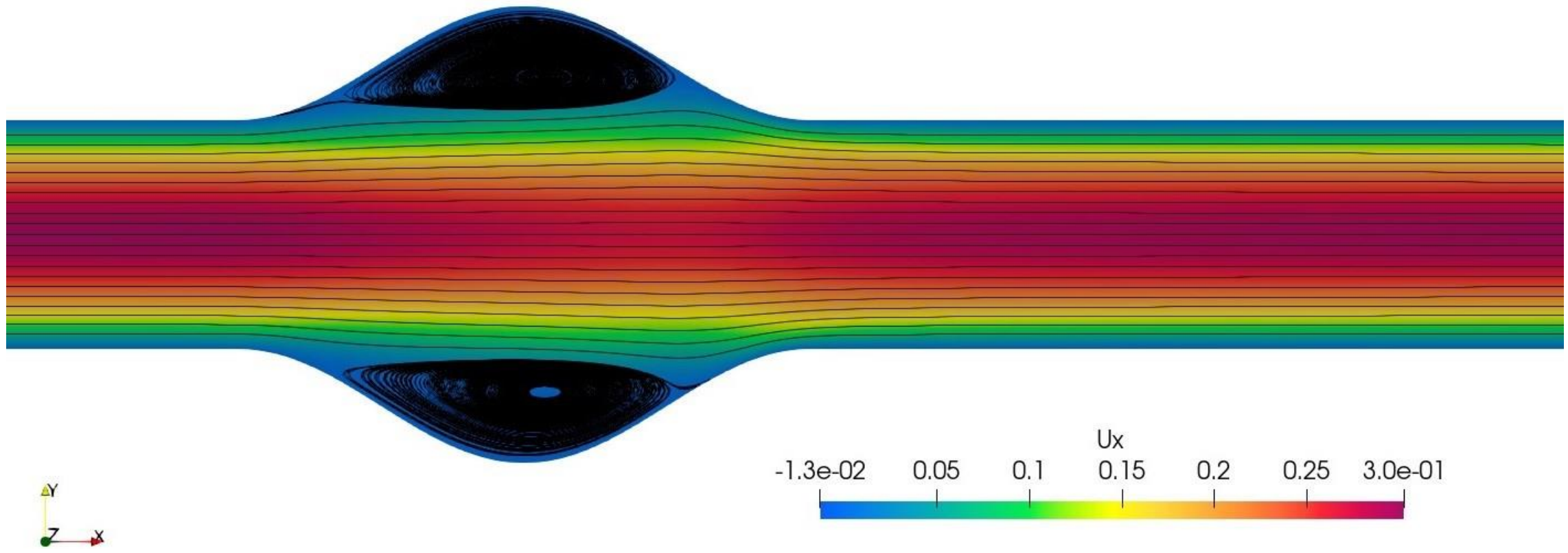

***(b)***

**Figure 24**. Streamlines for +blood flow through a 200% aneurysm using (a) Newtonian modelling, and (b) micropolar modelling without an applied magnetic field.

Figure 25 illustrates the vorticity contour plots for the non-magnetic Newtonian and micropolar blood flows through a 200% aneurysm. Compared to the previously examined 166% aneurysm, the increased sac diameter and length significantly intensify rotational dynamics across all models. The vorticity isoline reveals large and well-defined vortex pairs, similar to the streamlines' plot. The recirculation zones extend deep into the aneurysmal sac, with high rotational intensity concentrated at the aneurysm shoulders and gradual decay toward the core. When the micropolar effects are taken into account, vorticity remains symmetric but exhibits reduced magnitude and more diffuse spatial distribution, particularly near the aneurysm walls. Compared to the 166% aneurysmal micropolar blood flow, the larger aneurysm further attenuates rotational dynamics, increasing the vorticity penetration into the core region while reducing peak shear near the boundary layers.

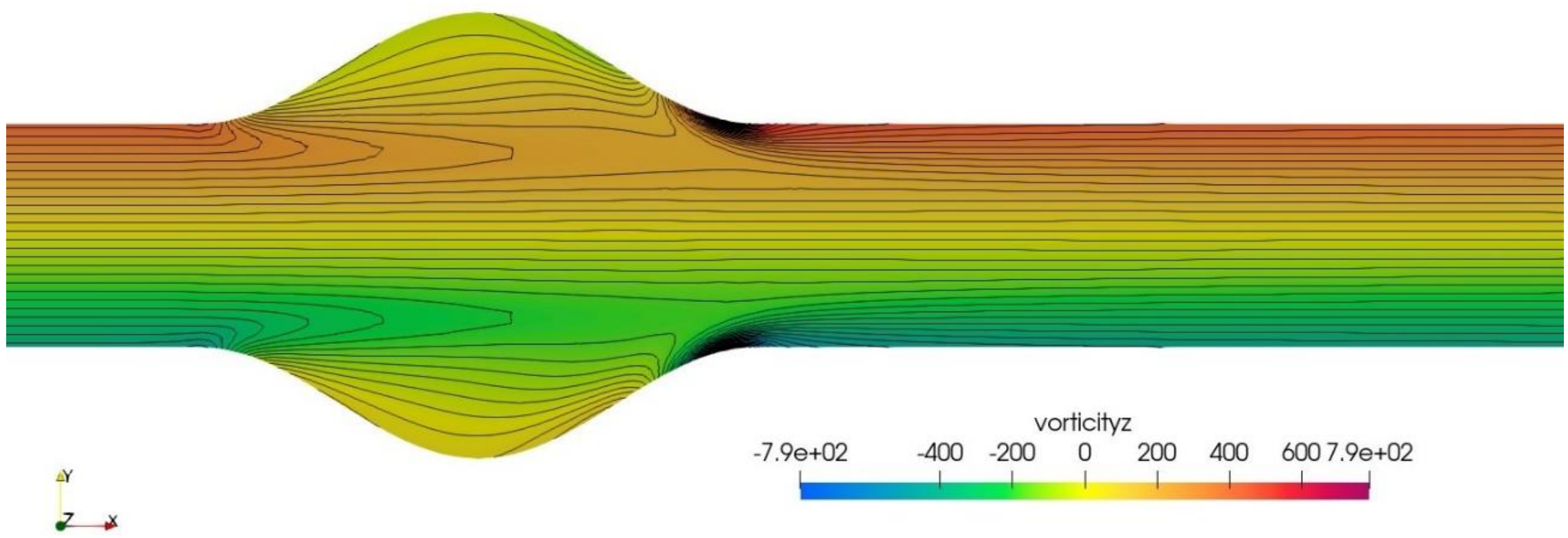

***(a)***

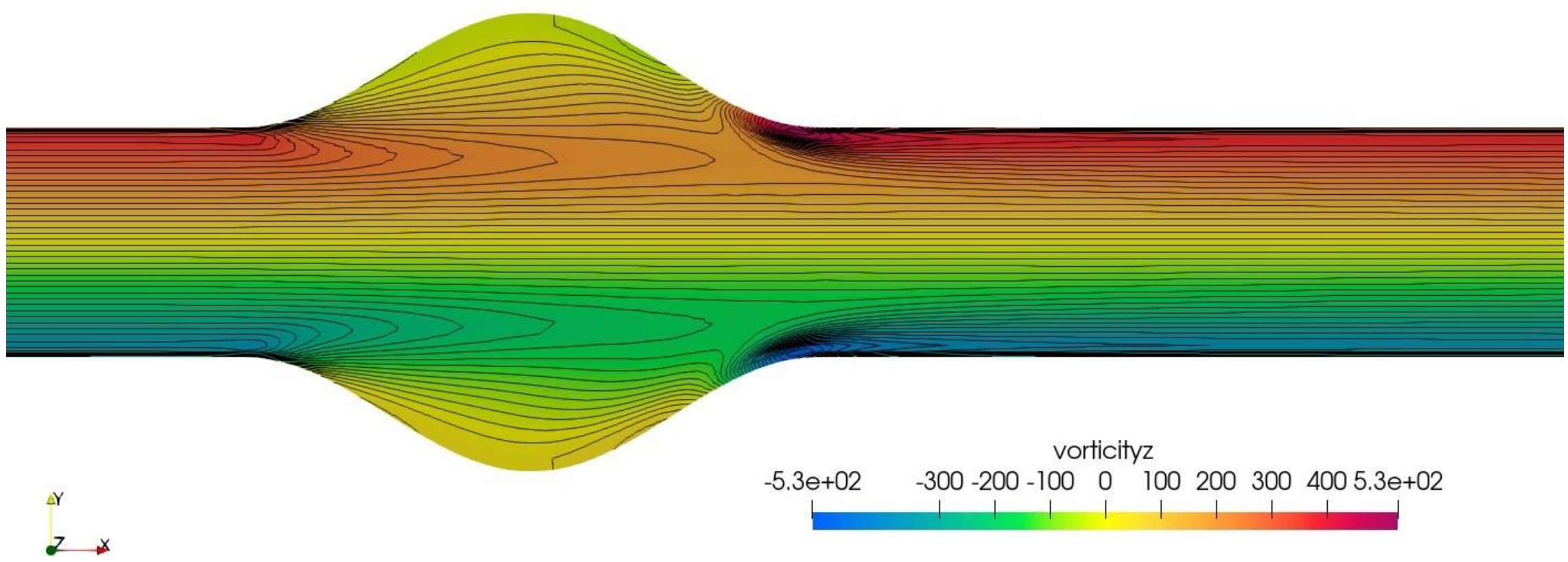

***(b)***

**Figure 25**. Vorticity contour plots for blood flow through a $200\%$ aneurysm using (a) Newtonian modelling, and (b) micropolar modelling without an applied magnetic field.

Figure 26 shows the microrotation contours for the non-magnetic micropolar blood flow through a 200% aneurysm. As in the 166% aneurysm, the microrotation isolines exhibit the expected antisymmetric structure about the vessel centerline with maximum microrotation magnitude appearing near the aneurysm shoulders. The spatial distribution is smooth and continuous, reflecting the diffusion of the erythrocytes' spin from the boundary layers into the flow core. Compared to the 166% aneurysm, the larger dilation in the 200% case results in broader microrotation zones and higher peak values, indicating more pronounced rotational activity as the flow navigates the expanded geometry.

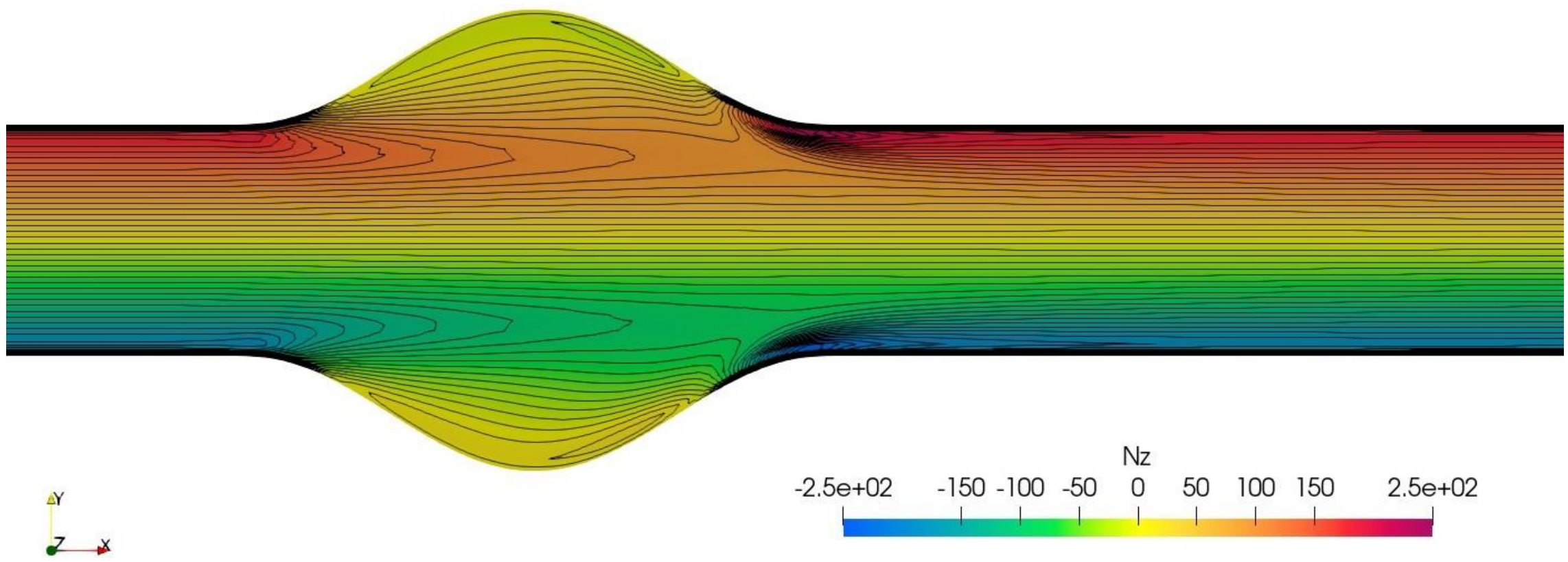

**Figure 26**. Microrotation contour plot for blood flow through a $200\%$ aneurysm using micropolar modelling without an applied magnetic field.

Figure 27 illustrates the streamlines for the 200% aneurysm for the MHD micropolar blood flow, both with and without accounting for the MMR effect. The applied magnetic field intensity was varied at $1\,T$, $3\,T$, and $8\,T$. As in the 166% aneurysm, the streamlines for the non-magnetic micropolar and MHD micropolar blood flows show negligible differences when the MMR effect is ignored, regardless of the intensity of the applied magnetic field. Once again, it is verified that the effect of the Lorentz force alone on an MHD micropolar blood flow is small due to the relatively low electrical conductivity of blood. In contrast, in the MMR case, the recirculation zones become highly localized and compressed, especially near the aneurysm walls. Compared to the same flow regime in the 166% aneurysm, the vortices in the 200% case are larger in diameter but more tightly confined, with greater streamline compression near the vortex cores and significantly dampened axial flow. The central core velocity is substantially reduced up to 37% at 8 $T$, and the streamlines appear more densely packed near the neck of the aneurysm.

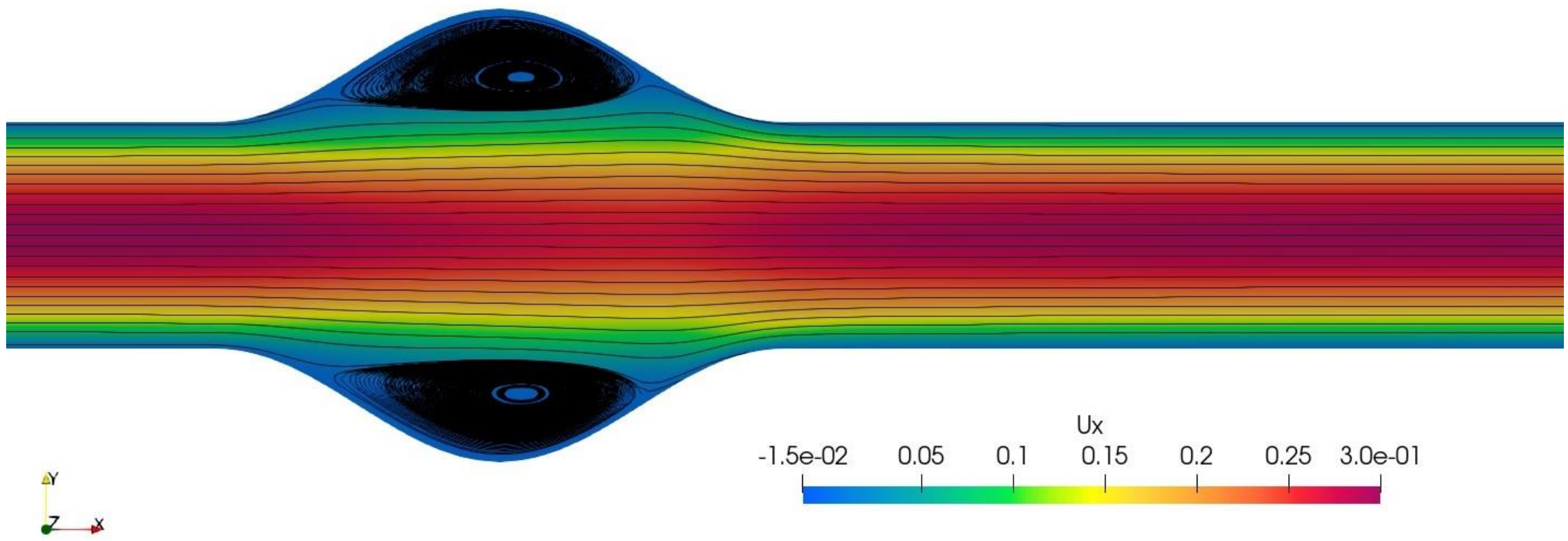

*(a.i)*

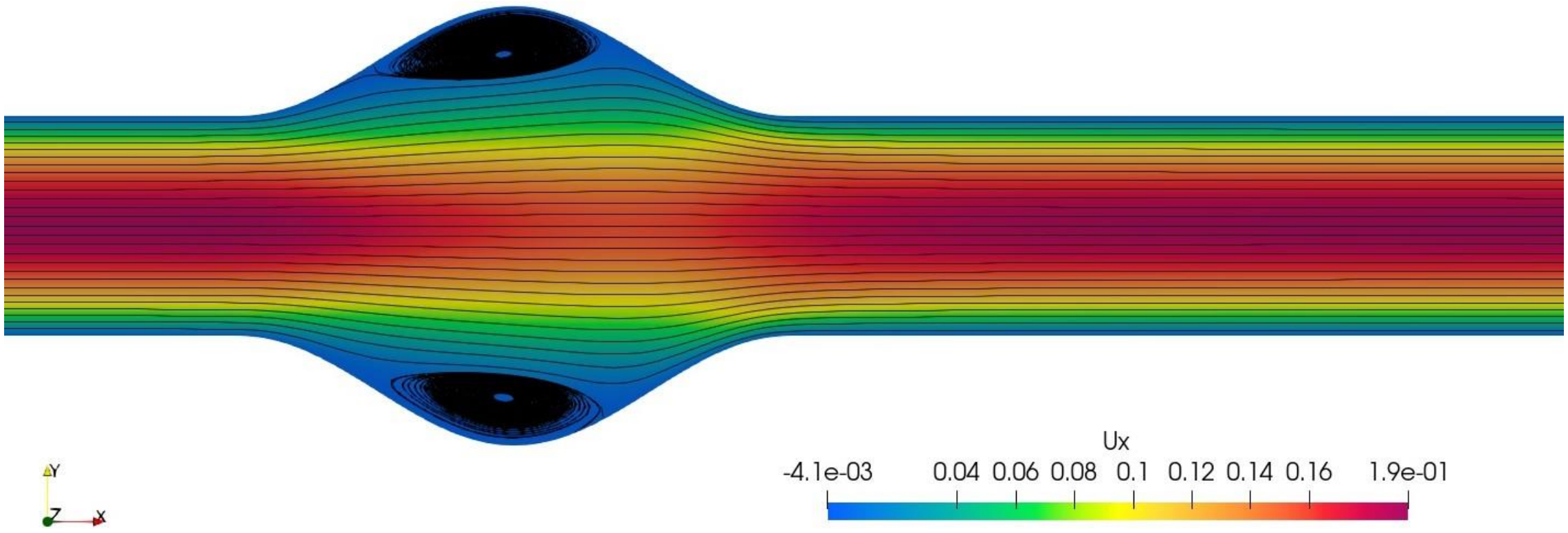
Ux
-4.1e-03 0.04 0.06 0.08 0.1 0.12 0.14 0.16 1.9e-01
Y
Z
X

***(a.ii)***

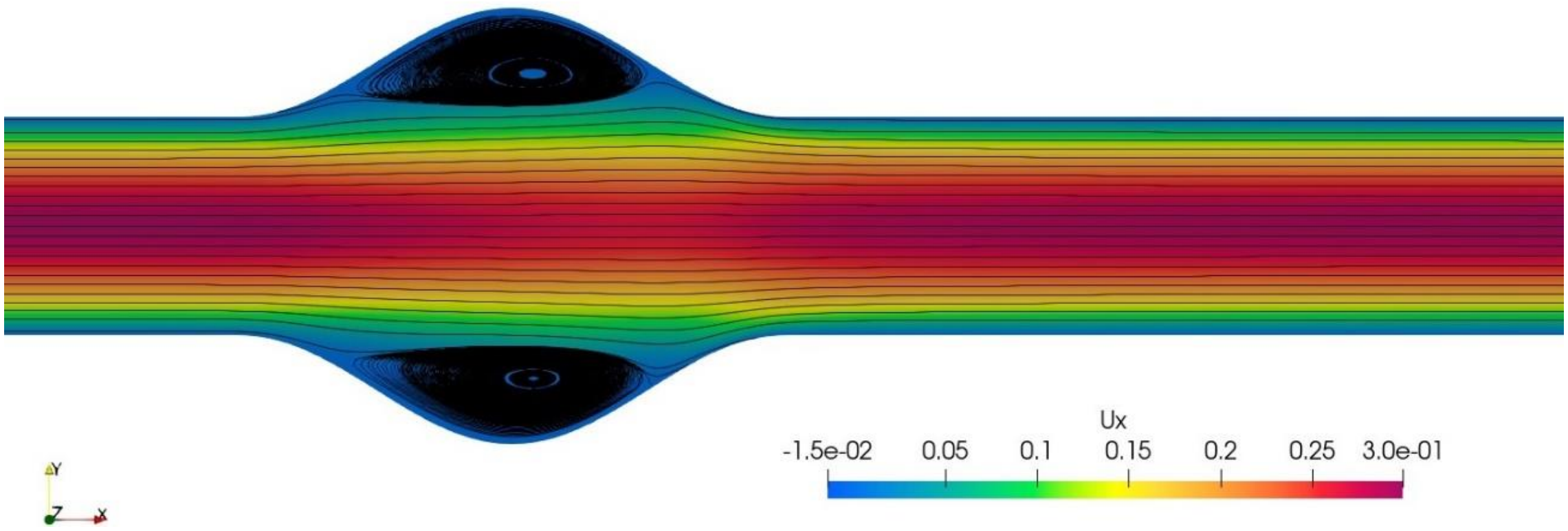
Ux
-1.5e-02 0.05 0.1 0.15 0.2 0.25 3.0e-01
Y
Z
X

***(b.i)***

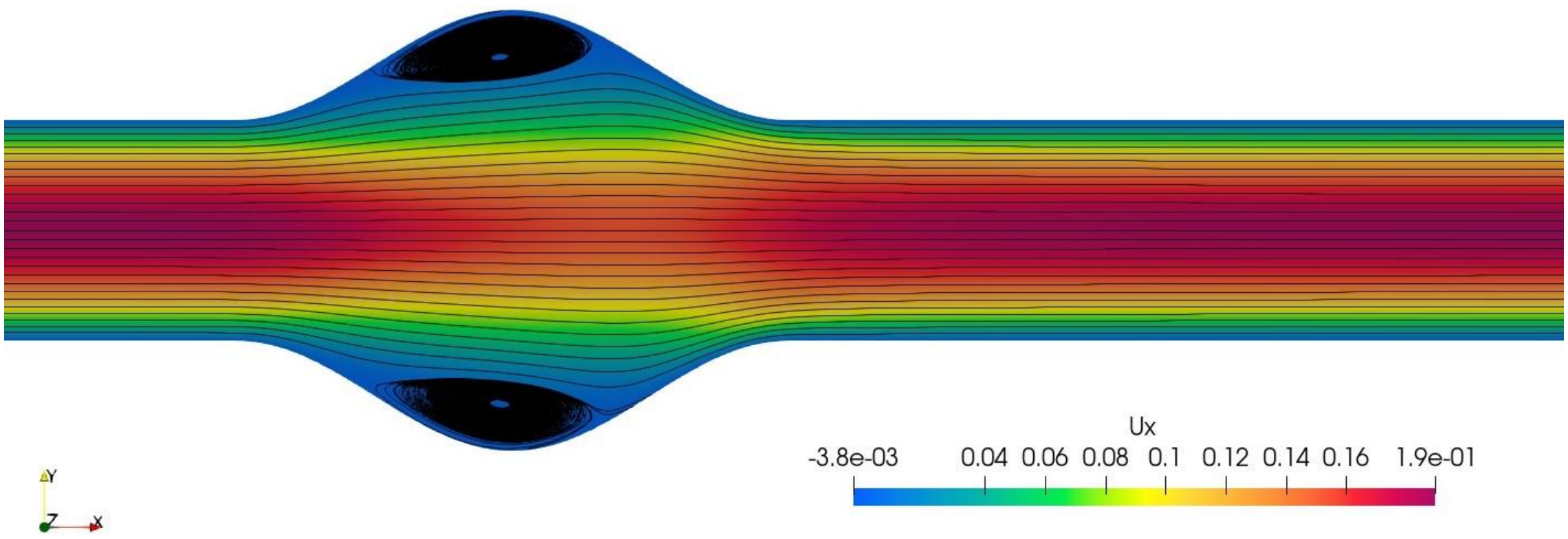
Ux
-3.8e-03 0.04 0.06 0.08 0.1 0.12 0.14 0.16 1.9e-01
Y
Z
X

***(b.ii)***

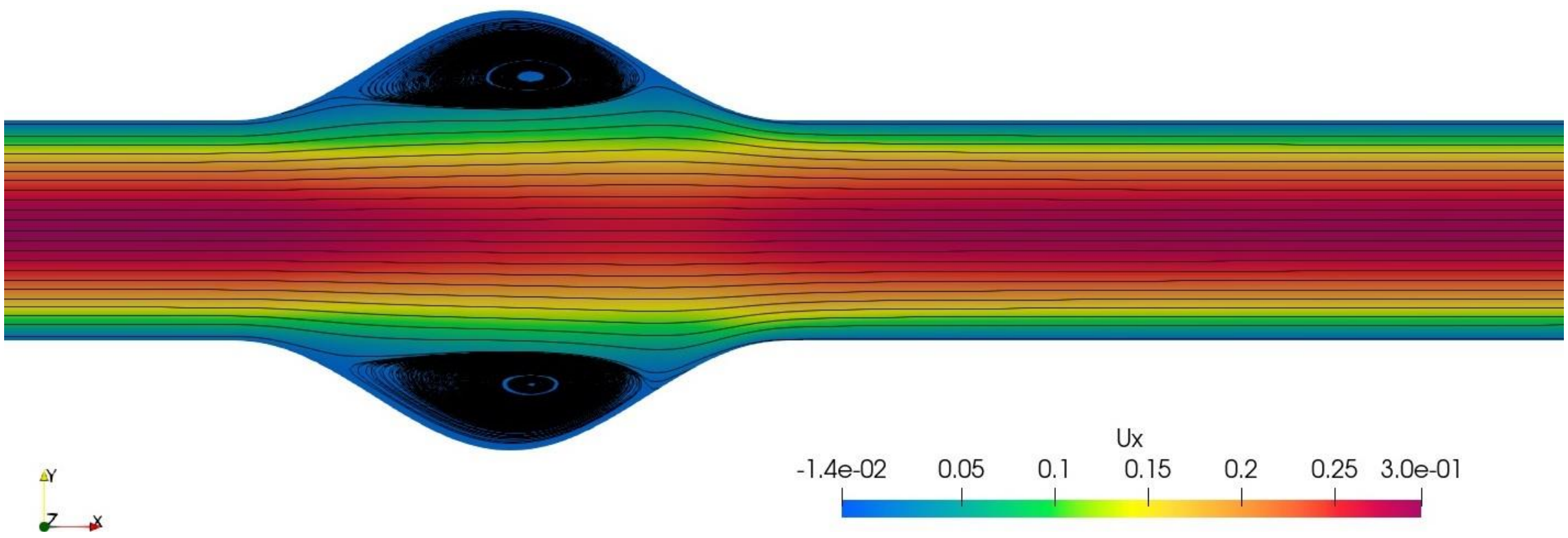
Ux
-1.4e-02 0.05 0.1 0.15 0.2 0.25 3.0e-01
Y
Z
X

***(c.i)***

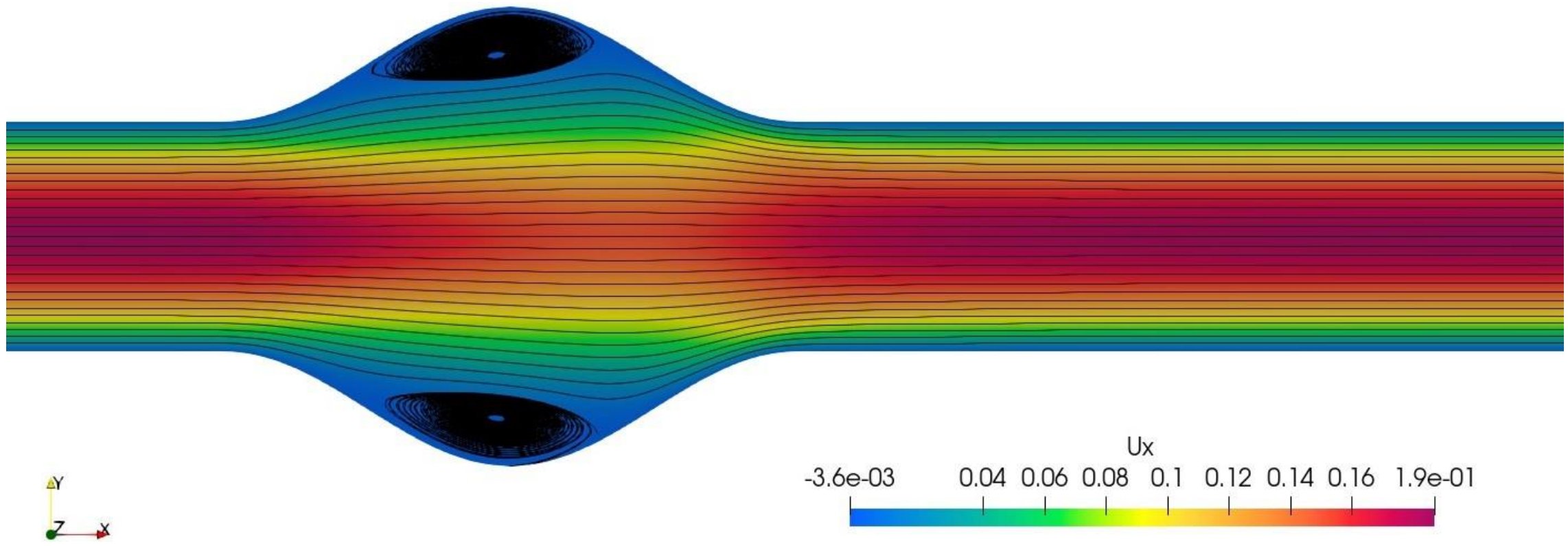

*(c.ii)*

**Figure 27**. Streamlines for blood flow through a 200% aneurysm using MHD micropolar modeling (i) without acknowledging MMR and (ii) considering MMR for an applied magnetic field of (a) 1 $T$, (b) 3 $T$, and (c) 8 $T$.

Figure 28 shows the vorticity contours for the 200% aneurysm for the MHD micropolar blood flow, considering both the absence and presence of the MMR term. As in all flow simulations, the intensity of the applied magnetic field is varied at 1 $T$, 3 $T$, and 8 $T$. Similar to the streamlines, the vorticity contours for the non-magnetic micropolar blood flow and the MHD micropolar blood flow through the aneurysm, when the impact of the MMR is ignored, do not show any significant differences, regardless of the magnetic field intensity. On the other hand, the MMR term strongly influences the vorticity isolines, while the latter are steepest near the wall. Moreover, the rotational structures are both compressed and less intensified. Compared to the 166% MMR case, the larger aneurysm sustains broader and more energetic vortical structures, yet with tighter confinement and sharper contours, reflecting a balance between the increased aneurysmatic sac and MMR damping.

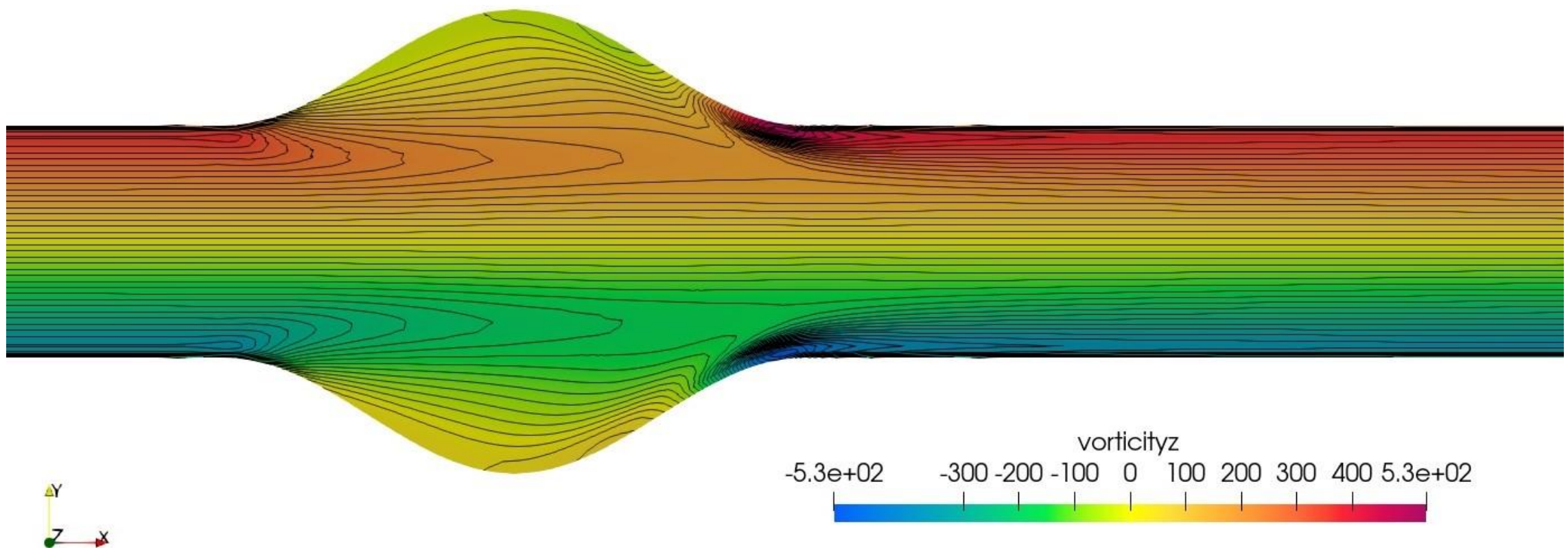

*(a.i)*

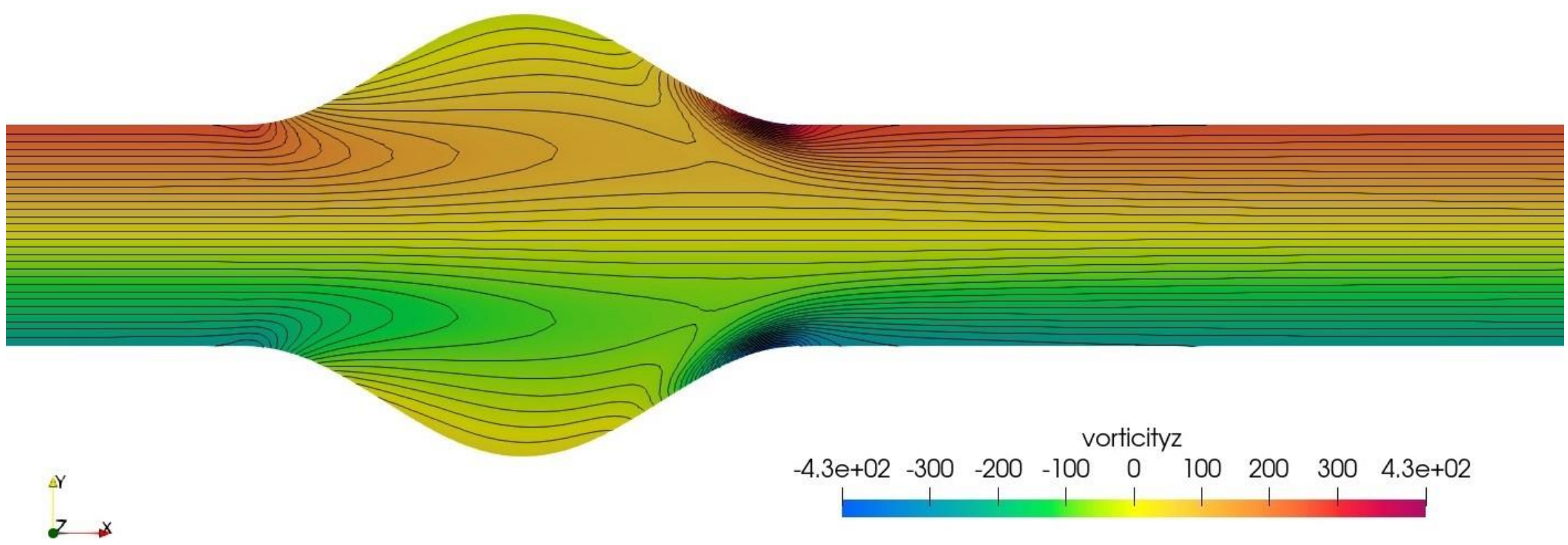

*(a.ii)*

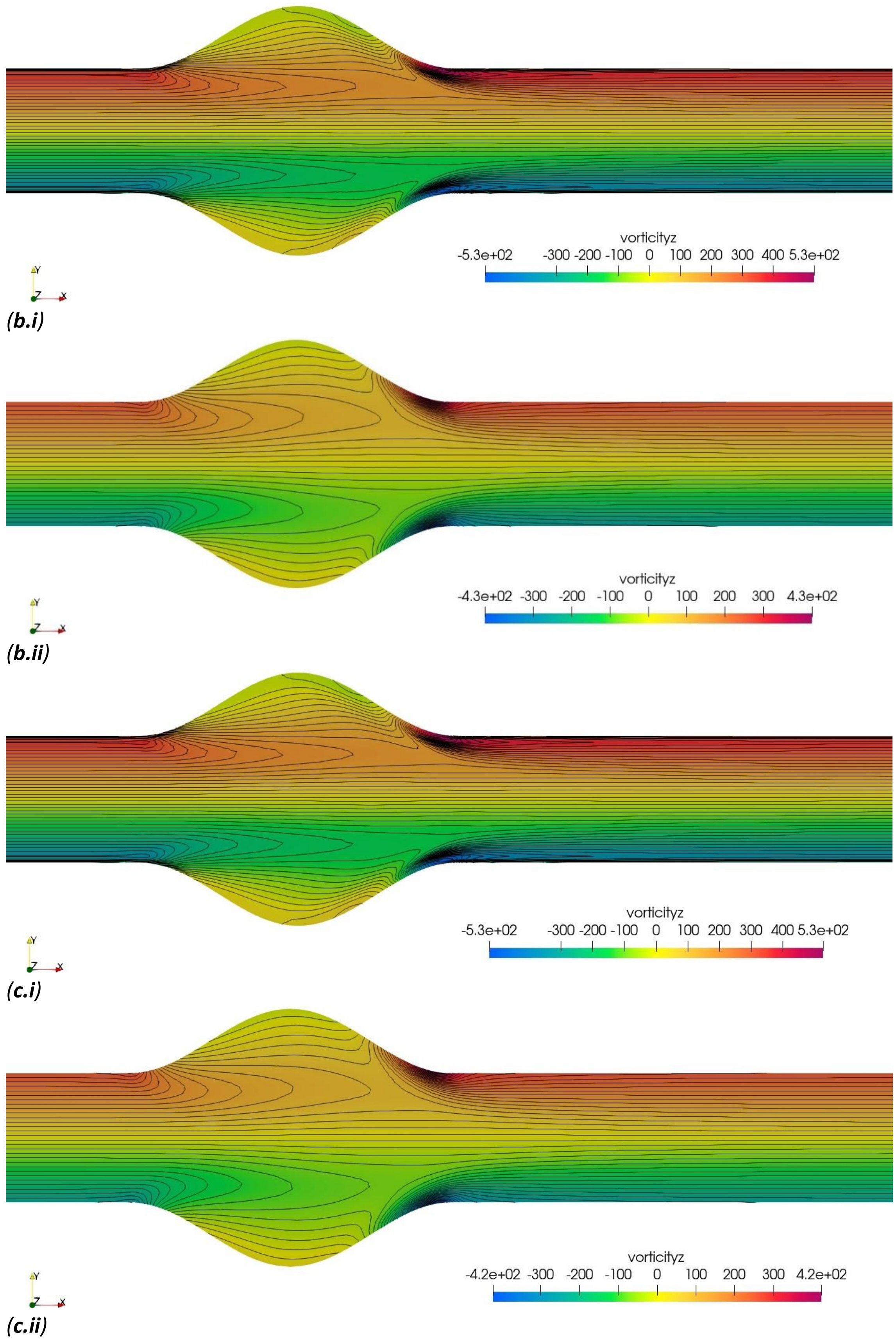

**Figure 28**. Vorticity contour plots for blood flow through a $200\%$ aneurysm using MHD micropolar modeling (i) without acknowledging MMR and (ii) considering MMR for an applied magnetic field of (a) 1 $T$, (b) 3 $T$, and (c) 8 $T$.

Figure 29 presents the microrotation contour plots for the 200% aneurysm for the MHD micropolar blood flow, both with and without considering the MMR effect. Again, the strength of the applied magnetic field is varied at 1 $T$, 3 $T$, and 8 $T$. As expected, the microrotation isolines for non-magnetic micropolar and MHD micropolar blood flow show insignificant differences when the MMR term is not taken into account, regardless of the magnetic field intensity. However, the MMR case shows a dramatic suppression of microrotation, with the peak values reduced to approximately 99.2% at $8\ T$. The spatial extent of the microrotation field is significantly compressed. The effect is more severe in the 200% aneurysm compared to the 166% case, as the enlarged geometry permits stronger vortex development that is then more forcefully damped by the MMR. These results confirm once again that the erythrocytes align with the externally applied magnetic field, preventing any internal rotation.

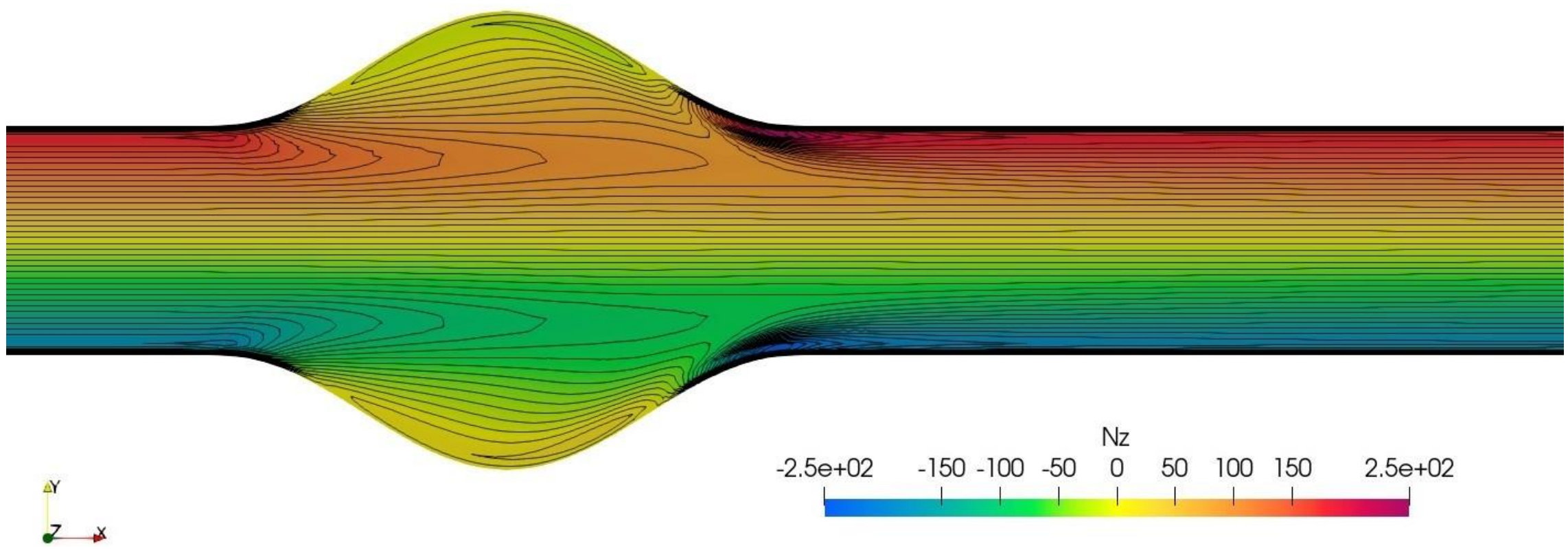

*(a.i)*

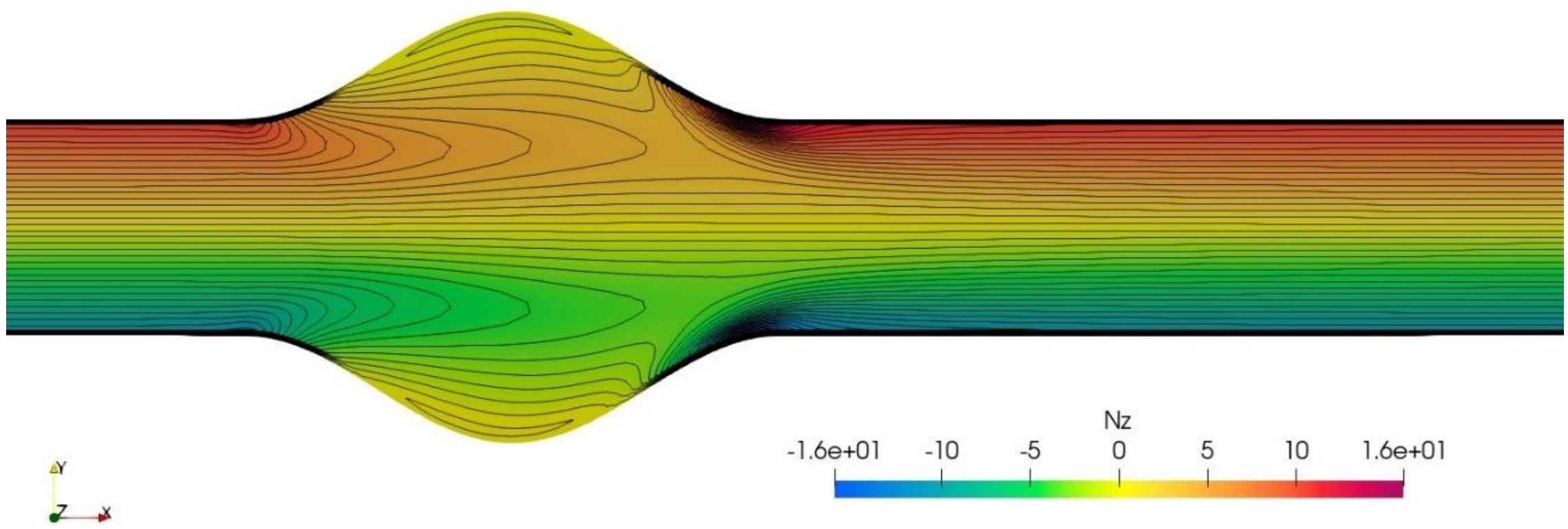

*(a.ii)*

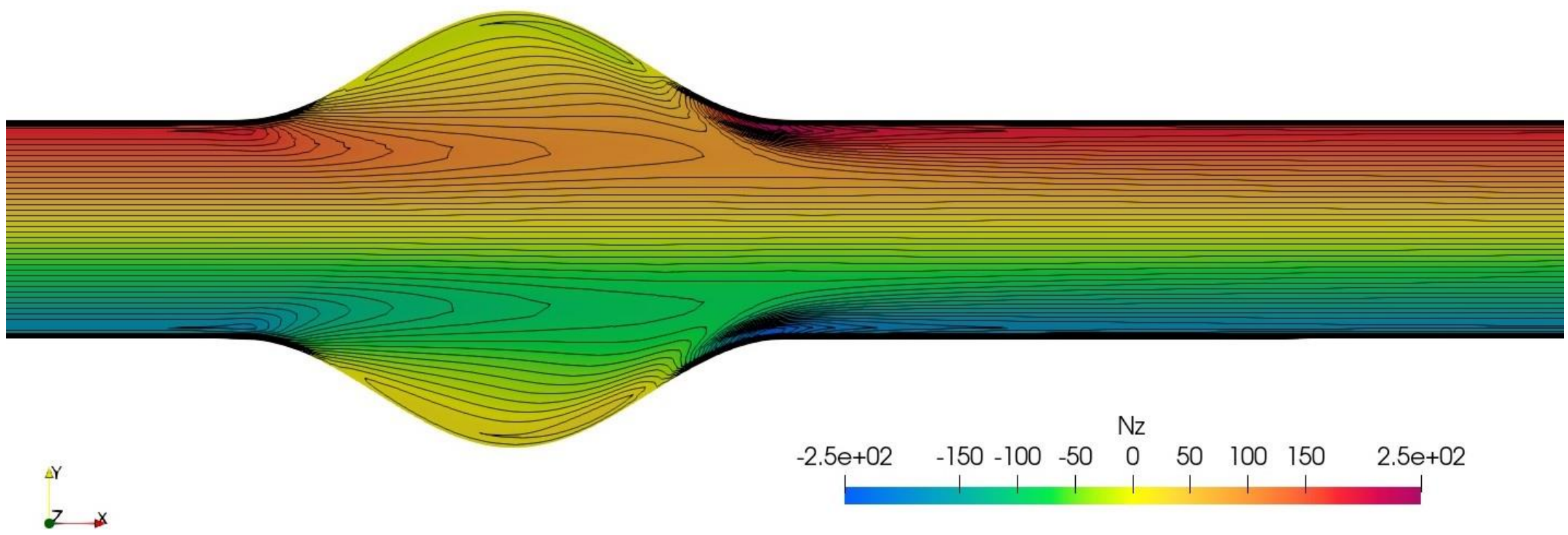

*(b.i)*

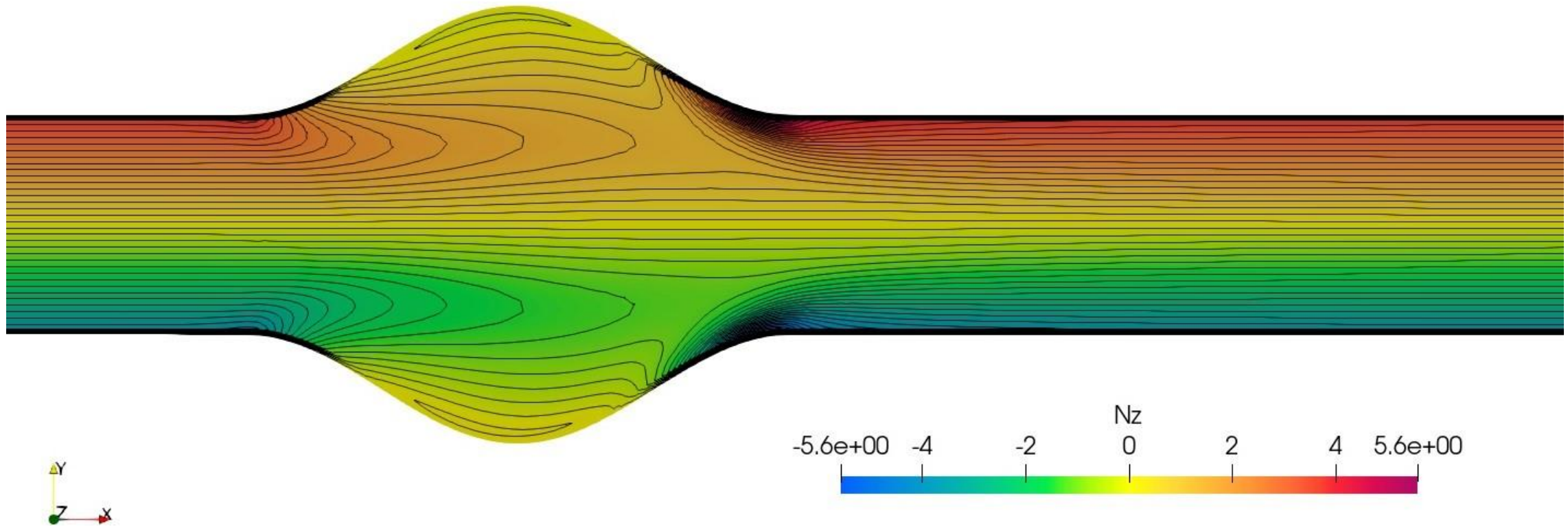

*(b.ii)*

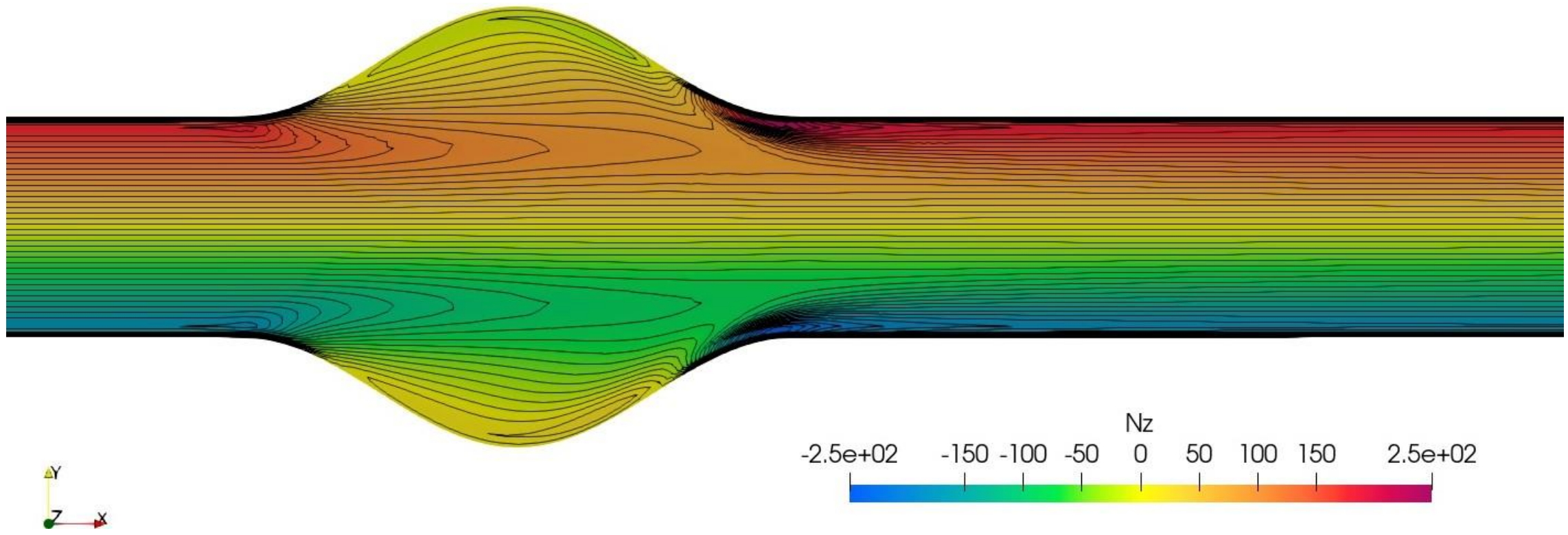

*(c.i)*

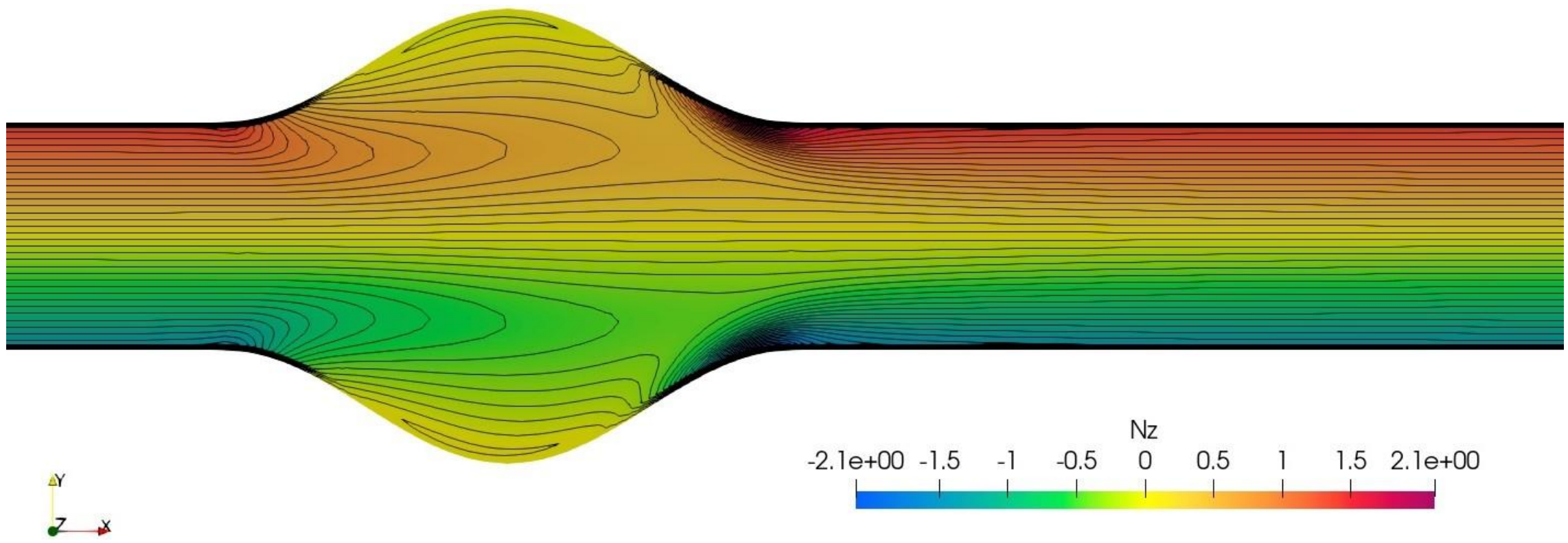

*(c.ii)*

**Figure 29**. Microrotation contour plots for blood flow through a $200\%$ aneurysm using MHD micropolar modeling (i) without acknowledging MMR and (ii) considering MMR for an applied magnetic field of (a) 1 $T$, (b) 3 $T$, and (c) 8 $T$.

Figure 30 illustrates the velocity profiles for the $200\%$ aneurysm, both within the aneurysm sac and downstream at $l = 0.05\ m$. The profiles correspond to four cases: Newtonian blood flow, micropolar blood flow, MHD micropolar blood flow without the MMR effect, and MHD micropolar blood flow with the MMR effect included. The strength of the applied magnetic field is varied 1 $T$, 3 $T$, and 8 $T$. Similar to the 166% aneurysm, the velocity exhibits an oscillatory profile due to the presence of the recirculating vortices inside the aneurysm sac. Downstream of the latter, the velocity profile recovers its parabolic shape. As with the 166% aneurysm, the differences in velocity between the non-magnetic Newtonian and micropolar profiles remain minor—both at the center and downstream of the aneurysm sac—with a maximum velocity reduction of 7%. Similarly, no notable differences are observed between the velocity profiles of the non-magnetic micropolar flow and the MHD micropolar flow without the MMR effect, across all applied magnetic field strengths. In contrast, when the MMR effect is taken into account, the velocity profiles are significantly dampened and structurally compressed, with a maximum velocity reduction of 42.9% observed at $8\ T$.

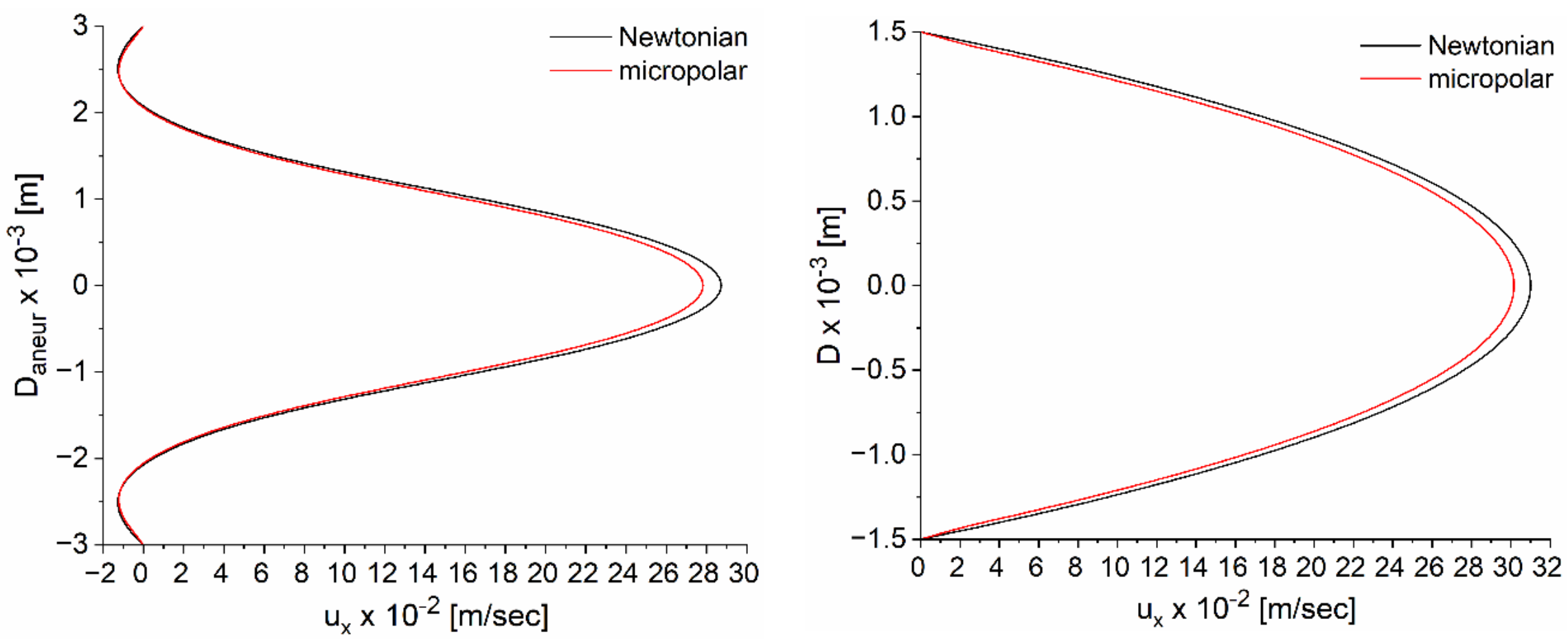

***(a)***

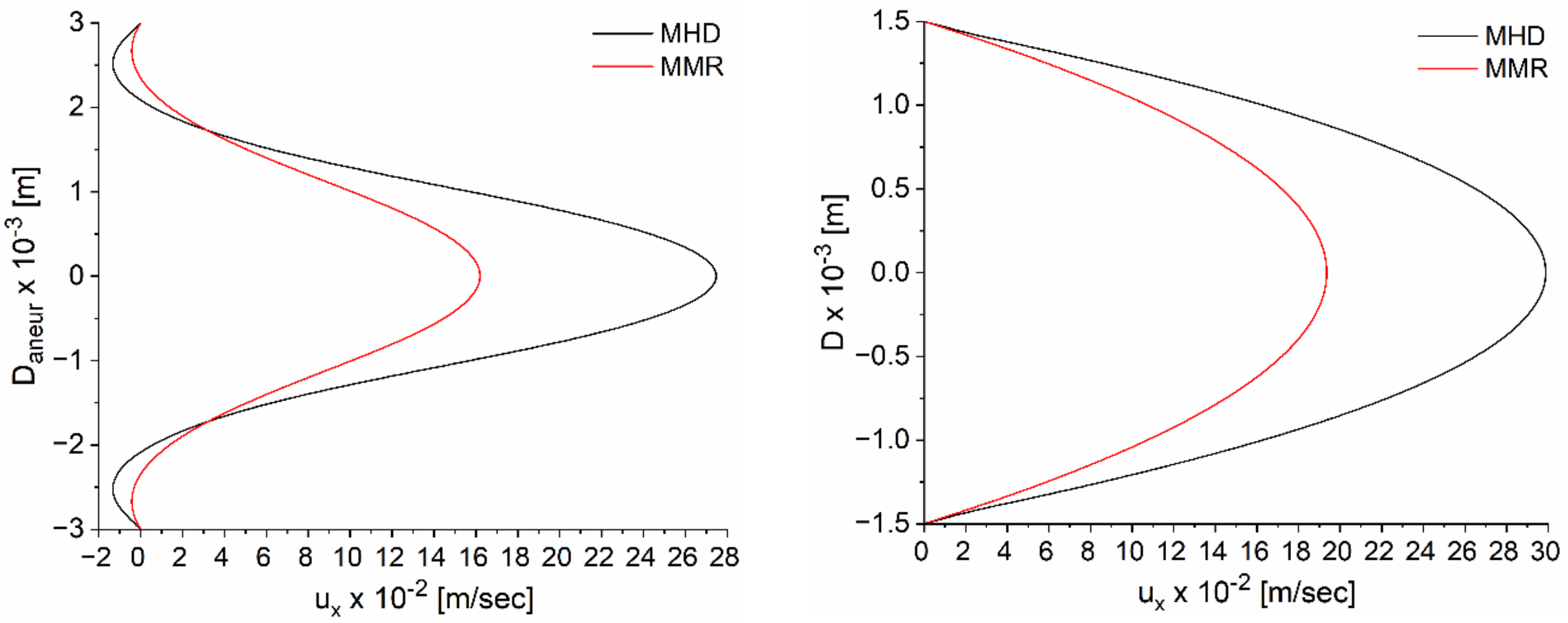

***(b)***

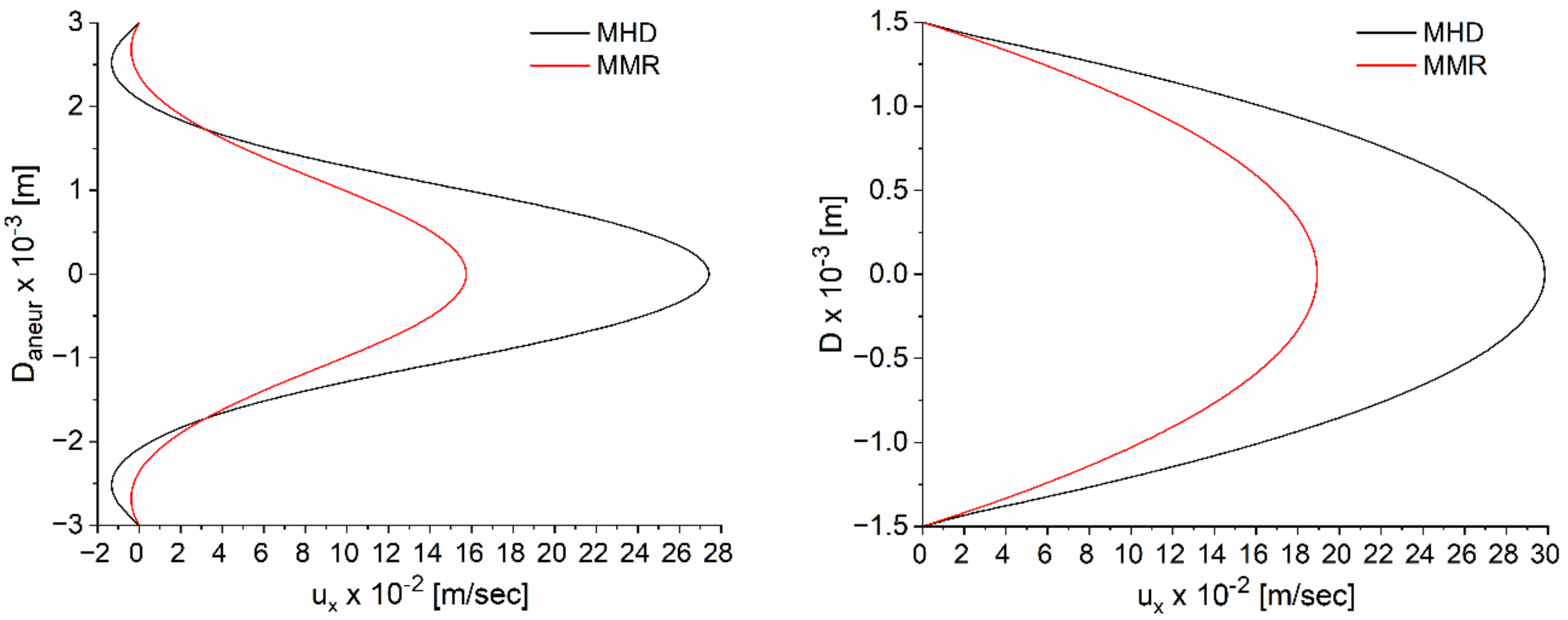

***(c)***

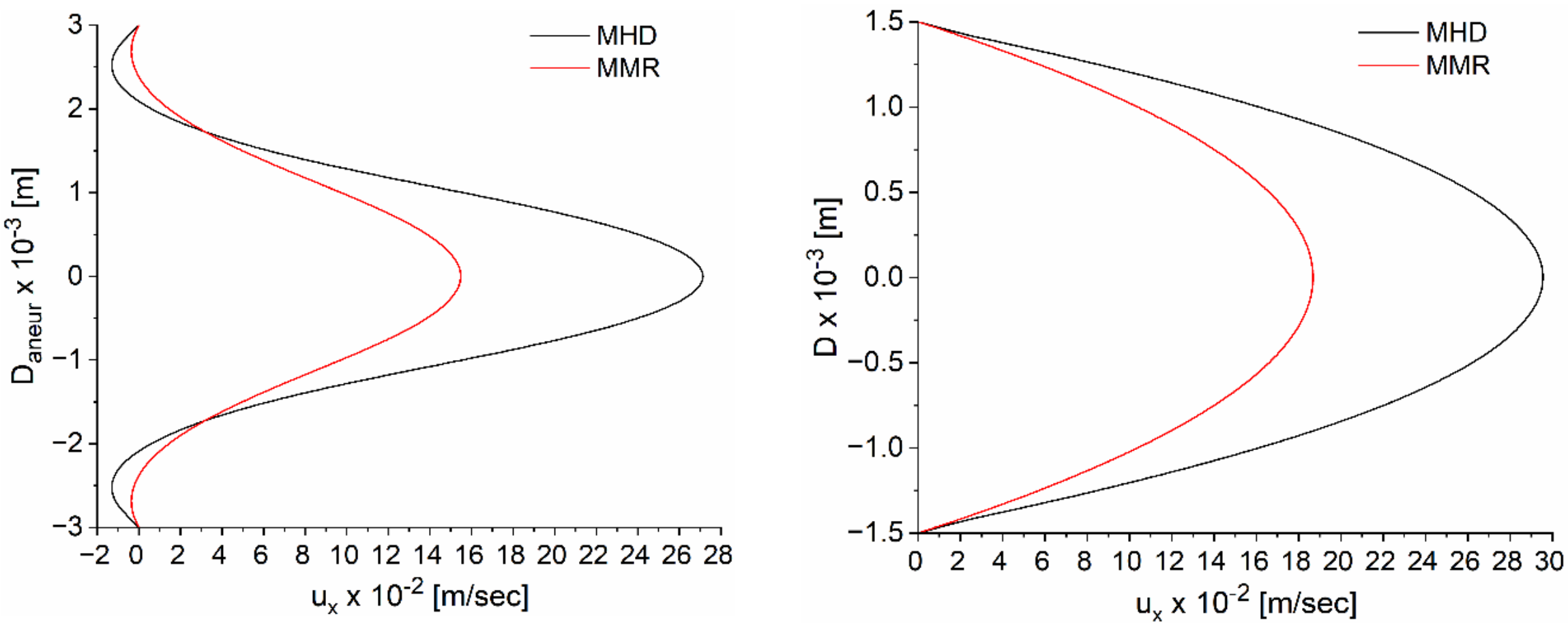

*(d)*

**Figure 30**. Velocity profiles at the center of the aneurysm sac (left) and downstream the latter (right) with a 200% aneurysm for (a) the non-magnetic Newtonian and micropolar blood flows, and (b) the MHD micropolar blood flow by ignoring and considering the MMR term with a magnetic field of 1 $T$ , (c) 3 $T$ and (d) 8 $T$.

Figure 31 presents the microrotation profiles for the 200% aneurysm, captured both within the aneurysm sac and downstream at $l = 0.05\ m$. As previously, the profiles correspond to four flow cases: Newtonian blood flow, micropolar blood flow, MHD micropolar blood flow without the MMR effect, and MHD micropolar blood flow including the MMR effect. The applied magnetic field strength is varied across three values: 1 $T$, 3 $T$, and 8 $T$. Similar to the 166% aneurysm, inside the 200% aneurysm sac, the microrotation profile exhibits a more oscillatory pattern, mirroring the behavior of the velocity profile, due to the presence of recirculating flow structures. Downstream of the sac, the microrotation profile returns to its characteristic shape, typically associated with parabolic flow profiles. As expected, the application of the external magnetic field without incorporating the MMR effect leads to no significant changes in the microrotation profile, owing to the minimal influence of the Lorentz force on the flow. In contrast, when the MMR effect is included, microrotation is drastically suppressed both within and beyond the aneurysm sac, with reductions reaching nearly 99.99% under the 8 $T$ magnetic field. Once again, it verified that the internal rotation of erythrocytes is almost entirely "frozen," as they tend to align parallel to the direction of the applied magnetic field.

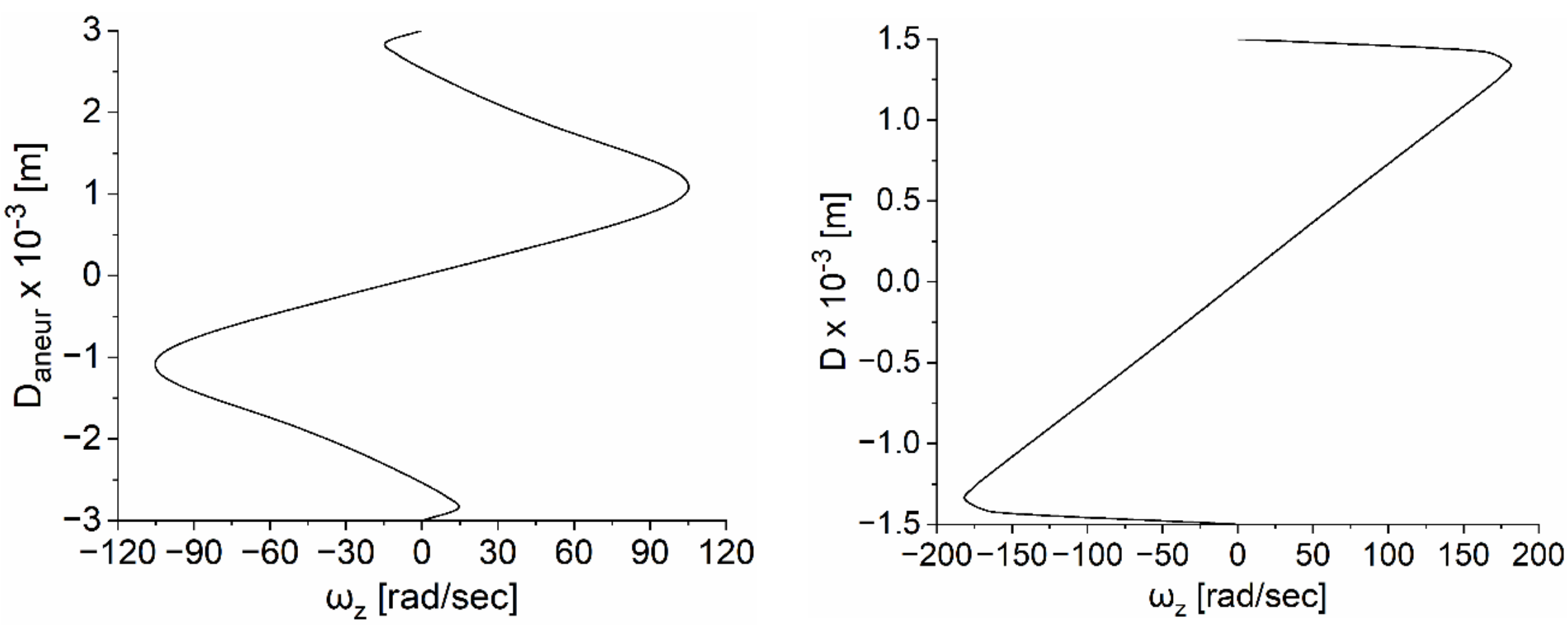

*(a)*

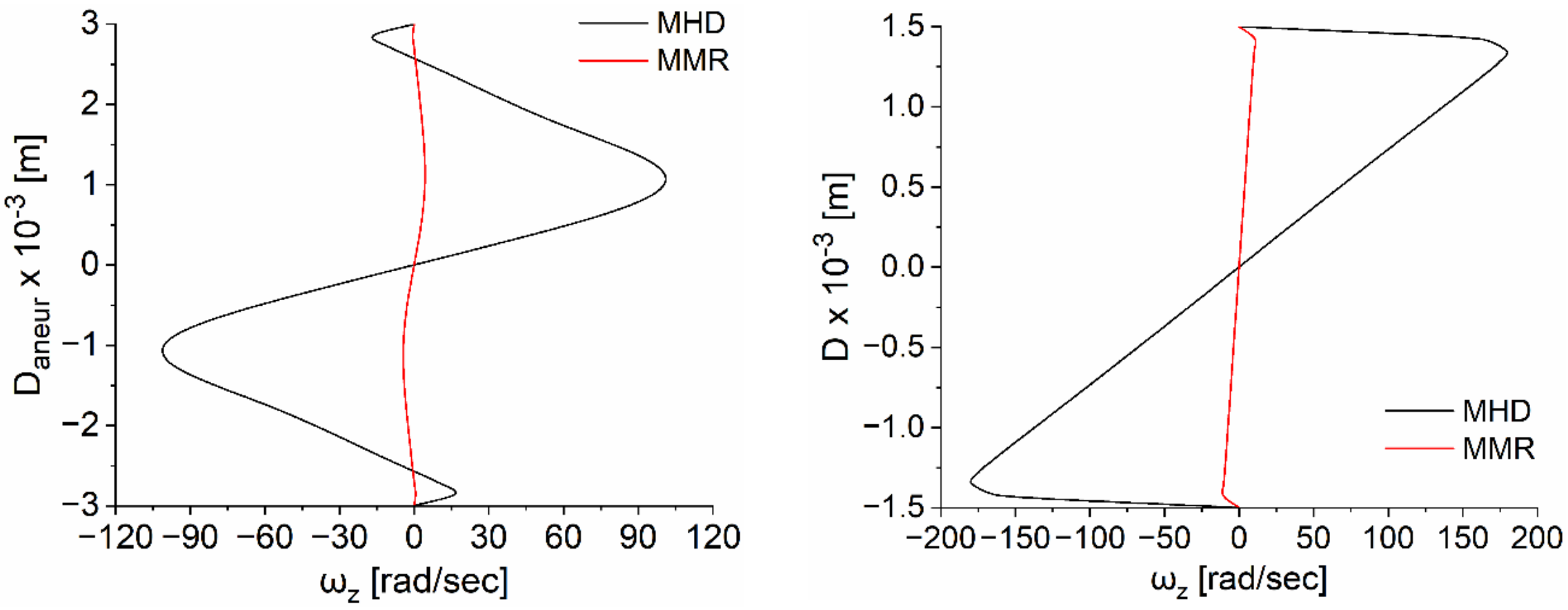

*(b)*

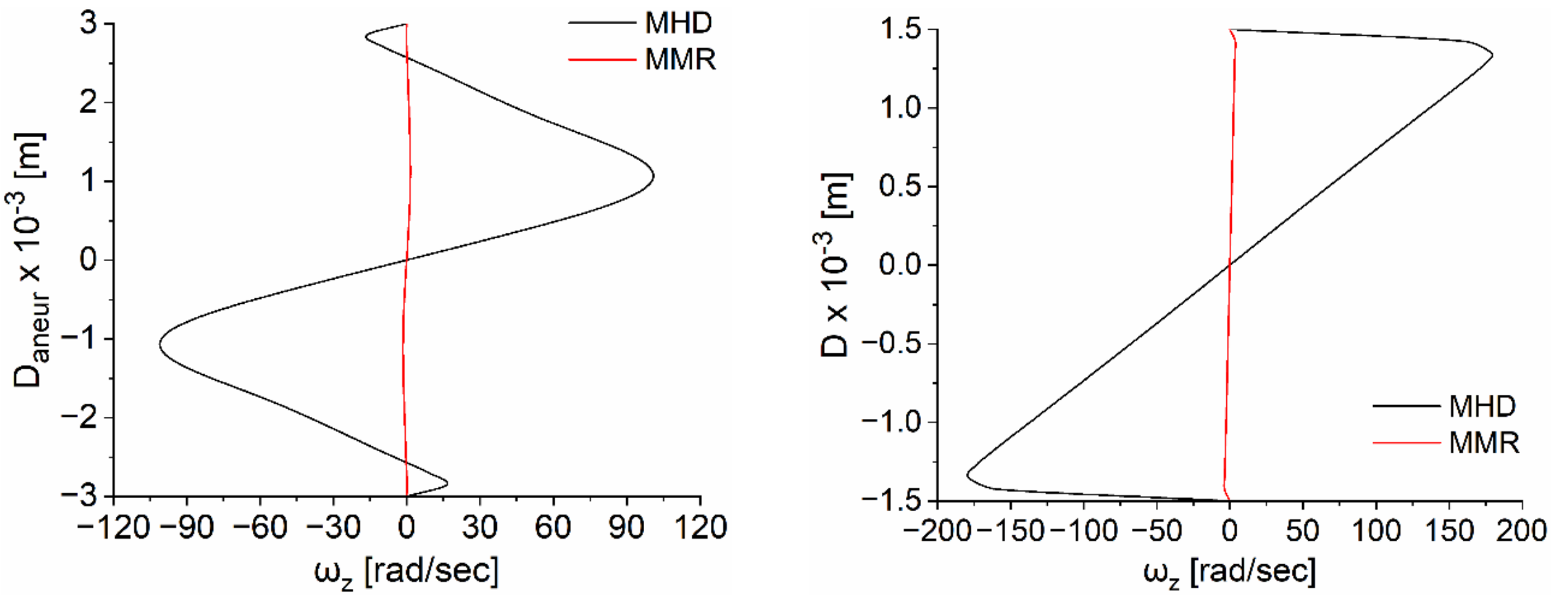

*(c)*

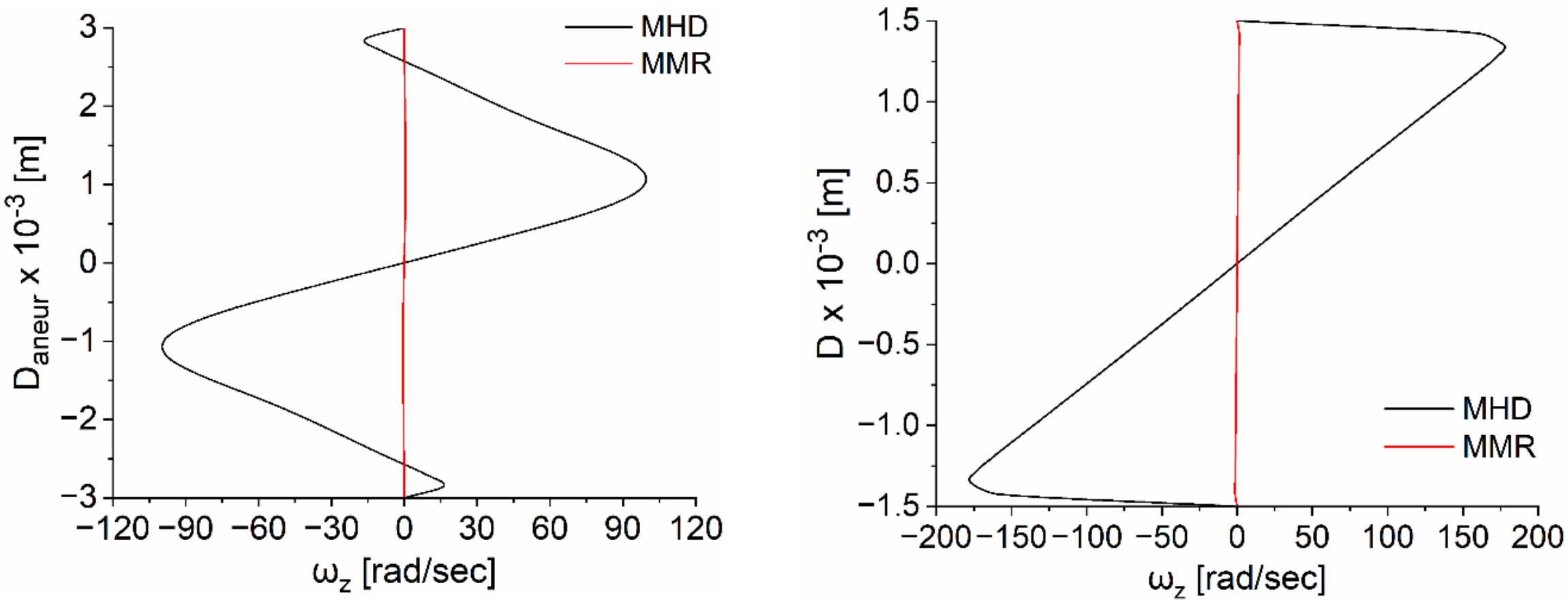

*(d)*

**Figure 31**. Microrotation profiles at the center of the aneurysm sac (left) and downstream the latter (right) with a 200% aneurysm for (a) the non-magnetic micropolar blood flow, and (b) the MHD micropolar blood flow by ignoring and considering the MMR term with a magnetic field of 1 $T$ , (c) 3 $T$ and (d) 8 $T$.

It should be mentioned that the results of this study highlight the strong influence of micromagnetorotation on MHD micropolar blood flows, especially under high magnetic field intensities and elevated hematocrit values. Physically, MMR suppresses the internal rotation of erythrocytes and increases the effective resistance of the flow, leading to noticeable reductions in velocity and microrotation. These findings may be important for biomedical applications such as targeted drug delivery, microfluidic manipulation of blood, and magnetic control of cellular suspensions. However, the present work also has several limitations. First, the simulations are restricted to laminar, steady-state flows; therefore, turbulent or pulsatile physiological conditions are not examined. Second, the model assumes rigid microparticles and does not account for erythrocyte deformability, aggregation, or cell–cell interactions. Third, the magnetic force term is neglected, which is a common assumption for weakly conducting fluids such as blood, but may not be accurate in cases involving contrast agents or magnetic nanoparticles. Finally, patient-specific geometries and non-Newtonian viscosity models were not considered. These aspects will be addressed in future studies using the developed solvers.

## 5. Conclusions and future perspectives

This work presented the development of two OpenFOAM solvers for simulating MHD micropolar flows with and without the effect of micromagnetorotation (MMR). The main findings of this study are summarized below:

- **Two new solvers were developed:** *epotMicropolarFoam* for classical MHD micropolar flows, *epotMMRFoam* for MHD micropolar flows including the MMR term and the magnetization equation.
  Both solvers use the PISO algorithm and the low-magnetic-Reynolds-number approximation.
- **Micropolar effects were incorporated** by adding the microrotation–vorticity difference in the momentum equation and solving the internal angular-momentum equation.
- **MMR was included** in *epotMMRFoam* through the magnetic torque term and the constitutive magnetization model of Shizawa and Tanahashi.
- **Validation against analytical solutions** of MHD micropolar Poiseuille blood flow showed excellent accuracy for all cases. The error for velocity and microrotation did not exceed 2%.
- **Physical behavior with MMR:** Including MMR resulted in strong reductions of velocity (up to 40%) and microrotation (up to 99.9%) at high magnetic fields and hematocrit values. Without MMR, the magnetic field had almost no influence, which agrees with the low electrical conductivity of blood.
- **3D artery results:** MMR caused a significant decrease in velocity and nearly complete suppression of microrotation. Without MMR, the Lorentz force remained weak and the magnetic field produced only small changes.
- **2D aneurysm results:** larger aneurysms and higher magnetic fields increased recirculation and vorticity in the absence of MMR. When MMR was included, these recirculation zones were strongly reduced, showing the stabilizing effect of magnetic alignment of the microparticles.
- **Biomedical relevance:** The solvers can be useful tools for studying magnetic control of blood flow, magnetic drug targeting, and other biomedical applications involving micropolar and magnetic fluids.

Future work will focus on extending the solvers to pulsatile flow, non-Newtonian viscosity models, deformable particles, and patient-specific geometries.

## CRediT authorship contribution statement

K.-E. Aslani: Writing -- original draft, Writing -- review & editing, Visualization, Validation, Software, Methodology, Conceptualization. I. Sarris: Supervision, Project administration. E. Tzirtzilakis: Writing -- review & editing, Project administration, Funding acquisition.

## Declaration of competing interest

The authors declare that they have no known competing financial interests or personal relationships that could have appeared to influence the work reported in this paper.

## Acknowledgments

This research was funded by the Action "Flagship actions in interdisciplinary scientific fields with a special focus on the productive fabric," implemented through the National Recovery and Resilience Fund Greece 2.0 and funded by the European Union–NextGenerationEU (Project ID: TAEDR-0535983).

## Data availability

Data will be made available on request.